\documentclass[aps,prb,reprint,twocolumn,showpacs,superscriptaddress]{revtex4-1}

\usepackage{amsmath, amssymb, braket, dsfont, units}
\usepackage{graphicx}
\usepackage{subfigure}		
\usepackage{epstopdf, times}

\usepackage[colorlinks=true,linkcolor=blue,citecolor=blue,breaklinks]{hyperref}
\usepackage{color}

\definecolor{Zcolour}{rgb}{0.992, 0.588, 0.22}
\definecolor{purple}{rgb}{0.5,0,0.5}
\definecolor{brown}{rgb}{0.6,0.2,0}
\definecolor{dkgreen}{rgb}{0,0.5,0}
\newcommand{\authorcomment}[3]{{ \color{#1} \footnotesize (\textsf{#2}) \textsf{\textsl{#3}} }}
\newcommand{\alttext}[2]{{\color{#1}#2}}

\newcommand{\roger}[1]{\authorcomment{purple}{RM}{#1}}
\newcommand{\maxm}[1]{\authorcomment{brown}{MM}{#1}}

\newcommand{\lesik}[1]{\authorcomment{red}{OM}{#1}}
\renewcommand{\authorcomment}[3]{} \renewcommand{\alttext}[2]{}

\newcommand{\norder}[1]{{\mkern1mu\colon\mkern-4mu{#1}\colon\mkern-3mu}}
\newcommand{\aHLR}{\ensuremath{\overline{\textrm{HLR}}}}
\newcommand{\PH}{\mathord{\mathcal{P}\mkern-4mu\mathcal{H}}}
\newcommand{\TR}{{\mathcal{T}}}		
\newcommand{\CC}{{\mathcal{C}}}		
\newcommand{\bq}{{\mathbf{q}}}
\newcommand{\bk}{{\mathbf{k}}}
\newcommand{\br}{{\mathbf{r}}}
\newcommand{\up}{\uparrow}
\newcommand{\dn}{\downarrow}
\newcommand{\Hc}{\mathrm{H.c.}}
\newcommand{\twokF}{\texorpdfstring{$2k_F$}{2k\_F} }		
\newcommand{\Norb}{N_{\text{orb}}}

\begin{document}

\title{The half-filled Landau level: the case for Dirac composite fermions}

\author{Scott D. Geraedts}
\affiliation{Department of Physics and Institute for Quantum Information and Matter, California Institute of Technology, Pasadena, CA 91125, USA}
\author{Michael P. Zaletel}
\affiliation{Station Q, Microsoft Research, Santa Barbara, CA 93106, USA}
\author{Roger S. K. Mong}
\affiliation{Department of Physics and Astronomy, University of Pittsburgh, Pittsburgh, PA 15260, USA}
\affiliation{Department of Physics and Institute for Quantum Information and Matter, California Institute of Technology, Pasadena, CA 91125, USA}
\affiliation{Walter Burke Institute for Theoretical Physics, Caltech, Pasadena, CA 91125 USA}
\author{\mbox{Max A. Metlitski}}
\affiliation{Kavli Institute for Theoretical Physics, UC Santa Barbara, CA 93106, USA}
\author{Ashvin Vishwanath}
\affiliation{University of California, Berkeley, CA 94720, USA}
\affiliation{Materials Science Division, Lawrence Berkeley National Laboratories, Berkeley, CA 94720, USA}
\author{Olexei I. Motrunich}
\affiliation{Department of Physics and Institute for Quantum Information and Matter, California Institute of Technology, Pasadena, CA 91125, USA}
\affiliation{Walter Burke Institute for Theoretical Physics, Caltech, Pasadena, CA 91125 USA}

\begin{abstract}
One of the most spectacular experimental findings in the fractional quantum Hall effect is evidence for an emergent Fermi surface when the electron density is nearly half the density of magnetic flux quanta ($\nu = 1/2$).
The seminal work of Halperin, Lee, and Read (HLR) first predicted that at $\nu = 1/2$ composite fermions---bound states of an electron and a pair of vortices---experience zero net magnetic field and can form a ``composite Fermi liquid'' with an emergent Fermi surface.
In this paper we use infinite cylinder DMRG to provide compelling numerical evidence for the existence of a Fermi sea of composite fermions for realistic interactions between electrons at $\nu = 1/2$.
Moreover, we show that the state is particle-hole symmetric, in contrast to the construction of HLR.
Instead, our findings are consistent if the composite fermions are massless Dirac particles, at finite density, similar to the surface state of a 3D topological insulator.
Exploiting this analogy we devise a numerical test and successfully observe the suppression of $2k_F$ backscattering characteristic of  Dirac particles.
\end{abstract}

\maketitle

\phantomsection
\addcontentsline{toc}{part}{Main text}

Electrons confined to a two-dimensional (2D) plane in a strong magnetic field organize into a plethora of remarkable phases in which electron correlations play the key role.
A global understanding of these phases is provided by topological excitations called composite fermions (CF).\cite{JainCF89, LopezFradkin91} 
One of the most intriguing phases occurs in the half-filled
($\nu = \frac12$)
Landau level, in which composite fermions were predicted by Halperin, Lee, and Read (HLR) to form a metallic state.\cite{HalperinLeeRead}
Experiments corroborated this understanding by identifying signs of a Fermi surface in spite of the intense magnetic field.\cite{WillettFS93, Kang93, Goldman94, Smet96}
Despite the tremendous success of the HLR theory, one aspect of the $\nu=\frac12$ phase has remained an enigma: a particle-hole (PH) symmetry of the quantum Hall problem projected into the spin-polarized lowest Landau level (LLL).\cite{RezayiRead:HalfFilledLL:1994, Kivelson1997, Lee1998, PasquierHaldane1998, Read1998, MurthyShankarRMP}
The HLR theory attaches fluxes to electrons rather than holes, and is not confined to the LLL,  breaking PH symmetry.
One possibility is that PH symmetry is spontaneously broken,\cite{Barkeshli2015}
	as is believed to occur in the 1\textsuperscript{st} excited Landau level (e.g. $\mathrm{GaAs}$ at filling $\nu = \frac52$),\cite{Eisenstein87, Morf1998, RezayiHaldane:PH:00, Papic2012}
	leading to the Moore-Read phase.\cite{MooreRead:Nonabelion:1991}

	After two decades of study, theorists have proposed a radical twist to this picture, forging deep connections with the physics of three-dimensional (3D) topological insulators (TI).\cite{DSon:CFL2015, CWangSenthilDDL2015, MetlitskiVishwanath2015}
In the new picture due to Son,\cite{DSon:CFL2015} composite fermions are massless Dirac particles, similar to electrons on the surface of a TI.
The CFs are coupled to an emergent gauge field, so the new theory retains the non-Fermi-liquid aspects of HLR---in fact, this ``Dirac-CFL'' is equivalent to finite-density QED$_3$.
In a TI, the masslessness of the Dirac fermions is protected by the Kramers time-reversal symmetry.
Here time-reversal is absent, but the PH symmetry that trades occupied and unoccupied electronic states at $\nu = \frac12$ plays an equivalent role.
It is proposed that when PH acts on the CFs, it behaves exactly like Kramers time-reversal symmetry, exchanging states on opposite points of the Fermi surface and forbidding a CF mass term.\cite{DSon:CFL2015}
Rather than an enigma, PH symmetry now plays a starring role.


We address the following two questions.
First, is the half-filled Landau level consistent with a composite Fermi liquid, either in its original form or the Dirac revision?
Second, is PH symmetry preserved, and what are the measurable effects if the composite fermions are indeed Dirac particles?
Using large scale  density matrix renormalization group (DMRG) numerical simulations,\cite{White92} we provide strong evidence for the formation of a CFL in the LLL with realistic Coulomb interactions.
We observe a single, nondegenerate Fermi surface consistent with a  Luttinger count of CFs, and signatures of an emergent gauge field.
In the idealized limit of a single Landau level,  particle-hole symmetry is preserved.
Combining these numerical observations with new theoretical insights into the microscopic action of PH, we argue this is logically sufficient to conclude that the CF is a Dirac fermion.
However, to demonstrate this point directly, we test for the analog of a classic signature of the TI surface---suppression of $2k_F$ backscattering off impurities that preserve time-reversal symmetry.\cite{RoushanHasan:TopSurface09, TZhangQKXue:QpInterference09}
We indeed find that backscattering off PH-symmetric potentials is suppressed in the CFL, reappearing only when PH-breaking perturbations are introduced.
We also find that the effect of the small PH-breaking perturbations present in experiments is very weak, suggesting that PH symmetry is experimentally relevant at $\nu = \frac12$.

\section{Model and methods}
\label{sec:methods}

We study electrons on an \emph{infinitely} long cylinder, in an external magnetic field $B$, interacting via Coulomb repulsion.
There are several advantages to the infinite cylinder geometry: 
there are no edge effects, PH symmetry can break spontaneously, and there can be algebraic correlations.
For our numerical simulations,\cite{ZaletelMixing} we truncate the interaction as $V(r) = \frac{1}{r} e^{-r^2/(2\lambda^2)}$ and project into the lowest Landau level.
We assume the system is spin-polarized, which is observed to occur in the high field regime.\cite{Kukushkin99}
We fix $\lambda = 6\ell_B$, which is large enough to capture the Coulomb interaction physics but small enough to avoid wrapping effects on a cylinder.
Throughout, the magnetic length $\ell_B \equiv \sqrt{\hbar/(e B)}$ is set to $1$; thus lengths are given in multiples of $\ell_B$ and momenta are measured in units of $\ell_B^{-1}$.
We use the coordinate $x$ along the infinite length and $y$ around the finite circumference $L_y$ of the cylinder.

The states we study are critical and have algebraic correlations along the cylinder, as well as infinite bipartite entanglement.
However, infinite-DMRG always introduces a finite length scale ``$\xi$'' to the system due to the finite bond-dimension $\chi$ used in the DMRG variational ansatz.
Instead of the usual ``finite size scaling,'' we perform the so-called ``finite entanglement scaling'' by running DMRG at various bond dimensions $\chi \approx 10^2\mbox{--}10^4$, and extrapolating the results as $\chi \to \infty$.\cite{Tagliacozzo:2008, Pollmann:FiniteES:2009}
This way we can analyze the $\xi \to \infty$ limit and extract the critical properties of the system on the infinite cylinder.


We describe further details of our DMRG setup in Appendices~\ref{app:cylinder}~and~\ref{app:numerical}.
Here, we only note that the momentum $K_y$ around the cylinder is conserved in the DMRG, so it must be chosen correctly in order to find the ground state.
We find that the momentum $K_y$ of the ground state depends on the circumference $L_y$.
Only two such momenta appear, and for reasons explained in App.~\ref{app:numerical} we label these momentum sectors by ``0110'' and ``0101.''
Interestingly, as we show below, these sectors allow us to access different boundary conditions for the CFs, even though the CFs are emergent fields.

\section{Mapping the Fermi surface of emergent fermions}
\label{sec:FermiSurf}

\begin{figure}[tb]
	\includegraphics[width=\linewidth]{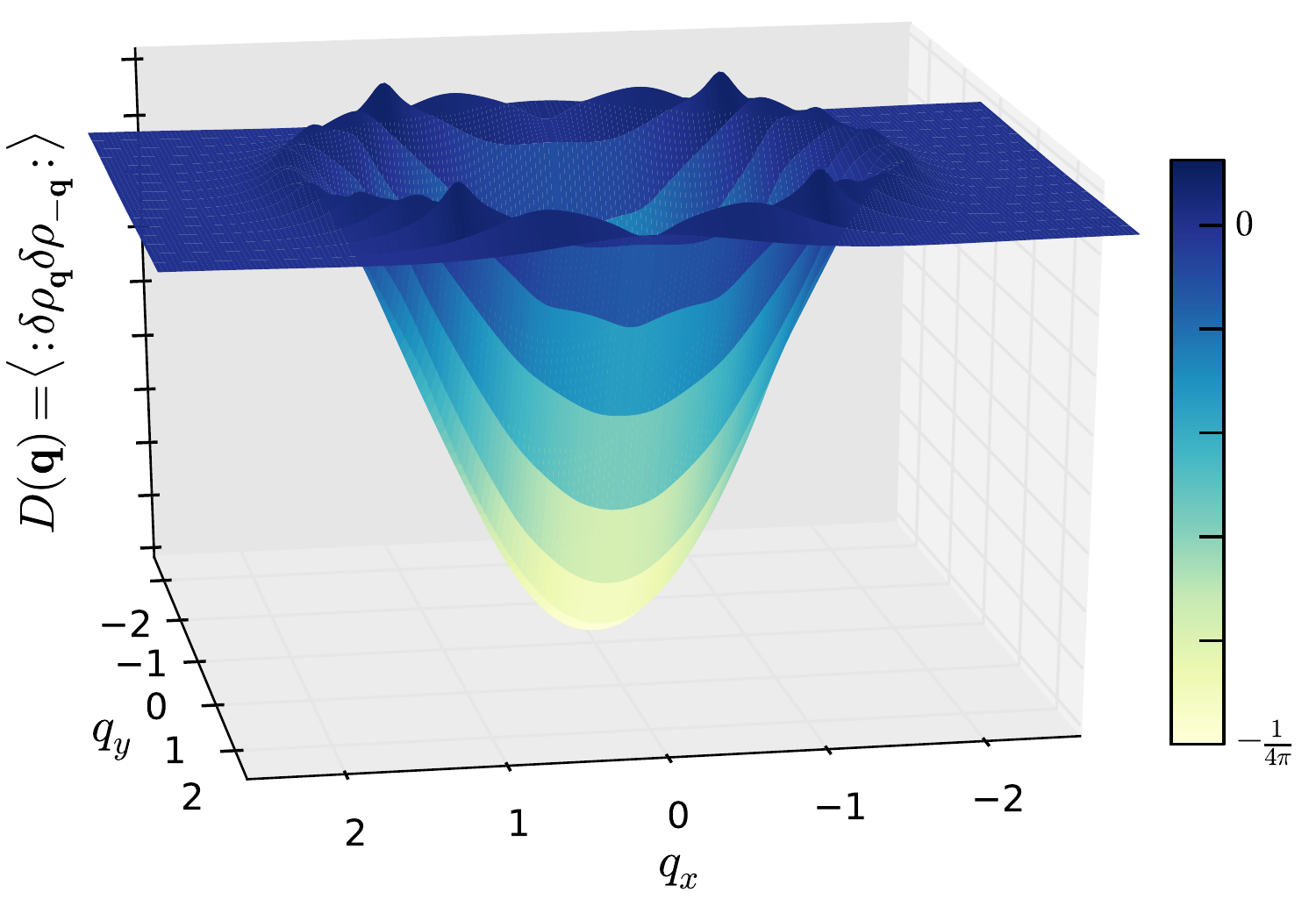}
	\caption{
		Density-density correlations $D(\bq) = \braket{\norder{\delta\rho_\bq \delta\rho_{-\bq}}}$ plotted as a function of 2D momenta $\bq = (q_x, q_y)$.
		The cylinder circumference is $L_y = 24$, where we see eight slices through the Fermi sea.
		Descendants of a singular $2k_F$ circle in $D(\bq)$ show up in cuts at $q_y$ between $-7 \big( \frac{2\pi}{L_y} \big)$ and $7 \big( \frac{2\pi}{L_y} \big)$ (see also Fig.~\ref{fig:boulder}).
		At this large circumference the correlations approach those of a 2D system.
	}
	\label{fig:cone}
\end{figure}

\begin{figure*}[ttt]
	\begin{minipage}{24mm}
		\includegraphics[width=24mm]{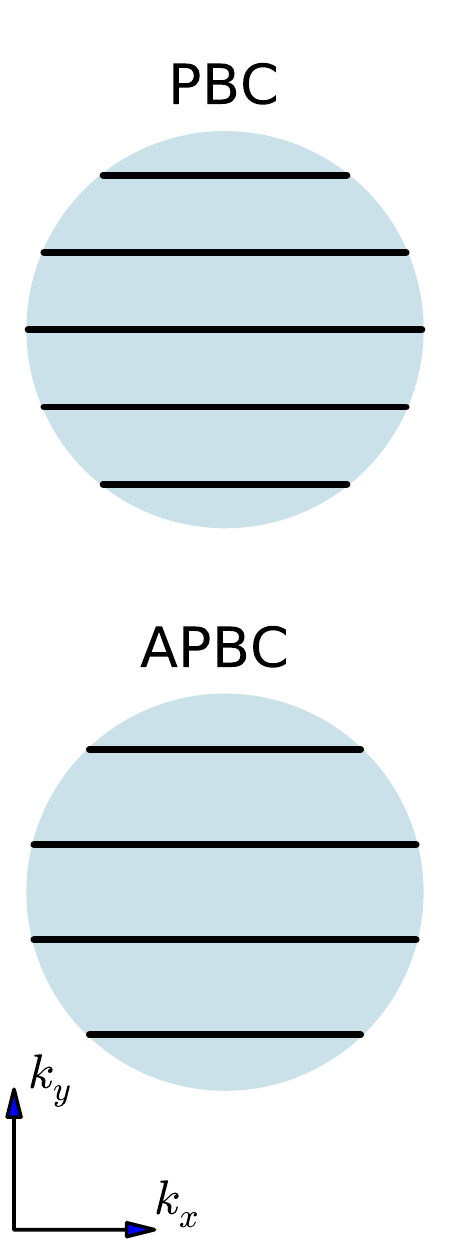}
		\\[-2ex]\hspace{-\textwidth}\hspace{0mm}\begin{minipage}{0mm}\vspace{-130mm}\subfigure[]{\label{fig:bc}}\hspace{-27mm}\end{minipage}
	\end{minipage}
	\hspace{1mm}
	\begin{minipage}{82mm}
		\includegraphics[width=82mm]{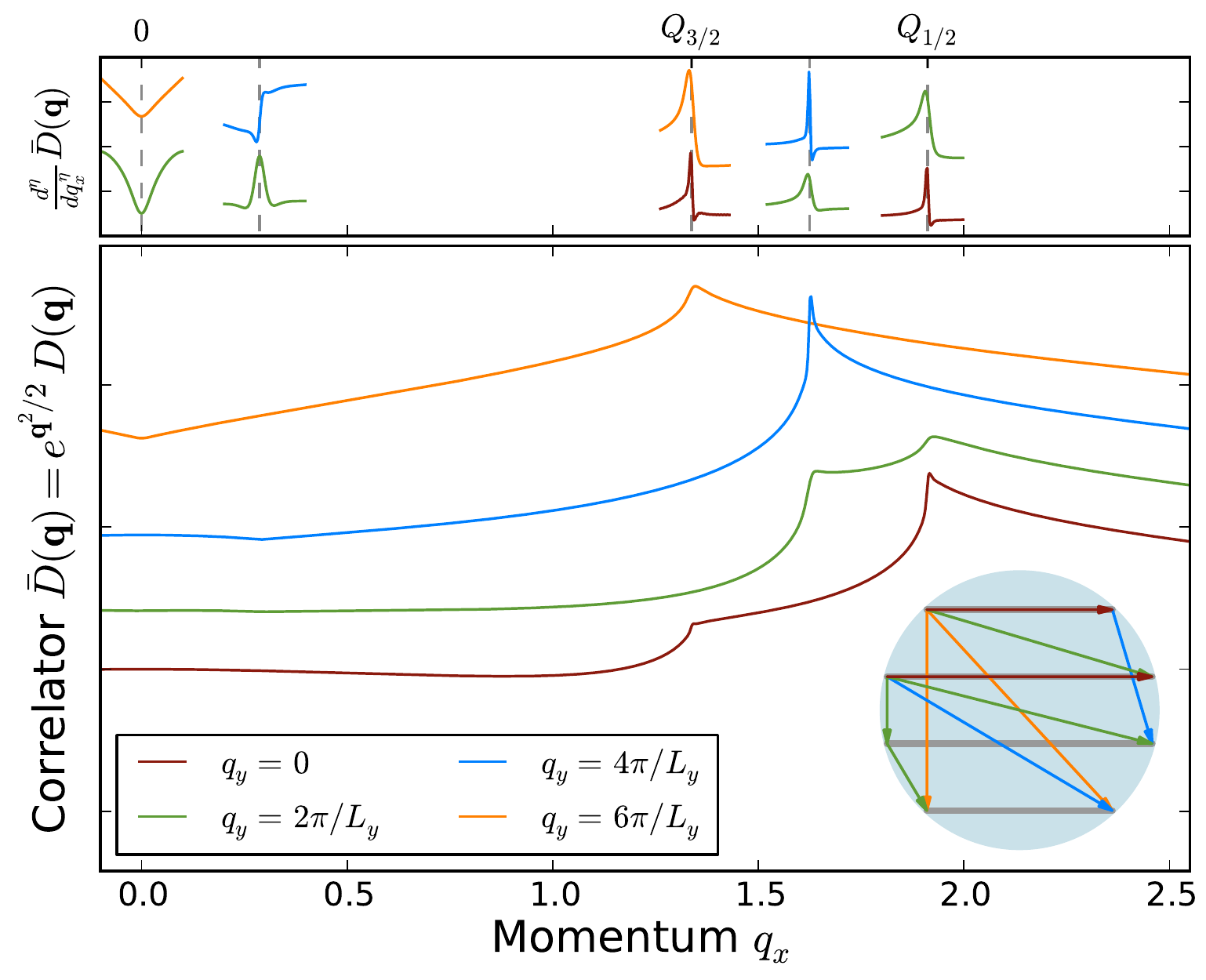}
		\\[-2ex]\hspace{-\textwidth}\hspace{1mm}\begin{minipage}{0mm}\vspace{-130mm}\subfigure[]{\label{fig:two_body}}\hspace{-27mm}\end{minipage}
	\end{minipage}
	\hspace{1mm}
	\begin{minipage}{64mm}
		\includegraphics[width=65mm]{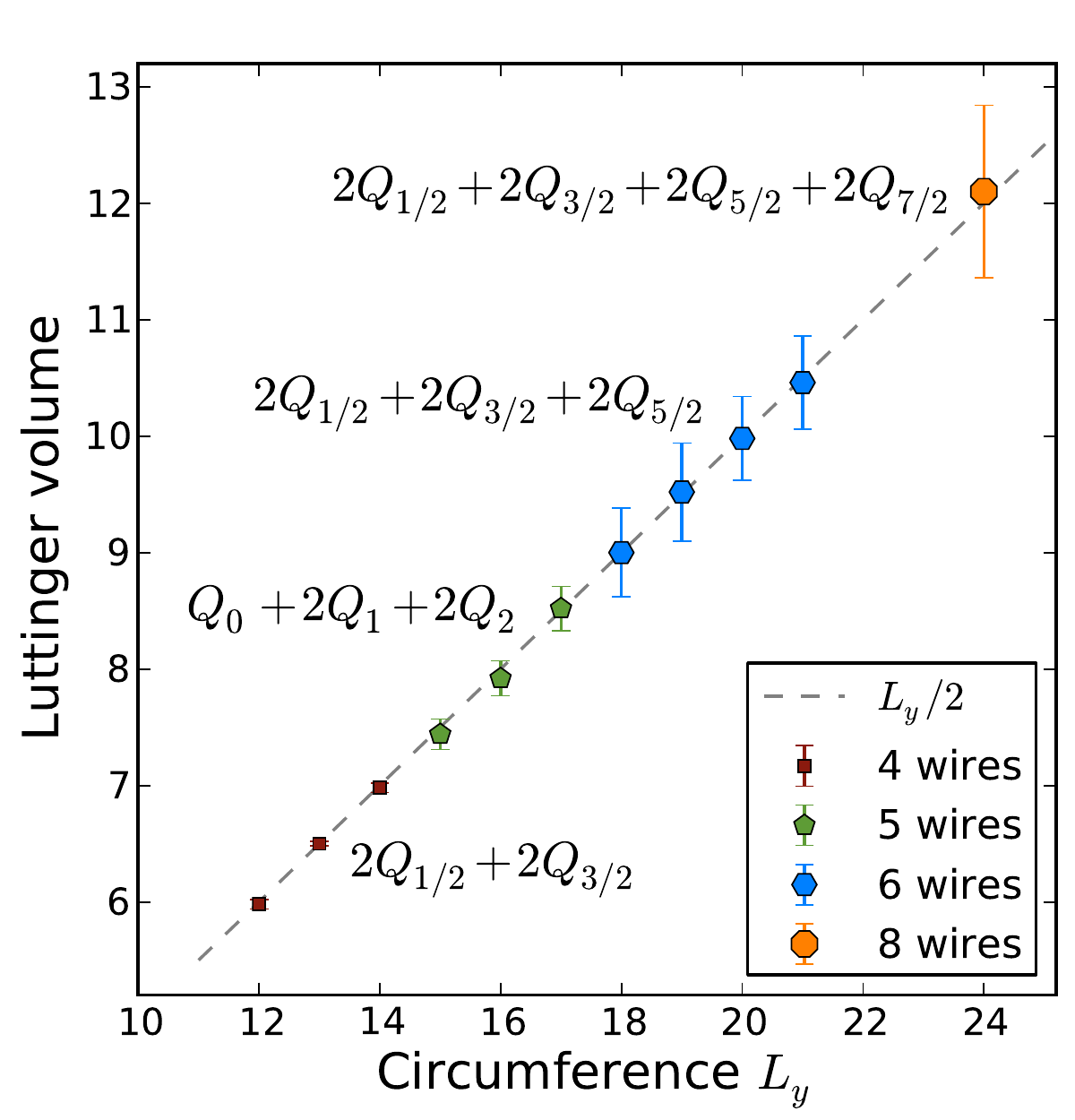}
		\\[-2ex]\hspace{-\textwidth}\hspace{2mm}\begin{minipage}{0mm}\vspace{-130mm}\subfigure[]{\label{fig:luttinger}}\hspace{-27mm}\end{minipage}
	\end{minipage}
	\caption{
	\textbf{(a)}~\textit{Boundary conditions on the composite fermions.}
		Our system can be described by a number of ``wires''---slices through the 2D Fermi sea---at various fixed $k_y$. 
		The number of wires is dictated by $L_y$ and the boundary condition (BC) of the composite fermions.
		An odd number of wires (upper panel) corresponds to PBC (periodic); an even number of wires (lower panel) corresponds to APBC (antiperiodic).
	\textbf{(b)}~\textit{Mapping the Fermi surface via the structure factor.}
		The lower panel shows the LLL density-density correlations
			$e^{\bq^2/2}D(\bq) = e^{\bq^2/2}\braket{\norder{\delta\rho_\bq \delta\rho_{-\bq}}}$ measured on a cylinder with $L_y = 13$.
		The singularities arise from CF scattering processes across the Fermi surface.
		The observed scatterings are illustrated in the inset, with colors corresponding to $q_y$.
		The upper panel shows the derivatives of the correlator, which aids in determining the location of the singularities.
	\textbf{(c)}~\textit{Testing Luttinger's Theorem.}
		Luttinger's theorem states that the area enclosed by the Fermi surface is related to the particle density.
		On a cylinder the ``area'' is given by the sum of the length of each wire in momentum space, which we determined from singularities in plots like Fig.~\ref{fig:two_body}.
		We define $Q_m$ to be the length of the Fermi sea slice at $k_y = \frac{2\pi}{L_y}m$, and plot the resulting sums for various circumferences against the Luttinger's  prediction.
		Note that we use the relation $Q_{-m} = Q_m$, a consequence of  rotation symmetry.
		There is excellent agreement between our data and the Luttinger count.
	}
	\label{fig:wires}
\end{figure*}

In the 2D limit, the composite fermions are expected to form a circular Fermi surface with $k_F = 1$ at $\nu = \frac12$, which we can study with our numerics.
While the cylinder is infinite in $x$, the finite $L_y$ quantizes the momenta $k_y$ into a discrete grid with spacing $2\pi/L_y$. 
Thus, instead of a circular Fermi surface, we expect a set of ``wires'' labeled by $k_y$, illustrated in Fig.~\ref{fig:bc}.
As $L_y$ is increased, the available $k_y$'s get closer together and the number of wires $N_w$ increases, better approximating the 2D system.

We get a sense of how well we approach the 2D limit in Fig.~\ref{fig:cone}, where we plot the electron density-density correlations $D(\bq) \equiv \braket{\norder{\delta\rho_\bq \delta\rho_{-\bq}}}$ as a function of momenta $\bq$ at $L_y = 24$.
[See App.~\ref{subapp:corr} for details on measuring $D(\bq)$.]
The density structure factor in the 2D limit is expected to be rotationally symmetric with a singularity on a circle of radius $2 k_F$.
While we only have access to data at discrete $q_y$, our measurements still approximate a circular shape, with distinct singular features near $|\bq| = 2k_F$.
One expects a singularity $D(\bq) \sim \big| |\bq|-2k_F \big|^{\alpha}$ with $\alpha = \frac32$ for free fermions and for the CFL with Coulomb interaction, but $\alpha$ is modified for the CFL with short-range interactions.
\cite{Polchinski94, Nayak_1, Nayak_2, Altshuler94, YBKim94, Mross2010}
Future studies approaching closer to the 2D regime should allow such predictions to be tested quantitatively.

We determine the number and lengths of the CF wires from $D(\bq)$.
The electron density will generically couple to CF scattering processes (allowed by the symmetries), i.e., 
$\delta\rho_\bq = \sum_\bk A_\bk \bar\psi_{\textrm{CF}; \bk} \psi_{\textrm{CF}; \bk + \bq} + \ldots$, where ellipses denote higher-body processes.
A transition of momentum $\bq$ across the Fermi surface will contribute a singularity to the structure factor $D(\bq)$.
Since $\bk$ and $\bk + \bq$ are restricted to the wires, the singularities in $D(\bq)$ can then be used to determine the configuration of the wires.
For example, the singularities in $D(q_x, q_y = 0)$ contain transitions within the wires (i.e., fixed $k_y$), and reveal the lengths of the wires inside the Fermi sea.
At other $q_y$, the singularities correspond to processes which connect the ends of the wires whose $k_y$ momenta differ by $q_y$.

Slices of the density structure factor for $L_y = 13$ are shown in the lower panel of Fig.~\ref{fig:two_body}.
Note that here we plot $\bar{D}(\bq) \equiv e^{\bq^2/2} D(\bq)$, which has the same singularities as $D(\bq)$ but is more conveniently scaled (see App.~\ref{subapp:corr}).
Visual inspection reveals some of the expected singularities, and we can increase the contrast by taking a ``fractional'' derivative with respect to $q_x$.\cite{VarjasZaletelMoore13}
We calculate an $\eta$\textsuperscript{th} order derivative by multiplying the real-space correlations by $|x-x'|^\eta$ before Fourier transforming.
In the upper panel we show the results for various $\eta \in (0.5, 1.5)$, with $\eta$ chosen for each singularity individually.
This method reveals many singularities 
which are barely visible in the raw data. 

Interestingly, while the physical electrons have periodic boundary conditions, the composite fermions need not.
An arbitrary flux $\Phi_{\rm int}$ of the \emph{emergent} internal gauge field can thread the cylinder, quantizing the CF momenta as $k_y \in \frac{2 \pi}{L_y} \big( \mathbb{Z} + \frac{\Phi_{\rm int}}{2 \pi} \big)$.
	\footnote{The momenta $\bk$ are in fact gauge-dependent quantities. 
		Here we choose a gauge in which $\int_0^{L_y} a_y dy = 0$ and the flux is implemented as a twist in the boundary condition for the composite fermions.
		With this choice $\bk \to -\bk$ corresponds to a 180$^\circ$ rotation.  
		The momentum transfers $\bq$ are gauge invariant and correspond to the momenta of physical observables.}
Being a dynamical degree of freedom, the emergent flux $\Phi_{\rm int}$ will adjust to lower the energy of the filled Fermi sea, depending on $L_y$.
Only two cases respect 180$^\circ$ rotational symmetry: periodic boundary condition (PBC) with $\Phi_{\rm int} = 0$ and  anti-periodic boundary condition (APBC) with $\Phi_{\rm int} = \pi$.
As shown in Fig.~\ref{fig:bc}, the boundary condition dictates the parity of the number of wires: PBC yields an odd $N_w$, while APBC yields an even $N_w$.

The observed singularities at $L_y = 13$ are in exceptional agreement with an $N_w = 4$ model with APBC.
The inset of Fig.~\ref{fig:two_body} shows arrows for the expected singularities, all of which are observed in the main plot.
We obtain similar data for other circumferences $L_y$ (see App.~\ref{app:more_data}), with Fermi surfaces cut by $N_w = 4, 5, 6$, and $8$ wires.
	\footnote{We do not have data for every system size for a given boundary condition because the simulations fail to converge when a wire is close to the edge of the Fermi sea.
 		The different ``root configurations'' 0110 and 0101 (i.e., $K_y$ momentum sectors) give different CF boundary conditions and allow us to access different sizes: the wires for one boundary condition are closest to the edge of the Fermi sea at the sizes when the wires for the other boundary condition are furthest.
		The omission of $N_w = 7$ was due only to our finite resources.}
We find the parity of $N_w$ depends only on the momentum sector used to initialize DMRG: $0101$ always yields $N_w$ odd, while $0110$ always gives an even $N_w$.
This correspondence between the momentum sector and the BC is consistent with the CF theory, where it arises from an anomaly in finite-density $\text{QED}_3$ (see App.~\ref{app:mom_bc}). 

A useful check on our conclusions so far is a comparison to Luttinger's theorem \cite{Luttinger60} for the composite fermions.
The electron density per unit-length of the cylinder is $\rho_{\textrm{1D}} = L_y \frac{\nu}{2 \pi \ell_B^2}$. 
The Luttinger count requires that $2 \pi \rho_{\textrm{1D}}$ be equal to the volume of the Fermi sea (here the sum of the lengths of the wires), as predicted by both the HLR and Dirac-CFL theories at $\nu = \frac12$.
We find excellent agreement with this prediction in Fig.~\ref{fig:luttinger}. 
Our findings are in contrast with a recent suggestion in Ref.~\onlinecite{BalramTokeJain2015} that the Luttinger count is violated by the CF model-wavefunctions.

\begin{figure}[tb]
	\includegraphics[width=72mm]{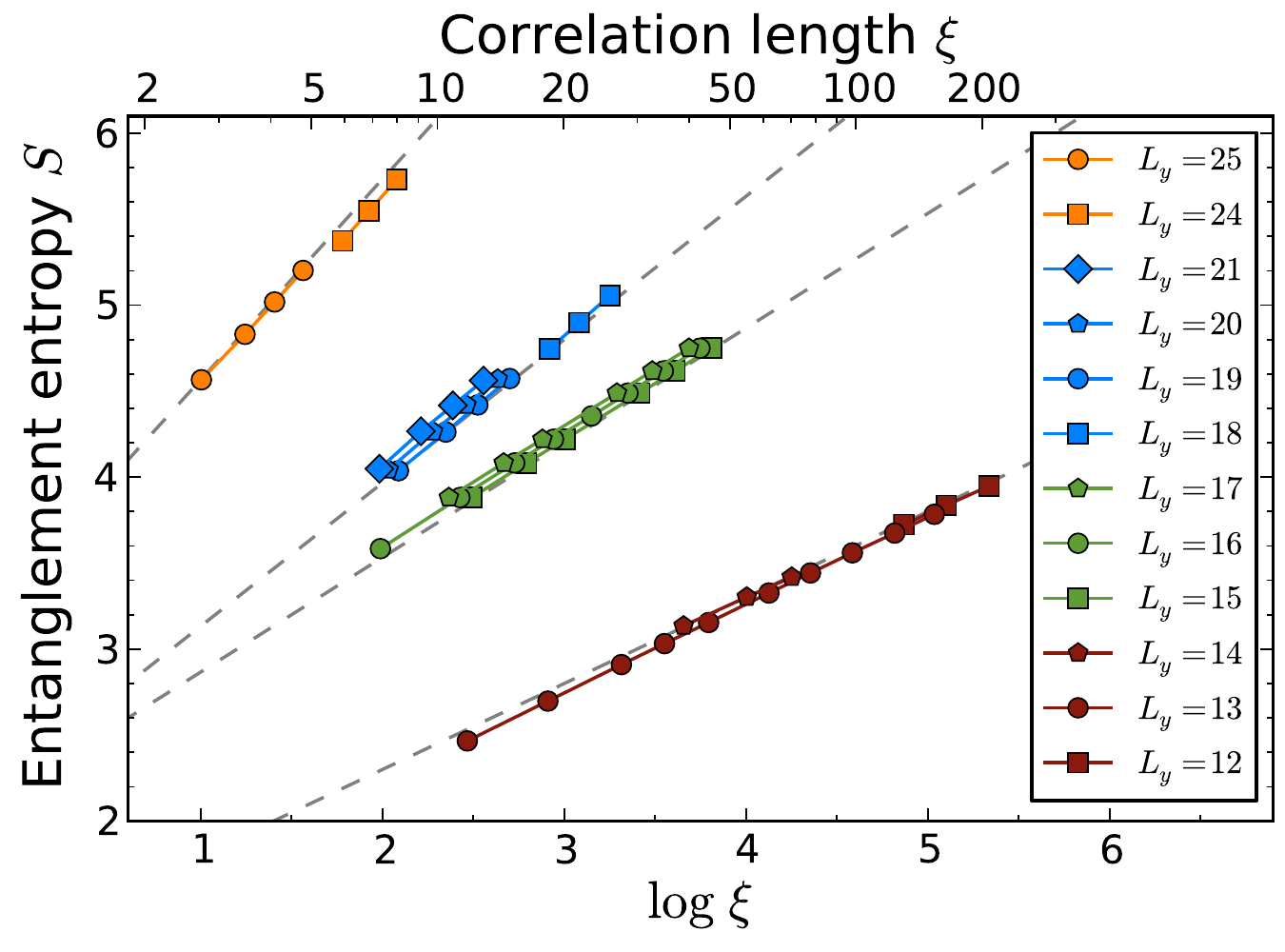}
	\caption{
		Entanglement entropy vs.\ the correlation length at a variety of circumferences $L_y$.
		Different data points at the same size correspond to different bond dimensions $\chi = 600\mbox{--}12000$.
		Both $S$ and $\xi$ increase as $\chi$ is increased.
		For a quasi-1D critical system, we expect a linear relationship between $S$ and $\log\xi$ with the slope proportional to the central charge $c$ [see Eq.~\eqref{eq:Sxi}].
		The dashed lines (from bottom to top) in the figure correspond to $c = 3$, $4$, $5$, and $7$.
	}
	\label{fig:central_charge}
\end{figure}

We can learn more about the composite fermions by measuring the central charge $c$.
This is determined from the scaling between the entanglement entropy $S$ and the DMRG ansatz correlation length $\xi$ (the effective cutoff length) as the bond dimension $\chi$ is increased.
While in a gapped state $S$ and $\xi$ converge with $\chi$, for a critical state both quantities diverge.
The central charge is extracted using the relation\cite{CalabreseCardy}
\begin{equation}
	S = \frac{c}{6}\log\xi + \textrm{const} ~.
	\label{eq:Sxi}
\end{equation}
Figure~\ref{fig:central_charge} shows $S$ vs.\ $\log\xi$ as $\chi$ is varied.
The dashed lines correspond to $c$ of $3$, $4$, $5$, and $7$.
Clearly $c$ increases with $L_y$, with each new wire adding $1$ to the central charge.
The preceding analysis found $N_w = 4$, $5$, $6$, and $8$ for these systems respectively, which allows us to deduce the relation $c = N_w - 1$.
Our data rules out an ordinary Fermi liquid, for which $c = N_w$.
Instead, it confirms the emergence of a gauge field in the composite fermion liquid;
in the quasi-1D limit, the effect of the gauge field is to gap out the total gauge charge mode, reducing $c$ by one (see App.~\ref{app:DCFLquasi1D}).



\section{Particle-hole symmetry and the absence of \twokF backscattering}
\label{sec:ph_sym}
Our numerical findings thus far are expected of both the HLR and Dirac-CFL phases, so to distinguish them we turn to particle-hole symmetry.
The symmetry is generated by $\PH$, an anti-unitary operator which swaps creation operators $c^\dagger$ with annihilation operators $c$:
\begin{align}
	\PH: \;
	c_j \leftrightarrow c_j^\dag , \quad i \leftrightarrow -i , \quad  \ket{000 \cdots} \rightarrow \ket{111 \cdots}.
\end{align}
Particle-hole is an exact symmetry of the Hamiltonian when the interaction is projected into the LLL, which is justified when the cyclotron energy $\hbar\omega_c \to \infty$.
In reality PH is weakly broken by the finite ratio of the Coulomb energy to the cyclotron energy.
In App.~\ref{app:LLmixing}, we provide quantitative evidence that this breaking is weak in typical $\mathrm{GaAs}$ samples.

The HLR construction breaks PH, while the Dirac-CFL explicitly preserves PH.
Our first test is for spontaneous PH symmetry breaking.
We note that PH can be spontaneously broken in our infinite cylinder numerics, as it is a discrete symmetry that we do not explicitly enforce.
Indeed, our numerical DMRG runs clearly break PH at $\nu = \frac52$, randomly selecting either the Pfaffian or anti-Pfaffian state.\cite{ZaletelMixing} 
Furthermore, our numerics will never produce a symmetric superposition (``cat state'') of the HLR and its PH-conjugate (see App.~\ref{app:numerical}).

\begin{figure}[tb]
	\includegraphics[width=86mm]{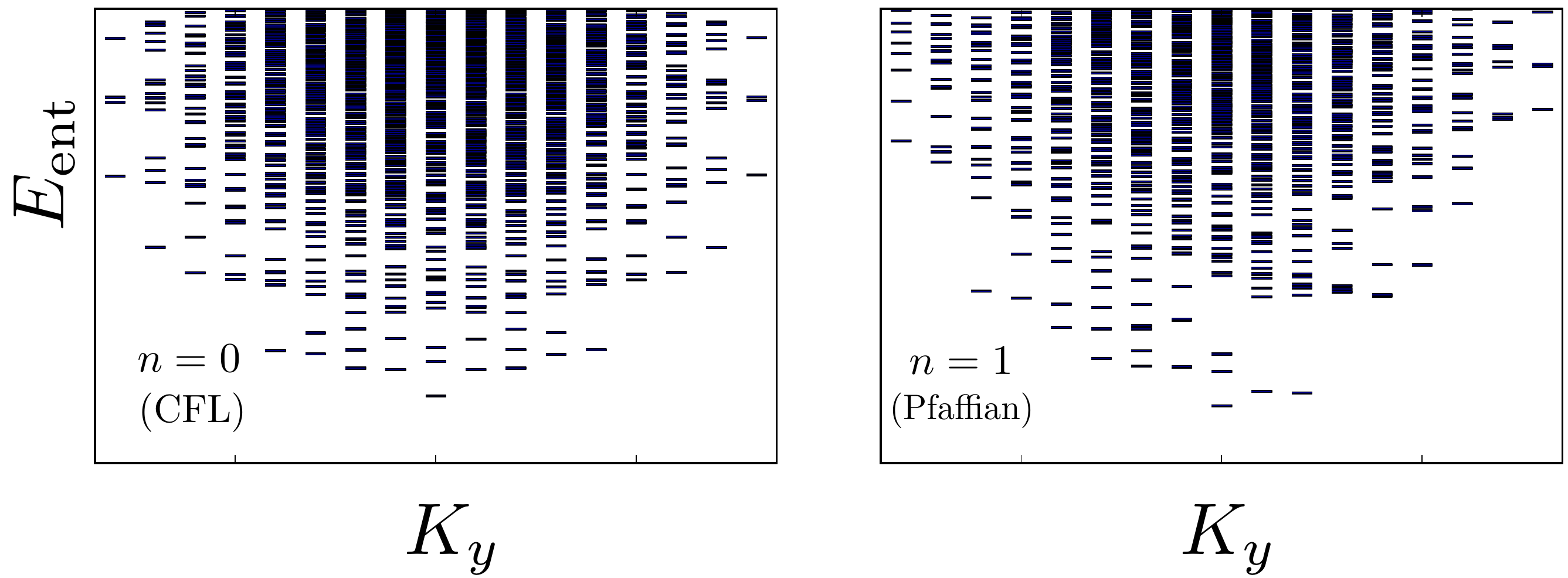}
	\caption{
		Contrasting orbital entanglement spectra of the $n=0$ and $n=1$ Landau levels at half-filling (corresponding to experiments at $\nu = \frac12$ and $\nu = \frac52$ respectively).
		Data is at $L_y = 19$; $n=0$ is the gapless CFL phase, and $n=1$ is consistent with the gapped Pfaffian phase.
		$E_{\textrm{ent}}$ corresponds to the eigenvalues of the reduced density matrix for the left half of the cylinder.
                Each eigenvalue is associated with a momentum around the cylinder $K_y$, just like the energy spectrum of a disk.
		PH acts as a reflection $K_y \leftrightarrow -K_y$. 
		The characteristic chiral ``dispersion'' at $n = 1$ clearly breaks PH, while $n = 0$ is PH-symmetric.
	}
	\label{haldane}
\end{figure}

To test for spontaneous symmetry breaking we compute the overlap between the ground state and its PH conjugate, $\bra{\Psi} \PH \ket{\Psi} = (1- \epsilon)^{\Norb}$, which should decrease exponentially with the number of orbitals (i.e., system size) $\Norb$.
DMRG directly computes the overlap per orbital $1 - \epsilon$;  non-zero $\epsilon > 0$ indicates symmetry breaking.
For comparison, we calculate the same overlap in the $n=1$ Landau level  (i.e., $\nu = \frac52$ mentioned above).\cite{RezayiHaldane:PH:00}
As an example, at $L_y = 16$ for $\nu = \frac12$ we find $\epsilon < 6 \times 10^{-5} $, while for $\nu = \frac52$ we find $\epsilon \approx 0.022(2)$.
At $\nu = \frac52$ this implies that the \emph{total overlap} on a $16 \times 16$ torus with $\Norb \simeq 16^2/(2\pi)$ is $0.978^{\Norb} \approx 0.4$, so the PH-breaking is in fact quite strong.
The difference between $\nu = \frac12$ and $\nu = \frac52$ is also manifest in qualitatively different ``entanglement spectra,''\cite{LiHaldane} as illustrated in Fig.~\ref{haldane}.
Similar results hold at $L_y = 13, 18$.
Therefore, barring some transition at even larger $L_y$, we conclude that the $\nu = \frac12$ CFL state is PH-symmetric.

Given the presence of PH symmetry, we can ask how the CFs transform under this symmetry.
According to Refs.~\onlinecite{DSon:CFL2015, CWangSenthilDDL2015, MetlitskiVishwanath2015}, PH acts like time-reversal symmetry on the CFs implying a twofold Kramers degeneracy whenever an odd number of CFs is present.
Indeed, this feature is already apparent by the very definition of $\PH$, since a system with $\Norb$ orbitals transforms as (cf.\ App.~\ref{app:PH})
\begin{align}
	\PH^2 = (-1)^{\Norb/2} \qquad \text{(with $\Norb$ even).}
\end{align}
Thus at filling $\nu = \frac12$, adding a pair of fluxes plus an electron---i.e., a composite fermion---changes $\PH^2$ by $-1$.

The Kramers degeneracy of the CFs leads to their twofold ``pseudospin'' degree of freedom, which are exchanged via PH.
Two kinds of low energy gapless theories are possible.
The first is a pair of CF Fermi surfaces which are exchanged by PH (or possibly Rashba split).
The other is a single nondegenerate Fermi surface in which the CF pseudospin is locked to the momentum, that is, a Dirac cone.
Our numerical finding of a single Fermi surface with the correct Luttinger volume implies the latter proposal is realized, consistent with Refs.~\onlinecite{DSon:CFL2015, CWangSenthilDDL2015, MetlitskiVishwanath2015}.

\begin{figure}[tb]
	\includegraphics[width=84mm]{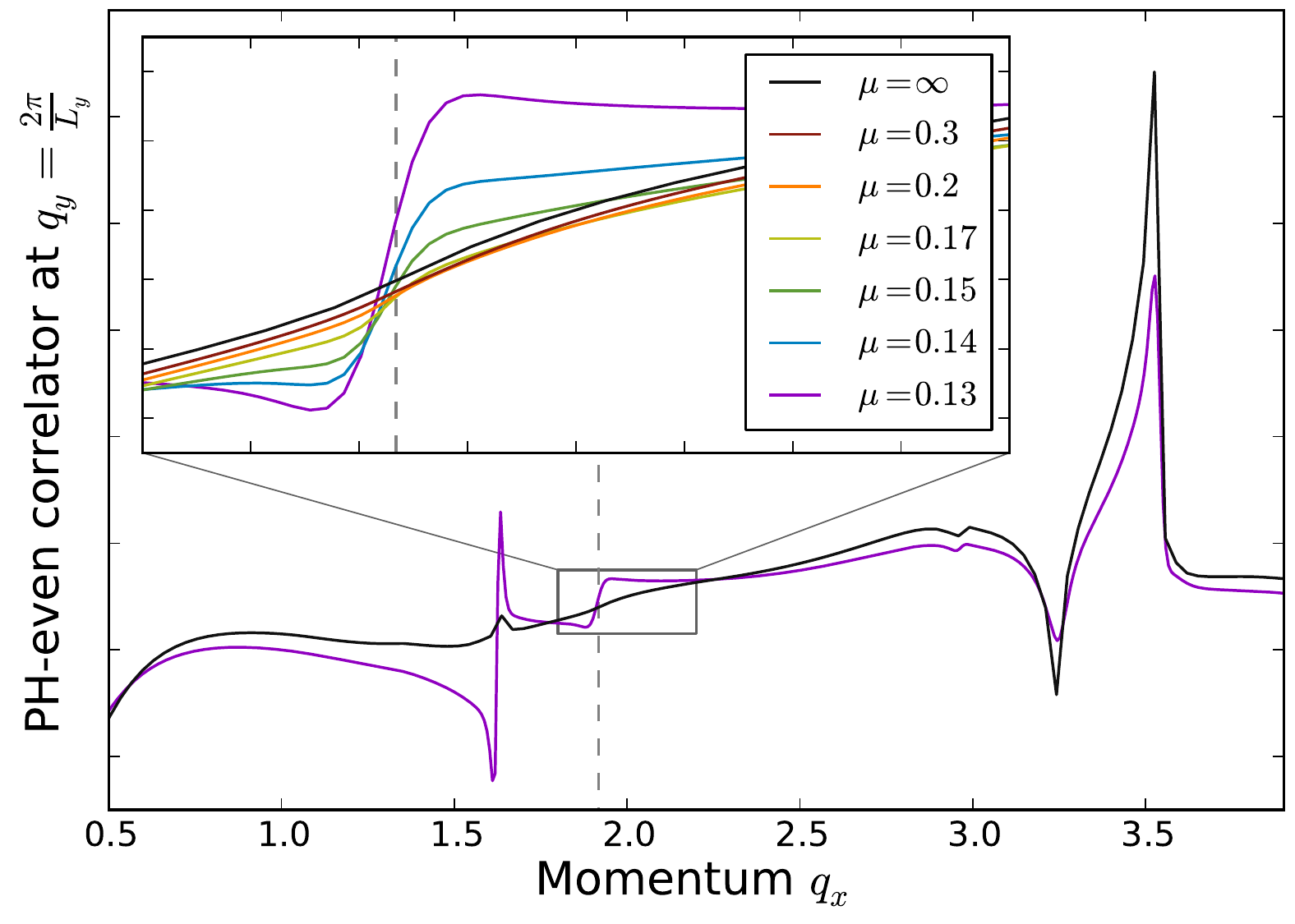}
	\caption{%
		The extinction of  $2 k_F$ backscattering off PH-symmetric impurities.
		A QH bilayer with chemical potential imbalance $\mu$ allows us to continuously tune from a PH-symmetric model ($\mu = \infty$) to a PH-broken one ($\mu$ finite).
		We compute the correlation function of a PH-even operator $\braket{ P_\bq P_{-\bq} }$ for $q_y = \frac{2\pi}{L_y}$ at $L_y = 13$ and plot its derivative with respect to $q_x$ to bring out singularities.
		At the PH-symmetric point, there are many singularities in $q_x$ (see App.~\ref{app:more_data} for analysis), but noticeably absent is any kink at $|\bq| = 2 k_F$.
		This demonstrates the Dirac structure of the CF Fermi surface: $P$ is even under PH, while scattering a CF across the Fermi surface to its antipode is PH-odd in the Dirac theory.
		At finite chemical potential $\mu$ the bilayer setup explicitly breaks PH symmetry, and a kink at $2 k_F$  continuously reappears.
	}
	\label{fig:bilayer}
\end{figure}

One of the most striking consequences of a Dirac cone is that composite fermions accumulate a $\pi$ Berry phase when circling the Dirac point.
In topological insulators, this Berry phase prevents precise $2k_F$ backscattering off time-reversal-symmetric impurities on the surface.
If our system is a Dirac CFL, we should observe the same effect, with $\PH$ playing the role of time-reversal operator.
At the level of equal-time correlation functions, we expect that a Hermitian, PH-even operator $P$ will \emph{not} have strong singularities in $\braket{ P_\bq P_{-\bq} }$ at $|\bq| = 2k_F$ in 2D (cf.~App.~\ref{app:DCFLquasi1D}).
The electron density studied earlier is odd under PH, so does not have this property.
A candidate PH-even observable is $P(\br) \equiv \delta\rho(\br) \nabla^2 \rho(\br)$.
A contribution to this correlation function (see App.~\ref{app:more_data} for details) is shown in Fig.~\ref{fig:bilayer}, measured for the same four-wire APBC CFL as in Fig.~\ref{fig:two_body}.
The measurements are done at $q_y = 2\pi/L_y$, so as to probe the CF scattering from the right Fermi point at $k_y = \pi/L_y$ to the left point at $k_y = -\pi/L_y$, which corresponds to exact $2k_F$ backscattering in 2D.

Despite showing many other singularities, all of which can be accounted for by various multiple-CF scattering processes (see App.~\ref{app:more_data}), the result is perfectly smooth in the vicinity of the (already measured) momentum corresponding to this exact $2 k_F$ backscattering.
We also confirm the absence of the exact backscattering in the PH-symmetric model for a wider cylinder with $L_y = 16$ which realizes five-wire CFL (see App.~\ref{app:more_data}).


If the absence of the $2k_F$ backscattering is truly due to the PH symmetry, rather than some peculiarity of $P$, we expect that the backscattering should return if PH symmetry is explicitly broken.
We break PH symmetry by adding a second quantum well at a distance $1\ell_B$ above the first, with Coulomb intra- and inter-well interactions.
The second well has a chemical potential $\mu$ relative to the first one, and electrons can tunnel between the wells. 
When $\mu = \infty$, electrons remain in the first well, and we recover PH symmetry.
As $\mu$ decreases, electrons tunnel to the second well, breaking PH symmetry, which should induce backscattering.

In Fig.~\ref{fig:bilayer} we show data confirming this hypothesis for $L_y = 13$. 
The tunneling strength is fixed at $t = 0.01$ in units of $e^2 / (4 \pi \epsilon \ell_B)$.
The system remains in an effectively one-well CFL phase for $\mu > 0.05$ [for smaller $\mu$ the system becomes a Halperin (331) state].
As $\mu$ is decreased within the CFL phase, the $2k_F$ singularity reappears.
Note that the measured central charge of the CFL phase remains unchanged; thus, the gapless Fermi surface is stable against PH-breaking perturbations (see also App.~\ref{app:DCFLquasi1D}).

Our findings appear to violate the famous 2D fermion doubling theorem: a single time-reversal-symmetric Dirac cone is anomalous, so cannot be realized in 2D.
However, as noted in Ref.~\onlinecite{MetlitskiVishwanath2015} our composite Dirac fermion is coupled to an emergent gauge-field with unusual compactification condition, which ``cures'' the anomaly.
The half-filled LLL has yet another anomalous property.
Just as the fully symmetric surface of a topological insulator must be nontrivial---that is either gapless or topologically ordered---a PH-symmetric state of the half-filled Landau level must be nontrivial.
Indeed, it must have a Hall conductance of $\sigma^{xy} = \frac{e^2}{2h}$, and if the state is gapped, this requires fractionally charged excitations and hence topological order.
In fact PH here behaves exactly like time-reversal symmetry on the surface of a 3D topological phase (in class $\mathrm{AIII}$---see App.~\ref{app:TSC}).
How does this occur in a purely 2D setting?
For any symmetry that is locally implemented, one can always obtain a symmetric and trivial product state.
The key observation resolving this apparent paradox is that PH symmetry of the LLL is special, in that its action is nonlocal.
The nonlocality is ultimately tied to the fact that the Landau level orbitals $\varphi_j(\br)$ cannot be localized due to the topological nature of the LLL.

We conclude with a number of open questions of relevance to both experiment and theory.
First, given the experimental success in observing the phenomenology of a Dirac cone in TI surfaces
and a CF Fermi surface in $\mathrm{GaAs}$,
it would be extremely interesting to find experimental probes of PH symmetry and the potential Dirac nature of CFs.
Existing experiments are already of some relevance, such as recent measurements in Ref.~\onlinecite{Kamburov2014} of CF ``geometric resonances'' induced by small deviations of $B$-field away from half-filling.
DMRG could be of use in guiding and interpreting such experiments, for instance by computing static structure factors, impurity responses, and the behavior of the CFL at $\nu = \tfrac{1}{2} + \delta$.

Second, Son has proposed a PH-symmetric version of a paired phase, the ``PH-Pfaffian,'' \cite{DSon:CFL2015} which  has previously been proposed as the surface  topological order of a PH-symmetric (class AIII) 3D topological superconductor.\cite{Fidkowski2013}
While our results appear to rule out this possibility in the $n = 1$ LL of $\mathrm{GaAs}$,
it would be interesting to search for such a phase in broader phase diagrams of PH-symmetric models. 

Last but not least, similar theories with a surface of emergent gapless fermions coupled to an emergent gauge field arise for other exotic phases with itinerant fractionalized excitations, such as spin liquids with a spinon Fermi sea, Bose-metals, and electron non-Fermi-liquid metals. 
Much recent theoretical effort has aimed to clarify the status of such field theories,\cite{SSLee2009, Mross2010, Metlitski2010a, *Metlitski2010b} though it remains not fully settled away from artificially controlled limits.
Unbiased numerical studies of the CFL thus bear directly on open questions for all these other non-Fermi liquids.
Recent numerical studies\cite{Fisher2008_gen, *Sheng2008_2legDBL, *Mishmash2011_4legDBL, *Block2011_4legSBM, *Jiang2013_dmetal} explored quasi-1D ladder descendants of various non-Fermi liquids.
Thanks to many innovations in the DMRG for FQHE, the present CFL work goes to effectively much wider strips and is much closer to the 2D physics than the previous studies.
It would be useful to push the numerical CFL study yet closer to 2D and develop scaling analysis tools for addressing 2D questions, such as detailed characterization of the $2k_F$ singularity in the structure factor.
Time-dependent DMRG could potentially study the dynamical properties of a non-Fermi liquid, which has not yet been investigated numerically.

\begin{acknowledgments}
We enjoyed conversations with Maissam Barkeshli, Matthew Fisher, Duncan Haldane, Ryan Mishmash, Chetan Nayak, Ed Rezayi, Dam Son, and Senthil Todadri.
We particularly thank Maissam Barkeshli for first bringing the enigma of PH symmetry to our attention.
SG was supported by National Science Engineering Research Council (NSERC) of Canada.
RM enjoyed support from the Sherman Fairchild Foundation.
MM was supported by the U.S.\ Army Research Office, grant No.~W911NF-14-1-0379.
AV was supported by NSF-DMR 1206728.
RM and AV acknowledge KITP for hospitality, supported by the National Science Foundation under Grant No.~NSF PHY11-25915.
SG and OIM were supported by NSF-DMR 1206096, and by the Caltech Institute for Quantum Information and Matter, an NSF Physics Frontiers Center with support of the Gordon and Betty Moore Foundation.
\end{acknowledgments}

\emph{During the completion of this work Ref.~\onlinecite{CWangSenthil:HalfLL2015} appeared, which also provides a theoretical discussion of the particle-hole symmetric CFL.}
\emph{Independently, Levin and Son also derived $\PH^2 = -1$ relation for CFs (unpublished).} 

\bibliography{bib12}

\begin{thebibliography}{61}%
\makeatletter
\providecommand \@ifxundefined [1]{%
 \@ifx{#1\undefined}
}%
\providecommand \@ifnum [1]{%
 \ifnum #1\expandafter \@firstoftwo
 \else \expandafter \@secondoftwo
 \fi
}%
\providecommand \@ifx [1]{%
 \ifx #1\expandafter \@firstoftwo
 \else \expandafter \@secondoftwo
 \fi
}%
\providecommand \natexlab [1]{#1}%
\providecommand \enquote  [1]{``#1''}%
\providecommand \bibnamefont  [1]{#1}%
\providecommand \bibfnamefont [1]{#1}%
\providecommand \citenamefont [1]{#1}%
\providecommand \href@noop [0]{\@secondoftwo}%
\providecommand \href [0]{\begingroup \@sanitize@url \@href}%
\providecommand \@href[1]{\@@startlink{#1}\@@href}%
\providecommand \@@href[1]{\endgroup#1\@@endlink}%
\providecommand \@sanitize@url [0]{\catcode `\\12\catcode `\$12\catcode
  `\&12\catcode `\#12\catcode `\^12\catcode `\_12\catcode `\%12\relax}%
\providecommand \@@startlink[1]{}%
\providecommand \@@endlink[0]{}%
\providecommand \url  [0]{\begingroup\@sanitize@url \@url }%
\providecommand \@url [1]{\endgroup\@href {#1}{\urlprefix }}%
\providecommand \urlprefix  [0]{URL }%
\providecommand \Eprint [0]{\href }%
\providecommand \doibase [0]{http://dx.doi.org/}%
\providecommand \selectlanguage [0]{\@gobble}%
\providecommand \bibinfo  [0]{\@secondoftwo}%
\providecommand \bibfield  [0]{\@secondoftwo}%
\providecommand \translation [1]{[#1]}%
\providecommand \BibitemOpen [0]{}%
\providecommand \bibitemStop [0]{}%
\providecommand \bibitemNoStop [0]{.\EOS\space}%
\providecommand \EOS [0]{\spacefactor3000\relax}%
\providecommand \BibitemShut  [1]{\csname bibitem#1\endcsname}%
\let\auto@bib@innerbib\@empty
\bibitem [{\citenamefont {Jain}(1989)}]{JainCF89}%
  \BibitemOpen
  \bibfield  {author} {\bibinfo {author} {\bibfnamefont {J.~K.}\ \bibnamefont
  {Jain}},\ }\href {\doibase 10.1103/PhysRevLett.63.199} {\bibfield  {journal}
  {\bibinfo  {journal} {Phys. Rev. Lett.}\ }\textbf {\bibinfo {volume} {63}},\
  \bibinfo {pages} {199} (\bibinfo {year} {1989})}\BibitemShut {NoStop}%
\bibitem [{\citenamefont {Lopez}\ and\ \citenamefont
  {Fradkin}(1991)}]{LopezFradkin91}%
  \BibitemOpen
  \bibfield  {author} {\bibinfo {author} {\bibfnamefont {A.}~\bibnamefont
  {Lopez}}\ and\ \bibinfo {author} {\bibfnamefont {E.}~\bibnamefont
  {Fradkin}},\ }\href {\doibase 10.1103/PhysRevB.44.5246} {\bibfield  {journal}
  {\bibinfo  {journal} {Phys. Rev. B}\ }\textbf {\bibinfo {volume} {44}},\
  \bibinfo {pages} {5246} (\bibinfo {year} {1991})}\BibitemShut {NoStop}%
\bibitem [{\citenamefont {Halperin}\ \emph {et~al.}(1993)\citenamefont
  {Halperin}, \citenamefont {Lee},\ and\ \citenamefont
  {Read}}]{HalperinLeeRead}%
  \BibitemOpen
  \bibfield  {author} {\bibinfo {author} {\bibfnamefont {B.~I.}\ \bibnamefont
  {Halperin}}, \bibinfo {author} {\bibfnamefont {P.~A.}\ \bibnamefont {Lee}},\
  and\ \bibinfo {author} {\bibfnamefont {N.}~\bibnamefont {Read}},\ }\href
  {\doibase 10.1103/PhysRevB.47.7312} {\bibfield  {journal} {\bibinfo
  {journal} {Phys. Rev. B}\ }\textbf {\bibinfo {volume} {47}},\ \bibinfo
  {pages} {7312} (\bibinfo {year} {1993})}\BibitemShut {NoStop}%
\bibitem [{\citenamefont {Willett}\ \emph {et~al.}(1993)\citenamefont
  {Willett}, \citenamefont {Ruel}, \citenamefont {West},\ and\ \citenamefont
  {Pfeiffer}}]{WillettFS93}%
  \BibitemOpen
  \bibfield  {author} {\bibinfo {author} {\bibfnamefont {R.~L.}\ \bibnamefont
  {Willett}}, \bibinfo {author} {\bibfnamefont {R.~R.}\ \bibnamefont {Ruel}},
  \bibinfo {author} {\bibfnamefont {K.~W.}\ \bibnamefont {West}}, and\
  \bibinfo {author} {\bibfnamefont {L.~N.}\ \bibnamefont {Pfeiffer}},\ }\href
  {\doibase 10.1103/PhysRevLett.71.3846} {\bibfield  {journal} {\bibinfo
  {journal} {Phys. Rev. Lett.}\ }\textbf {\bibinfo {volume} {71}},\ \bibinfo
  {pages} {3846} (\bibinfo {year} {1993})}\BibitemShut {NoStop}%
\bibitem [{\citenamefont {Kang}\ \emph {et~al.}(1993)\citenamefont {Kang},
  \citenamefont {Stormer}, \citenamefont {Pfeiffer}, \citenamefont {Baldwin},\
  and\ \citenamefont {West}}]{Kang93}%
  \BibitemOpen
  \bibfield  {author} {\bibinfo {author} {\bibfnamefont {W.}~\bibnamefont
  {Kang}}, \bibinfo {author} {\bibfnamefont {H.~L.}\ \bibnamefont {Stormer}},
  \bibinfo {author} {\bibfnamefont {L.~N.}\ \bibnamefont {Pfeiffer}}, \bibinfo
  {author} {\bibfnamefont {K.~W.}\ \bibnamefont {Baldwin}}, and\ \bibinfo
  {author} {\bibfnamefont {K.~W.}\ \bibnamefont {West}},\ }\href {\doibase%
  10.1103/PhysRevLett.71.3850} {\bibfield  {journal} {\bibinfo  {journal}
  {Phys. Rev. Lett.}\ }\textbf {\bibinfo {volume} {71}},\ \bibinfo {pages}
  {3850} (\bibinfo {year} {1993})}\BibitemShut {NoStop}%
\bibitem [{\citenamefont {Goldman}\ \emph {et~al.}(1994)\citenamefont
  {Goldman}, \citenamefont {Su},\ and\ \citenamefont {Jain}}]{Goldman94}%
  \BibitemOpen
  \bibfield  {author} {\bibinfo {author} {\bibfnamefont {V.~J.}\ \bibnamefont
  {Goldman}}, \bibinfo {author} {\bibfnamefont {B.}~\bibnamefont {Su}}, and\
  \bibinfo {author} {\bibfnamefont {J.~K.}\ \bibnamefont {Jain}},\ }\href
  {\doibase 10.1103/PhysRevLett.72.2065} {\bibfield  {journal} {\bibinfo
  {journal} {Phys. Rev. Lett.}\ }\textbf {\bibinfo {volume} {72}},\ \bibinfo
  {pages} {2065} (\bibinfo {year} {1994})}\BibitemShut {NoStop}%
\bibitem [{\citenamefont {Smet}\ \emph {et~al.}(1996)\citenamefont {Smet},
  \citenamefont {Weiss}, \citenamefont {Blick}, \citenamefont {L\"utjering},
  \citenamefont {von Klitzing}, \citenamefont {Fleischmann}, \citenamefont
  {Ketzmerick}, \citenamefont {Geisel},\ and\ \citenamefont
  {Weimann}}]{Smet96}%
  \BibitemOpen
  \bibfield  {author} {\bibinfo {author} {\bibfnamefont {J.~H.}\ \bibnamefont
  {Smet}}, \bibinfo {author} {\bibfnamefont {D.}~\bibnamefont {Weiss}},
  \bibinfo {author} {\bibfnamefont {R.~H.}\ \bibnamefont {Blick}}, \bibinfo
  {author} {\bibfnamefont {G.}~\bibnamefont {L\"utjering}}, \bibinfo {author}
  {\bibfnamefont {K.}~\bibnamefont {von Klitzing}}, \bibinfo {author}
  {\bibfnamefont {R.}~\bibnamefont {Fleischmann}}, \bibinfo {author}
  {\bibfnamefont {R.}~\bibnamefont {Ketzmerick}}, \bibinfo {author}
  {\bibfnamefont {T.}~\bibnamefont {Geisel}}, and\ \bibinfo {author}
  {\bibfnamefont {G.}~\bibnamefont {Weimann}},\ }\href {\doibase%
  10.1103/PhysRevLett.77.2272} {\bibfield  {journal} {\bibinfo  {journal}
  {Phys. Rev. Lett.}\ }\textbf {\bibinfo {volume} {77}},\ \bibinfo {pages}
  {2272} (\bibinfo {year} {1996})}\BibitemShut {NoStop}%
\bibitem [{\citenamefont {Rezayi}\ and\ \citenamefont
  {Read}(1994)}]{RezayiRead:HalfFilledLL:1994}%
  \BibitemOpen
  \bibfield  {author} {\bibinfo {author} {\bibfnamefont {E.}~\bibnamefont
  {Rezayi}}\ and\ \bibinfo {author} {\bibfnamefont {N.}~\bibnamefont {Read}},\
  }\href {\doibase 10.1103/PhysRevLett.72.900} {\bibfield  {journal} {\bibinfo
  {journal} {Phys. Rev. Lett.}\ }\textbf {\bibinfo {volume} {72}},\ \bibinfo
  {pages} {900} (\bibinfo {year} {1994})}\BibitemShut {NoStop}%
\bibitem [{\citenamefont {Kivelson}\ \emph {et~al.}(1997)\citenamefont
  {Kivelson}, \citenamefont {Lee}, \citenamefont {Krotov},\ and\ \citenamefont
  {Gan}}]{Kivelson1997}%
  \BibitemOpen
  \bibfield  {author} {\bibinfo {author} {\bibfnamefont {S.~A.}\ \bibnamefont
  {Kivelson}}, \bibinfo {author} {\bibfnamefont {D.-H.}\ \bibnamefont {Lee}},
  \bibinfo {author} {\bibfnamefont {Y.}~\bibnamefont {Krotov}}, and\ \bibinfo
  {author} {\bibfnamefont {J.}~\bibnamefont {Gan}},\ }\href {\doibase%
  10.1103/PhysRevB.55.15552} {\bibfield  {journal} {\bibinfo  {journal} {Phys.
  Rev. B}\ }\textbf {\bibinfo {volume} {55}},\ \bibinfo {pages} {15552}
  (\bibinfo {year} {1997})}\BibitemShut {NoStop}%
\bibitem [{\citenamefont {Lee}(1998)}]{Lee1998}%
  \BibitemOpen
  \bibfield  {author} {\bibinfo {author} {\bibfnamefont {D.-H.}\ \bibnamefont
  {Lee}},\ }\href {\doibase 10.1103/PhysRevLett.80.4745} {\bibfield  {journal}
  {\bibinfo  {journal} {Phys. Rev. Lett.}\ }\textbf {\bibinfo {volume} {80}},\
  \bibinfo {pages} {4745} (\bibinfo {year} {1998})}\BibitemShut {NoStop}%
\bibitem [{\citenamefont {Pasquier}\ and\ \citenamefont
  {Haldane}(1998)}]{PasquierHaldane1998}%
  \BibitemOpen
  \bibfield  {author} {\bibinfo {author} {\bibfnamefont {V.}~\bibnamefont
  {Pasquier}}\ and\ \bibinfo {author} {\bibfnamefont {F.~D.~M.}\ \bibnamefont
  {Haldane}},\ }\href {\doibase 10.1016/S0550-3213(98)00069-8} {\bibfield
  {journal} {\bibinfo  {journal} {Nucl. Phys. B}\ }\textbf {\bibinfo {volume}
  {516}},\ \bibinfo {pages} {719 } (\bibinfo {year} {1998})}\BibitemShut
  {NoStop}%
\bibitem [{\citenamefont {Read}(1998)}]{Read1998}%
  \BibitemOpen
  \bibfield  {author} {\bibinfo {author} {\bibfnamefont {N.}~\bibnamefont
  {Read}},\ }\href {\doibase 10.1103/PhysRevB.58.16262} {\bibfield  {journal}
  {\bibinfo  {journal} {Phys. Rev. B}\ }\textbf {\bibinfo {volume} {58}},\
  \bibinfo {pages} {16262} (\bibinfo {year} {1998})}\BibitemShut {NoStop}%
\bibitem [{\citenamefont {Murthy}\ and\ \citenamefont
  {Shankar}(2003)}]{MurthyShankarRMP}%
  \BibitemOpen
  \bibfield  {author} {\bibinfo {author} {\bibfnamefont {G.}~\bibnamefont
  {Murthy}}\ and\ \bibinfo {author} {\bibfnamefont {R.}~\bibnamefont
  {Shankar}},\ }\href {\doibase 10.1103/RevModPhys.75.1101} {\bibfield
  {journal} {\bibinfo  {journal} {Rev. Mod. Phys.}\ }\textbf {\bibinfo {volume}
  {75}},\ \bibinfo {pages} {1101} (\bibinfo {year} {2003})}\BibitemShut
  {NoStop}%
\bibitem [{\citenamefont {{Barkeshli}}\ \emph {et~al.}(2015)\citenamefont
  {{Barkeshli}}, \citenamefont {{Mulligan}},\ and\ \citenamefont
  {{Fisher}}}]{Barkeshli2015}%
  \BibitemOpen
  \bibfield  {author} {\bibinfo {author} {\bibfnamefont {M.}~\bibnamefont
  {{Barkeshli}}}, \bibinfo {author} {\bibfnamefont {M.}~\bibnamefont
  {{Mulligan}}}, and\ \bibinfo {author} {\bibfnamefont {M.~P.~A.}\
  \bibnamefont {{Fisher}}},\ }\href@noop {} {\enquote {\bibinfo {title}
  {{Particle-Hole Symmetry and the Composite Fermi Liquid}},}\ } (\bibinfo
  {year} {2015}),\ \bibinfo {note} {unpublished},\ \Eprint
  {http://arxiv.org/abs/1502.05404} {arXiv:1502.05404} \BibitemShut {NoStop}%
\bibitem [{\citenamefont {Willett}\ \emph {et~al.}(1987)\citenamefont
  {Willett}, \citenamefont {Eisenstein}, \citenamefont {St\"ormer},
  \citenamefont {Tsui}, \citenamefont {Gossard},\ and\ \citenamefont
  {English}}]{Eisenstein87}%
  \BibitemOpen
  \bibfield  {author} {\bibinfo {author} {\bibfnamefont {R.}~\bibnamefont
  {Willett}}, \bibinfo {author} {\bibfnamefont {J.~P.}\ \bibnamefont
  {Eisenstein}}, \bibinfo {author} {\bibfnamefont {H.~L.}\ \bibnamefont
  {St\"ormer}}, \bibinfo {author} {\bibfnamefont {D.~C.}\ \bibnamefont {Tsui}},
  \bibinfo {author} {\bibfnamefont {A.~C.}\ \bibnamefont {Gossard}}, and\
  \bibinfo {author} {\bibfnamefont {J.~H.}\ \bibnamefont {English}},\ }\href
  {\doibase 10.1103/PhysRevLett.59.1776} {\bibfield  {journal} {\bibinfo
  {journal} {Phys. Rev. Lett.}\ }\textbf {\bibinfo {volume} {59}},\ \bibinfo
  {pages} {1776} (\bibinfo {year} {1987})}\BibitemShut {NoStop}%
\bibitem [{\citenamefont {Morf}(1998)}]{Morf1998}%
  \BibitemOpen
  \bibfield  {author} {\bibinfo {author} {\bibfnamefont {R.~H.}\ \bibnamefont
  {Morf}},\ }\href {\doibase 10.1103/PhysRevLett.80.1505} {\bibfield  {journal}
  {\bibinfo  {journal} {Phys. Rev. Lett.}\ }\textbf {\bibinfo {volume} {80}},\
  \bibinfo {pages} {1505} (\bibinfo {year} {1998})}\BibitemShut {NoStop}%
\bibitem [{\citenamefont {Rezayi}\ and\ \citenamefont
  {Haldane}(2000)}]{RezayiHaldane:PH:00}%
  \BibitemOpen
  \bibfield  {author} {\bibinfo {author} {\bibfnamefont {E.~H.}\ \bibnamefont
  {Rezayi}}\ and\ \bibinfo {author} {\bibfnamefont {F.~D.~M.}\ \bibnamefont
  {Haldane}},\ }\href {\doibase 10.1103/PhysRevLett.84.4685} {\bibfield
  {journal} {\bibinfo  {journal} {Phys. Rev. Lett.}\ }\textbf {\bibinfo
  {volume} {84}},\ \bibinfo {pages} {4685} (\bibinfo {year}
  {2000})}\BibitemShut {NoStop}%
\bibitem [{\citenamefont {Papi\ifmmode~\acute{c}\else \'{c}\fi{}}\ \emph
  {et~al.}(2012)\citenamefont {Papi\ifmmode~\acute{c}\else \'{c}\fi{}},
  \citenamefont {Haldane},\ and\ \citenamefont {Rezayi}}]{Papic2012}%
  \BibitemOpen
  \bibfield  {author} {\bibinfo {author} {\bibfnamefont {Z.}~\bibnamefont
  {Papi\ifmmode~\acute{c}\else \'{c}\fi{}}}, \bibinfo {author} {\bibfnamefont
  {F.~D.~M.}\ \bibnamefont {Haldane}}, and\ \bibinfo {author} {\bibfnamefont
  {E.~H.}\ \bibnamefont {Rezayi}},\ }\href {\doibase%
  10.1103/PhysRevLett.109.266806} {\bibfield  {journal} {\bibinfo  {journal}
  {Phys. Rev. Lett.}\ }\textbf {\bibinfo {volume} {109}},\ \bibinfo {pages}
  {266806} (\bibinfo {year} {2012})}\BibitemShut {NoStop}%
\bibitem [{\citenamefont {Moore}\ and\ \citenamefont
  {Read}(1991)}]{MooreRead:Nonabelion:1991}%
  \BibitemOpen
  \bibfield  {author} {\bibinfo {author} {\bibfnamefont {G.}~\bibnamefont
  {Moore}}\ and\ \bibinfo {author} {\bibfnamefont {N.}~\bibnamefont {Read}},\
  }\href {\doibase 10.1016/0550-3213(91)90407-O} {\bibfield  {journal}
  {\bibinfo  {journal} {Nucl. Phys. B}\ }\textbf {\bibinfo {volume} {360}},\
  \bibinfo {pages} {362} (\bibinfo {year} {1991})}\BibitemShut {NoStop}%
\bibitem [{\citenamefont {Son}(2015)}]{DSon:CFL2015}%
  \BibitemOpen
  \bibfield  {author} {\bibinfo {author} {\bibfnamefont {D.~T.}\ \bibnamefont
  {Son}},\ }\href@noop {} {\enquote {\bibinfo {title} {Is the composite fermion
  a {D}irac particle?}}\ } (\bibinfo {year} {2015}),\ \bibinfo {note}
  {unpublished},\ \Eprint {http://arxiv.org/abs/1502.03446} {arXiv:1502.03446}
  \BibitemShut {NoStop}%
\bibitem [{\citenamefont {{Wang}}\ and\ \citenamefont
  {{Senthil}}(2015{\natexlab{a}})}]{CWangSenthilDDL2015}%
  \BibitemOpen
  \bibfield  {author} {\bibinfo {author} {\bibfnamefont {C.}~\bibnamefont
  {{Wang}}}\ and\ \bibinfo {author} {\bibfnamefont {T.}~\bibnamefont
  {{Senthil}}},\ }\href@noop {} {\enquote {\bibinfo {title} {Dual {D}irac
  liquid on the surface of the electron topological insulator},}\ } (\bibinfo
  {year} {2015}{\natexlab{a}}),\ \bibinfo {note} {unpublished},\ \Eprint
  {http://arxiv.org/abs/1505.05141} {arXiv:1505.05141} \BibitemShut {NoStop}%
\bibitem [{\citenamefont {{Metlitski}}\ and\ \citenamefont
  {{Vishwanath}}(2015)}]{MetlitskiVishwanath2015}%
  \BibitemOpen
  \bibfield  {author} {\bibinfo {author} {\bibfnamefont {M.~A.}\ \bibnamefont
  {{Metlitski}}}\ and\ \bibinfo {author} {\bibfnamefont {A.}~\bibnamefont
  {{Vishwanath}}},\ }\href@noop {} {\enquote {\bibinfo {title} {Particle-vortex
  duality of 2d {D}irac fermion from electric-magnetic duality of 3d
  topological insulators},}\ } (\bibinfo {year} {2015}),\ \bibinfo {note}
  {unpublished},\ \Eprint {http://arxiv.org/abs/1505.05142} {arXiv:1505.05142}
  \BibitemShut {NoStop}%
\bibitem [{\citenamefont {White}(1992)}]{White92}%
  \BibitemOpen
  \bibfield  {author} {\bibinfo {author} {\bibfnamefont {S.~R.}\ \bibnamefont
  {White}},\ }\href {\doibase 10.1103/PhysRevLett.69.2863} {\bibfield
  {journal} {\bibinfo  {journal} {Phys. Rev. Lett.}\ }\textbf {\bibinfo
  {volume} {69}},\ \bibinfo {pages} {2863} (\bibinfo {year}
  {1992})}\BibitemShut {NoStop}%
\bibitem [{\citenamefont {Roushan}\ \emph {et~al.}(2009)\citenamefont
  {Roushan}, \citenamefont {Seo}, \citenamefont {Parker}, \citenamefont {Hor},
  \citenamefont {Hsieh}, \citenamefont {Qian}, \citenamefont {Richardella},
  \citenamefont {Hasan}, \citenamefont {Cava},\ and\ \citenamefont
  {Yazdani}}]{RoushanHasan:TopSurface09}%
  \BibitemOpen
  \bibfield  {author} {\bibinfo {author} {\bibfnamefont {P.}~\bibnamefont
  {Roushan}}, \bibinfo {author} {\bibfnamefont {J.}~\bibnamefont {Seo}},
  \bibinfo {author} {\bibfnamefont {C.~V.}\ \bibnamefont {Parker}}, \bibinfo
  {author} {\bibfnamefont {Y.~S.}\ \bibnamefont {Hor}}, \bibinfo {author}
  {\bibfnamefont {D.}~\bibnamefont {Hsieh}}, \bibinfo {author} {\bibfnamefont
  {D.}~\bibnamefont {Qian}}, \bibinfo {author} {\bibfnamefont {A.}~\bibnamefont
  {Richardella}}, \bibinfo {author} {\bibfnamefont {M.~Z.}\ \bibnamefont
  {Hasan}}, \bibinfo {author} {\bibfnamefont {R.~J.}\ \bibnamefont {Cava}},\
  and\ \bibinfo {author} {\bibfnamefont {A.}~\bibnamefont {Yazdani}},\ }\href
  {\doibase 10.1038/nature08308} {\bibfield  {journal} {\bibinfo  {journal}
  {Nature}\ }\textbf {\bibinfo {volume} {460}},\ \bibinfo {pages} {1106}
  (\bibinfo {year} {2009})}\BibitemShut {NoStop}%
\bibitem [{\citenamefont {Zhang}\ \emph {et~al.}(2009)\citenamefont {Zhang},
  \citenamefont {Cheng}, \citenamefont {Chen}, \citenamefont {Jia},
  \citenamefont {Ma}, \citenamefont {He}, \citenamefont {Wang}, \citenamefont
  {Zhang}, \citenamefont {Dai}, \citenamefont {Fang}, \citenamefont {Xie},\
  and\ \citenamefont {Xue}}]{TZhangQKXue:QpInterference09}%
  \BibitemOpen
  \bibfield  {author} {\bibinfo {author} {\bibfnamefont {T.}~\bibnamefont
  {Zhang}}, \bibinfo {author} {\bibfnamefont {P.}~\bibnamefont {Cheng}},
  \bibinfo {author} {\bibfnamefont {X.}~\bibnamefont {Chen}}, \bibinfo {author}
  {\bibfnamefont {J.-F.}\ \bibnamefont {Jia}}, \bibinfo {author} {\bibfnamefont
  {X.}~\bibnamefont {Ma}}, \bibinfo {author} {\bibfnamefont {K.}~\bibnamefont
  {He}}, \bibinfo {author} {\bibfnamefont {L.}~\bibnamefont {Wang}}, \bibinfo
  {author} {\bibfnamefont {H.}~\bibnamefont {Zhang}}, \bibinfo {author}
  {\bibfnamefont {X.}~\bibnamefont {Dai}}, \bibinfo {author} {\bibfnamefont
  {Z.}~\bibnamefont {Fang}}, \bibinfo {author} {\bibfnamefont {X.}~\bibnamefont
  {Xie}}, and\ \bibinfo {author} {\bibfnamefont {Q.-K.}\ \bibnamefont
  {Xue}},\ }\href {\doibase 10.1103/PhysRevLett.103.266803} {\bibfield
  {journal} {\bibinfo  {journal} {Phys. Rev. Lett.}\ }\textbf {\bibinfo
  {volume} {103}},\ \bibinfo {pages} {266803} (\bibinfo {year}
  {2009})}\BibitemShut {NoStop}%
\bibitem [{\citenamefont {Zaletel}\ \emph {et~al.}(2015)\citenamefont
  {Zaletel}, \citenamefont {Mong}, \citenamefont {Pollmann},\ and\
  \citenamefont {Rezayi}}]{ZaletelMixing}%
  \BibitemOpen
  \bibfield  {author} {\bibinfo {author} {\bibfnamefont {M.~P.}\ \bibnamefont
  {Zaletel}}, \bibinfo {author} {\bibfnamefont {R.~S.~K.}\ \bibnamefont
  {Mong}}, \bibinfo {author} {\bibfnamefont {F.}~\bibnamefont {Pollmann}},\
  and\ \bibinfo {author} {\bibfnamefont {E.~H.}\ \bibnamefont {Rezayi}},\
  }\href {\doibase 10.1103/PhysRevB.91.045115} {\bibfield  {journal} {\bibinfo
  {journal} {Phys. Rev. B}\ }\textbf {\bibinfo {volume} {91}},\ \bibinfo
  {pages} {045115} (\bibinfo {year} {2015})}\BibitemShut {NoStop}%
\bibitem [{\citenamefont {Kukushkin}\ \emph {et~al.}(1999)\citenamefont
  {Kukushkin}, \citenamefont {von Klitzing},\ and\ \citenamefont
  {Eberl}}]{Kukushkin99}%
  \BibitemOpen
  \bibfield  {author} {\bibinfo {author} {\bibfnamefont {I.~V.}\ \bibnamefont
  {Kukushkin}}, \bibinfo {author} {\bibfnamefont {K.}~\bibnamefont {von
  Klitzing}}, and\ \bibinfo {author} {\bibfnamefont {K.}~\bibnamefont
  {Eberl}},\ }\href {\doibase 10.1103/PhysRevLett.82.3665} {\bibfield
  {journal} {\bibinfo  {journal} {Phys. Rev. Lett.}\ }\textbf {\bibinfo
  {volume} {82}},\ \bibinfo {pages} {3665} (\bibinfo {year}
  {1999})}\BibitemShut {NoStop}%
\bibitem [{\citenamefont {Tagliacozzo}\ \emph {et~al.}(2008)\citenamefont
  {Tagliacozzo}, \citenamefont {de~Oliveira}, \citenamefont {Iblisdir},\ and\
  \citenamefont {Latorre}}]{Tagliacozzo:2008}%
  \BibitemOpen
  \bibfield  {author} {\bibinfo {author} {\bibfnamefont {L.}~\bibnamefont
  {Tagliacozzo}}, \bibinfo {author} {\bibfnamefont {T.~R.}\ \bibnamefont
  {de~Oliveira}}, \bibinfo {author} {\bibfnamefont {S.}~\bibnamefont
  {Iblisdir}}, and\ \bibinfo {author} {\bibfnamefont {J.~I.}\ \bibnamefont
  {Latorre}},\ }\href {\doibase 10.1103/PhysRevB.78.024410} {\bibfield
  {journal} {\bibinfo  {journal} {Phys. Rev. B}\ }\textbf {\bibinfo {volume}
  {78}},\ \bibinfo {pages} {024410} (\bibinfo {year} {2008})}\BibitemShut
  {NoStop}%
\bibitem [{\citenamefont {Pollmann}\ \emph {et~al.}(2009)\citenamefont
  {Pollmann}, \citenamefont {Mukerjee}, \citenamefont {Turner},\ and\
  \citenamefont {Moore}}]{Pollmann:FiniteES:2009}%
  \BibitemOpen
  \bibfield  {author} {\bibinfo {author} {\bibfnamefont {F.}~\bibnamefont
  {Pollmann}}, \bibinfo {author} {\bibfnamefont {S.}~\bibnamefont {Mukerjee}},
  \bibinfo {author} {\bibfnamefont {A.~M.}\ \bibnamefont {Turner}}, and\
  \bibinfo {author} {\bibfnamefont {J.~E.}\ \bibnamefont {Moore}},\ }\href
  {\doibase 10.1103/PhysRevLett.102.255701} {\bibfield  {journal} {\bibinfo
  {journal} {Phys. Rev. Lett.}\ }\textbf {\bibinfo {volume} {102}},\ \bibinfo
  {pages} {255701} (\bibinfo {year} {2009})}\BibitemShut {NoStop}%
\bibitem [{\citenamefont {Polchinski}(1994)}]{Polchinski94}%
  \BibitemOpen
  \bibfield  {author} {\bibinfo {author} {\bibfnamefont {J.}~\bibnamefont
  {Polchinski}},\ }\href {\doibase 10.1016/0550-3213(94)90449-9} {\bibfield
  {journal} {\bibinfo  {journal} {Nucl. Phys. B}\ }\textbf {\bibinfo {volume}
  {422}},\ \bibinfo {pages} {617} (\bibinfo {year} {1994})}\BibitemShut
  {NoStop}%
\bibitem [{\citenamefont {Nayak}\ and\ \citenamefont
  {Wilczek}(1994{\natexlab{a}})}]{Nayak_1}%
  \BibitemOpen
  \bibfield  {author} {\bibinfo {author} {\bibfnamefont {C.}~\bibnamefont
  {Nayak}}\ and\ \bibinfo {author} {\bibfnamefont {F.}~\bibnamefont
  {Wilczek}},\ }\href {\doibase 10.1016/0550-3213(94)90477-4} {\bibfield
  {journal} {\bibinfo  {journal} {Nucl. Phys. B}\ }\textbf {\bibinfo {volume}
  {417}},\ \bibinfo {pages} {359} (\bibinfo {year}
  {1994}{\natexlab{a}})}\BibitemShut {NoStop}%
\bibitem [{\citenamefont {Nayak}\ and\ \citenamefont
  {Wilczek}(1994{\natexlab{b}})}]{Nayak_2}%
  \BibitemOpen
  \bibfield  {author} {\bibinfo {author} {\bibfnamefont {C.}~\bibnamefont
  {Nayak}}\ and\ \bibinfo {author} {\bibfnamefont {F.}~\bibnamefont
  {Wilczek}},\ }\href {\doibase 10.1016/0550-3213(94)90158-9} {\bibfield
  {journal} {\bibinfo  {journal} {Nucl. Phys. B}\ }\textbf {\bibinfo {volume}
  {430}},\ \bibinfo {pages} {534} (\bibinfo {year}
  {1994}{\natexlab{b}})}\BibitemShut {NoStop}%
\bibitem [{\citenamefont {Altshuler}\ \emph {et~al.}(1994)\citenamefont
  {Altshuler}, \citenamefont {Ioffe},\ and\ \citenamefont
  {Millis}}]{Altshuler94}%
  \BibitemOpen
  \bibfield  {author} {\bibinfo {author} {\bibfnamefont {B.~L.}\ \bibnamefont
  {Altshuler}}, \bibinfo {author} {\bibfnamefont {L.~B.}\ \bibnamefont
  {Ioffe}}, and\ \bibinfo {author} {\bibfnamefont {A.~J.}\ \bibnamefont
  {Millis}},\ }\href {\doibase 10.1103/PhysRevB.50.14048} {\bibfield  {journal}
  {\bibinfo  {journal} {Phys. Rev. B}\ }\textbf {\bibinfo {volume} {50}},\
  \bibinfo {pages} {14048} (\bibinfo {year} {1994})}\BibitemShut {NoStop}%
\bibitem [{\citenamefont {Kim}\ \emph {et~al.}(1994)\citenamefont {Kim},
  \citenamefont {Furusaki}, \citenamefont {Wen},\ and\ \citenamefont
  {Lee}}]{YBKim94}%
  \BibitemOpen
  \bibfield  {author} {\bibinfo {author} {\bibfnamefont {Y.~B.}\ \bibnamefont
  {Kim}}, \bibinfo {author} {\bibfnamefont {A.}~\bibnamefont {Furusaki}},
  \bibinfo {author} {\bibfnamefont {X.-G.}\ \bibnamefont {Wen}}, and\
  \bibinfo {author} {\bibfnamefont {P.~A.}\ \bibnamefont {Lee}},\ }\href
  {\doibase 10.1103/PhysRevB.50.17917} {\bibfield  {journal} {\bibinfo
  {journal} {Phys. Rev. B}\ }\textbf {\bibinfo {volume} {50}},\ \bibinfo
  {pages} {17917} (\bibinfo {year} {1994})}\BibitemShut {NoStop}%
\bibitem [{\citenamefont {Mro{\ss}}\ \emph {et~al.}(2010)\citenamefont
  {Mro{\ss}}, \citenamefont {McGreevy}, \citenamefont {Liu},\ and\
  \citenamefont {Senthil}}]{Mross2010}%
  \BibitemOpen
  \bibfield  {author} {\bibinfo {author} {\bibfnamefont {D.~F.}\ \bibnamefont
  {Mro{\ss}}}, \bibinfo {author} {\bibfnamefont {J.}~\bibnamefont {McGreevy}},
  \bibinfo {author} {\bibfnamefont {H.}~\bibnamefont {Liu}}, and\ \bibinfo
  {author} {\bibfnamefont {T.}~\bibnamefont {Senthil}},\ }\href {\doibase%
  10.1103/PhysRevB.82.045121} {\bibfield  {journal} {\bibinfo  {journal} {Phys.
  Rev. B}\ }\textbf {\bibinfo {volume} {82}},\ \bibinfo {pages} {045121}
  (\bibinfo {year} {2010})}\BibitemShut {NoStop}%
\bibitem [{\citenamefont {Varjas}\ \emph {et~al.}(2013)\citenamefont {Varjas},
  \citenamefont {Zaletel},\ and\ \citenamefont {Moore}}]{VarjasZaletelMoore13}%
  \BibitemOpen
  \bibfield  {author} {\bibinfo {author} {\bibfnamefont {D.}~\bibnamefont
  {Varjas}}, \bibinfo {author} {\bibfnamefont {M.~P.}\ \bibnamefont {Zaletel}},
  and\ \bibinfo {author} {\bibfnamefont {J.~E.}\ \bibnamefont {Moore}},\
  }\href {\doibase 10.1103/PhysRevB.88.155314} {\bibfield  {journal} {\bibinfo
  {journal} {Phys. Rev. B}\ }\textbf {\bibinfo {volume} {88}},\ \bibinfo
  {pages} {155314} (\bibinfo {year} {2013})}\BibitemShut {NoStop}%
\bibitem [{Note1()}]{Note1}%
  \BibitemOpen
  \bibinfo {note} {The momenta ${\protect \mathbf {k}}$ are in fact
  gauge-dependent quantities. Here we choose a gauge in which $\DOTSI \intop
  \ilimits@ _0^{L_y} a_y dy = 0$ and the flux is implemented as a twist in the
  boundary condition for the composite fermions. With this choice ${\protect
  \mathbf {k}}\to -{\protect \mathbf {k}}$ corresponds to a 180$^\circ $
  rotation. The momentum transfers ${\protect \mathbf {q}}$ are gauge invariant
  and correspond to the momenta of physical observables.}\BibitemShut {Stop}%
\bibitem [{Note2()}]{Note2}%
  \BibitemOpen
  \bibinfo {note} {We do not have data for every system size for a given
  boundary condition because the simulations fail to converge when a wire is
  close to the edge of the Fermi sea. The different ``root configurations''
  0110 and 0101 (i.e., $K_y$ momentum sectors) give different CF boundary
  conditions and allow us to access different sizes: the wires for one boundary
  condition are closest to the edge of the Fermi sea at the sizes when the
  wires for the other boundary condition are furthest. The omission of $N_w =
  7$ was due only to our finite resources.}\BibitemShut {Stop}%
\bibitem [{\citenamefont {Luttinger}(1960)}]{Luttinger60}%
  \BibitemOpen
  \bibfield  {author} {\bibinfo {author} {\bibfnamefont {J.~M.}\ \bibnamefont
  {Luttinger}},\ }\href {\doibase 10.1103/PhysRev.119.1153} {\bibfield
  {journal} {\bibinfo  {journal} {Phys. Rev.}\ }\textbf {\bibinfo {volume}
  {119}},\ \bibinfo {pages} {1153} (\bibinfo {year} {1960})}\BibitemShut
  {NoStop}%
\bibitem [{\citenamefont {{Balram}}\ \emph {et~al.}(2015)\citenamefont
  {{Balram}}, \citenamefont {{T{\H o}ke}},\ and\ \citenamefont
  {{Jain}}}]{BalramTokeJain2015}%
  \BibitemOpen
  \bibfield  {author} {\bibinfo {author} {\bibfnamefont {A.~C.}\ \bibnamefont
  {{Balram}}}, \bibinfo {author} {\bibfnamefont {C.}~\bibnamefont {{T{\H
  o}ke}}}, and\ \bibinfo {author} {\bibfnamefont {J.~K.}\ \bibnamefont
  {{Jain}}},\ }\href@noop {} {\enquote {\bibinfo {title} {{Do Composite
  Fermions Satisfy Luttinger's Theorem?}}}\ } (\bibinfo {year} {2015}),\
  \bibinfo {note} {unpublished},\ \Eprint {http://arxiv.org/abs/1506.02747}
  {arXiv:1506.02747} \BibitemShut {NoStop}%
\bibitem [{\citenamefont {Calabrese}\ and\ \citenamefont
  {Cardy}(2009)}]{CalabreseCardy}%
  \BibitemOpen
  \bibfield  {author} {\bibinfo {author} {\bibfnamefont {P.}~\bibnamefont
  {Calabrese}}\ and\ \bibinfo {author} {\bibfnamefont {J.}~\bibnamefont
  {Cardy}},\ }\href {\doibase 10.1088/1751-8113/42/50/504005} {\bibfield
  {journal} {\bibinfo  {journal} {J. Phys. A: Math. Theor.}\ }\textbf {\bibinfo
  {volume} {42}},\ \bibinfo {pages} {504005} (\bibinfo {year}
  {2009})}\BibitemShut {NoStop}%
\bibitem [{\citenamefont {Li}\ and\ \citenamefont {Haldane}(2008)}]{LiHaldane}%
  \BibitemOpen
  \bibfield  {author} {\bibinfo {author} {\bibfnamefont {H.}~\bibnamefont
  {Li}}\ and\ \bibinfo {author} {\bibfnamefont {F.}~\bibnamefont {Haldane}},\
  }\href {\doibase 10.1103/PhysRevLett.101.010504} {\bibfield  {journal}
  {\bibinfo  {journal} {Phys. Rev. Lett.}\ }\textbf {\bibinfo {volume} {101}},\
  \bibinfo {pages} {010504} (\bibinfo {year} {2008})}\BibitemShut {NoStop}%
\bibitem [{\citenamefont {Kamburov}\ \emph {et~al.}(2014)\citenamefont
  {Kamburov}, \citenamefont {Liu}, \citenamefont {Mueed}, \citenamefont
  {Shayegan}, \citenamefont {Pfeiffer}, \citenamefont {West},\ and\
  \citenamefont {Baldwin}}]{Kamburov2014}%
  \BibitemOpen
  \bibfield  {author} {\bibinfo {author} {\bibfnamefont {D.}~\bibnamefont
  {Kamburov}}, \bibinfo {author} {\bibfnamefont {Y.}~\bibnamefont {Liu}},
  \bibinfo {author} {\bibfnamefont {M.~A.}\ \bibnamefont {Mueed}}, \bibinfo
  {author} {\bibfnamefont {M.}~\bibnamefont {Shayegan}}, \bibinfo {author}
  {\bibfnamefont {L.~N.}\ \bibnamefont {Pfeiffer}}, \bibinfo {author}
  {\bibfnamefont {K.~W.}\ \bibnamefont {West}}, and\ \bibinfo {author}
  {\bibfnamefont {K.~W.}\ \bibnamefont {Baldwin}},\ }\href {\doibase%
  10.1103/PhysRevLett.113.196801} {\bibfield  {journal} {\bibinfo  {journal}
  {Phys. Rev. Lett.}\ }\textbf {\bibinfo {volume} {113}},\ \bibinfo {pages}
  {196801} (\bibinfo {year} {2014})}\BibitemShut {NoStop}%
\bibitem [{\citenamefont {Fidkowski}\ \emph {et~al.}(2013)\citenamefont
  {Fidkowski}, \citenamefont {Chen},\ and\ \citenamefont
  {Vishwanath}}]{Fidkowski2013}%
  \BibitemOpen
  \bibfield  {author} {\bibinfo {author} {\bibfnamefont {L.}~\bibnamefont
  {Fidkowski}}, \bibinfo {author} {\bibfnamefont {X.}~\bibnamefont {Chen}},\
  and\ \bibinfo {author} {\bibfnamefont {A.}~\bibnamefont {Vishwanath}},\
  }\href {\doibase 10.1103/PhysRevX.3.041016} {\bibfield  {journal} {\bibinfo
  {journal} {Phys. Rev. X}\ }\textbf {\bibinfo {volume} {3}},\ \bibinfo {pages}
  {041016} (\bibinfo {year} {2013})}\BibitemShut {NoStop}%
\bibitem [{\citenamefont {Lee}(2009)}]{SSLee2009}%
  \BibitemOpen
  \bibfield  {author} {\bibinfo {author} {\bibfnamefont {S.-S.}\ \bibnamefont
  {Lee}},\ }\href {\doibase 10.1103/PhysRevB.80.165102} {\bibfield  {journal}
  {\bibinfo  {journal} {Phys. Rev. B}\ }\textbf {\bibinfo {volume} {80}},\
  \bibinfo {pages} {165102} (\bibinfo {year} {2009})}\BibitemShut {NoStop}%
\bibitem [{\citenamefont {Metlitski}\ and\ \citenamefont
  {Sachdev}(2010{\natexlab{a}})}]{Metlitski2010a}%
  \BibitemOpen
  \bibfield  {author} {\bibinfo {author} {\bibfnamefont {M.~A.}\ \bibnamefont
  {Metlitski}}\ and\ \bibinfo {author} {\bibfnamefont {S.}~\bibnamefont
  {Sachdev}},\ }\href {\doibase 10.1103/PhysRevB.82.075127} {\bibfield
  {journal} {\bibinfo  {journal} {Phys. Rev. B}\ }\textbf {\bibinfo {volume}
  {82}},\ \bibinfo {pages} {075127} (\bibinfo {year}
  {2010}{\natexlab{a}})}\BibitemShut {NoStop}%
\bibitem [{\citenamefont {Metlitski}\ and\ \citenamefont
  {Sachdev}(2010{\natexlab{b}})}]{Metlitski2010b}%
  \BibitemOpen
  \bibfield  {author} {\bibinfo {author} {\bibfnamefont {M.~A.}\ \bibnamefont
  {Metlitski}}\ and\ \bibinfo {author} {\bibfnamefont {S.}~\bibnamefont
  {Sachdev}},\ }\href {\doibase 10.1103/PhysRevB.82.075128} {\bibfield
  {journal} {\bibinfo  {journal} {Phys. Rev. B}\ }\textbf {\bibinfo {volume}
  {82}},\ \bibinfo {pages} {075128} (\bibinfo {year}
  {2010}{\natexlab{b}})}\BibitemShut {NoStop}%
\bibitem [{\citenamefont {{Fisher}}\ \emph {et~al.}(2008)\citenamefont
  {{Fisher}}, \citenamefont {{Motrunich}},\ and\ \citenamefont
  {{Sheng}}}]{Fisher2008_gen}%
  \BibitemOpen
  \bibfield  {author} {\bibinfo {author} {\bibfnamefont {M.~P.~A.}\
  \bibnamefont {{Fisher}}}, \bibinfo {author} {\bibfnamefont {O.~I.}\
  \bibnamefont {{Motrunich}}}, and\ \bibinfo {author} {\bibfnamefont {D.~N.}\
  \bibnamefont {{Sheng}}},\ }\href@noop {} {\enquote {\bibinfo {title} {{Spin,
  Bose, and Non-Fermi Liquid Metals in Two Dimensions: Accessing via Multi-Leg
  Ladders}},}\ } (\bibinfo {year} {2008}),\ \bibinfo {note} {unpublished},\
  \Eprint {http://arxiv.org/abs/0812.2955} {arXiv:0812.2955} \BibitemShut
  {NoStop}%
\bibitem [{\citenamefont {Sheng}\ \emph {et~al.}(2008)\citenamefont {Sheng},
  \citenamefont {Motrunich}, \citenamefont {Trebst}, \citenamefont {Gull},\
  and\ \citenamefont {Fisher}}]{Sheng2008_2legDBL}%
  \BibitemOpen
  \bibfield  {author} {\bibinfo {author} {\bibfnamefont {D.~N.}\ \bibnamefont
  {Sheng}}, \bibinfo {author} {\bibfnamefont {O.~I.}\ \bibnamefont
  {Motrunich}}, \bibinfo {author} {\bibfnamefont {S.}~\bibnamefont {Trebst}},
  \bibinfo {author} {\bibfnamefont {E.}~\bibnamefont {Gull}}, and\ \bibinfo
  {author} {\bibfnamefont {M.~P.~A.}\ \bibnamefont {Fisher}},\ }\href {\doibase%
  10.1103/PhysRevB.78.054520} {\bibfield  {journal} {\bibinfo  {journal} {Phys.
  Rev. B}\ }\textbf {\bibinfo {volume} {78}},\ \bibinfo {pages} {054520}
  (\bibinfo {year} {2008})}\BibitemShut {NoStop}%
\bibitem [{\citenamefont {Mishmash}\ \emph {et~al.}(2011)\citenamefont
  {Mishmash}, \citenamefont {Block}, \citenamefont {Kaul}, \citenamefont
  {Sheng}, \citenamefont {Motrunich},\ and\ \citenamefont
  {Fisher}}]{Mishmash2011_4legDBL}%
  \BibitemOpen
  \bibfield  {author} {\bibinfo {author} {\bibfnamefont {R.~V.}\ \bibnamefont
  {Mishmash}}, \bibinfo {author} {\bibfnamefont {M.~S.}\ \bibnamefont {Block}},
  \bibinfo {author} {\bibfnamefont {R.~K.}\ \bibnamefont {Kaul}}, \bibinfo
  {author} {\bibfnamefont {D.~N.}\ \bibnamefont {Sheng}}, \bibinfo {author}
  {\bibfnamefont {O.~I.}\ \bibnamefont {Motrunich}}, and\ \bibinfo {author}
  {\bibfnamefont {M.~P.~A.}\ \bibnamefont {Fisher}},\ }\href {\doibase%
  10.1103/PhysRevB.84.245127} {\bibfield  {journal} {\bibinfo  {journal} {Phys.
  Rev. B}\ }\textbf {\bibinfo {volume} {84}},\ \bibinfo {pages} {245127}
  (\bibinfo {year} {2011})}\BibitemShut {NoStop}%
\bibitem [{\citenamefont {Block}\ \emph {et~al.}(2011)\citenamefont {Block},
  \citenamefont {Sheng}, \citenamefont {Motrunich},\ and\ \citenamefont
  {Fisher}}]{Block2011_4legSBM}%
  \BibitemOpen
  \bibfield  {author} {\bibinfo {author} {\bibfnamefont {M.~S.}\ \bibnamefont
  {Block}}, \bibinfo {author} {\bibfnamefont {D.~N.}\ \bibnamefont {Sheng}},
  \bibinfo {author} {\bibfnamefont {O.~I.}\ \bibnamefont {Motrunich}}, and\
  \bibinfo {author} {\bibfnamefont {M.~P.~A.}\ \bibnamefont {Fisher}},\ }\href
  {\doibase 10.1103/PhysRevLett.106.157202} {\bibfield  {journal} {\bibinfo
  {journal} {Phys. Rev. Lett.}\ }\textbf {\bibinfo {volume} {106}},\ \bibinfo
  {pages} {157202} (\bibinfo {year} {2011})}\BibitemShut {NoStop}%
\bibitem [{\citenamefont {{Jiang}}\ \emph {et~al.}(2013)\citenamefont
  {{Jiang}}, \citenamefont {{Block}}, \citenamefont {{Mishmash}}, \citenamefont
  {{Garrison}}, \citenamefont {{Sheng}}, \citenamefont {{Motrunich}},\ and\
  \citenamefont {{Fisher}}}]{Jiang2013_dmetal}%
  \BibitemOpen
  \bibfield  {author} {\bibinfo {author} {\bibfnamefont {H.-C.}\ \bibnamefont
  {{Jiang}}}, \bibinfo {author} {\bibfnamefont {M.~S.}\ \bibnamefont
  {{Block}}}, \bibinfo {author} {\bibfnamefont {R.~V.}\ \bibnamefont
  {{Mishmash}}}, \bibinfo {author} {\bibfnamefont {J.~R.}\ \bibnamefont
  {{Garrison}}}, \bibinfo {author} {\bibfnamefont {D.~N.}\ \bibnamefont
  {{Sheng}}}, \bibinfo {author} {\bibfnamefont {O.~I.}\ \bibnamefont
  {{Motrunich}}}, and\ \bibinfo {author} {\bibfnamefont {M.~P.~A.}\
  \bibnamefont {{Fisher}}},\ }\href {\doibase 10.1038/nature11732} {\bibfield
  {journal} {\bibinfo  {journal} {Nature}\ }\textbf {\bibinfo {volume} {493}},\
  \bibinfo {pages} {39} (\bibinfo {year} {2013})}\BibitemShut {NoStop}%
\bibitem [{\citenamefont {{Wang}}\ and\ \citenamefont
  {{Senthil}}(2015{\natexlab{b}})}]{CWangSenthil:HalfLL2015}%
  \BibitemOpen
  \bibfield  {author} {\bibinfo {author} {\bibfnamefont {C.}~\bibnamefont
  {{Wang}}}\ and\ \bibinfo {author} {\bibfnamefont {T.}~\bibnamefont
  {{Senthil}}},\ }\href@noop {} {\enquote {\bibinfo {title} {Half-filled
  {L}andau level, topological insulator surfaces, and three dimensional quantum
  spin liquids},}\ } (\bibinfo {year} {2015}{\natexlab{b}}),\ \bibinfo {note}
  {unpublished},\ \Eprint {http://arxiv.org/abs/1507.08290} {arXiv:1507.08290}
  \BibitemShut {NoStop}%
\bibitem [{\citenamefont {Fidkowski}\ and\ \citenamefont
  {Kitaev}(2010)}]{FidkowskiKitaev:10}%
  \BibitemOpen
  \bibfield  {author} {\bibinfo {author} {\bibfnamefont {L.}~\bibnamefont
  {Fidkowski}}\ and\ \bibinfo {author} {\bibfnamefont {A.}~\bibnamefont
  {Kitaev}},\ }\href {\doibase 10.1103/PhysRevB.81.134509} {\bibfield
  {journal} {\bibinfo  {journal} {Phys. Rev. B}\ }\textbf {\bibinfo {volume}
  {81}},\ \bibinfo {pages} {134509} (\bibinfo {year} {2010})}\BibitemShut
  {NoStop}%
\bibitem [{\citenamefont {Turner}\ \emph {et~al.}(2011)\citenamefont {Turner},
  \citenamefont {Pollmann},\ and\ \citenamefont
  {Berg}}]{TurnerPollmannBerg:11}%
  \BibitemOpen
  \bibfield  {author} {\bibinfo {author} {\bibfnamefont {A.~M.}\ \bibnamefont
  {Turner}}, \bibinfo {author} {\bibfnamefont {F.}~\bibnamefont {Pollmann}},\
  and\ \bibinfo {author} {\bibfnamefont {E.}~\bibnamefont {Berg}},\ }\href
  {\doibase 10.1103/PhysRevB.83.075102} {\bibfield  {journal} {\bibinfo
  {journal} {Phys. Rev. B}\ }\textbf {\bibinfo {volume} {83}},\ \bibinfo
  {pages} {075102} (\bibinfo {year} {2011})}\BibitemShut {NoStop}%
\bibitem [{\citenamefont {Fidkowski}\ and\ \citenamefont
  {Kitaev}(2011)}]{FidkowskiKitaev:11}%
  \BibitemOpen
  \bibfield  {author} {\bibinfo {author} {\bibfnamefont {L.}~\bibnamefont
  {Fidkowski}}\ and\ \bibinfo {author} {\bibfnamefont {A.}~\bibnamefont
  {Kitaev}},\ }\href {\doibase 10.1103/PhysRevB.83.075103} {\bibfield
  {journal} {\bibinfo  {journal} {Phys. Rev. B}\ }\textbf {\bibinfo {volume}
  {83}},\ \bibinfo {pages} {075103} (\bibinfo {year} {2011})}\BibitemShut
  {NoStop}%
\bibitem [{\citenamefont {Fischler}\ \emph {et~al.}(1979)\citenamefont
  {Fischler}, \citenamefont {Kogut},\ and\ \citenamefont {Susskind}}]{FKS79}%
  \BibitemOpen
  \bibfield  {author} {\bibinfo {author} {\bibfnamefont {W.}~\bibnamefont
  {Fischler}}, \bibinfo {author} {\bibfnamefont {J.}~\bibnamefont {Kogut}},\
  and\ \bibinfo {author} {\bibfnamefont {L.}~\bibnamefont {Susskind}},\ }\href
  {\doibase 10.1103/PhysRevD.19.1188} {\bibfield  {journal} {\bibinfo
  {journal} {Phys. Rev. D}\ }\textbf {\bibinfo {volume} {19}},\ \bibinfo
  {pages} {1188} (\bibinfo {year} {1979})}\BibitemShut {NoStop}%
\bibitem [{\citenamefont {Metlitski}(2007)}]{Metlitski07}%
  \BibitemOpen
  \bibfield  {author} {\bibinfo {author} {\bibfnamefont {M.~A.}\ \bibnamefont
  {Metlitski}},\ }\href {\doibase 10.1103/PhysRevD.75.045004} {\bibfield
  {journal} {\bibinfo  {journal} {Phys. Rev. D}\ }\textbf {\bibinfo {volume}
  {75}},\ \bibinfo {pages} {045004} (\bibinfo {year} {2007})}\BibitemShut
  {NoStop}%
\bibitem [{\citenamefont {Kim}\ and\ \citenamefont {Lee}(1999)}]{KimLee99}%
  \BibitemOpen
  \bibfield  {author} {\bibinfo {author} {\bibfnamefont {D.~H.}\ \bibnamefont
  {Kim}}\ and\ \bibinfo {author} {\bibfnamefont {P.~A.}\ \bibnamefont {Lee}},\
  }\href {\doibase 10.1006/aphy.1998.5888} {\bibfield  {journal} {\bibinfo
  {journal} {Ann. Phys. (N.Y.)}\ }\textbf {\bibinfo {volume} {272}},\ \bibinfo
  {pages} {130} (\bibinfo {year} {1999})}\BibitemShut {NoStop}%
\bibitem [{\citenamefont {Sheng}\ \emph {et~al.}(2009)\citenamefont {Sheng},
  \citenamefont {Motrunich},\ and\ \citenamefont
  {Fisher}}]{Sheng2009_zigzagSBM}%
  \BibitemOpen
  \bibfield  {author} {\bibinfo {author} {\bibfnamefont {D.~N.}\ \bibnamefont
  {Sheng}}, \bibinfo {author} {\bibfnamefont {O.~I.}\ \bibnamefont
  {Motrunich}}, and\ \bibinfo {author} {\bibfnamefont {M.~P.~A.}\
  \bibnamefont {Fisher}},\ }\href {\doibase 10.1103/PhysRevB.79.205112}
  {\bibfield  {journal} {\bibinfo  {journal} {Phys. Rev. B}\ }\textbf {\bibinfo
  {volume} {79}},\ \bibinfo {pages} {205112} (\bibinfo {year}
  {2009})}\BibitemShut {NoStop}%
\bibitem [{\citenamefont {Bergholtz}\ and\ \citenamefont
  {Karlhede}(2005)}]{Bergholtz2005}%
  \BibitemOpen
  \bibfield  {author} {\bibinfo {author} {\bibfnamefont {E.~J.}\ \bibnamefont
  {Bergholtz}}\ and\ \bibinfo {author} {\bibfnamefont {A.}~\bibnamefont
  {Karlhede}},\ }\href {\doibase 10.1103/PhysRevLett.94.026802} {\bibfield
  {journal} {\bibinfo  {journal} {Phys. Rev. Lett.}\ }\textbf {\bibinfo
  {volume} {94}},\ \bibinfo {pages} {026802} (\bibinfo {year}
  {2005})}\BibitemShut {NoStop}%
\end{thebibliography}%

\clearpage
\appendix
\begin{widetext}
\phantomsection
\addcontentsline{toc}{part}{Appendix}

\tableofcontents

\section{Particle-hole symmetry and Kramers degeneracy of composite fermions}
\label{app:PH}

Here we discuss important formal properties of the particle-hole symmetry.
One can define the following particle-hole (PH) transformation: 
\begin{align}
	\PH:
	\begin{array}{ccc}
		c(\br) & \to & c^\dag(\br) \;,
	\\	c^\dag(\br) & \to & c(\br) \;,
	\\	\quad i & \to & -i \;,
	\end{array}
\end{align}
where $c(\br)$ is the electron destruction operator in the continuum.
Note that $\PH$ is an anti-unitary transformation.
Now consider restricting to the lowest Landau level.
Then, if $c_j^\dagger$ creates a particle in the lowest Landau level orbital with wavefunction $\varphi_j(\br)$, thus 
\begin{align}
	c_j^\dag = \int_\br \varphi_j(\br)\, c^\dag(\br) ~,
	\qquad
	c_j = \int_\br \varphi_j^*(\br)\, c(\br) ~,
	\label{eq:c_continuum}
\end{align}
we have that
\begin{align}
	\PH: c_j \leftrightarrow c_j^\dagger ~,
	\quad  i \leftrightarrow -i ~.
	\label{eq:PHorb}
\end{align}
Note that even though $c_j$'s depend on the choice of orbitals, the PH transformation is basis-independent.

We also need to specify the action of particle-hole transformation on the vacuum state $\ket{0}$, which is defined as $c_j\ket{0} = 0$ for all $j$ (i.e.,~the empty state).
Let us denote $\ket{0'} \equiv \PH \ket{0}$.
Under particle-hole the appropriate condition is $c^\dagger_j\ket{0'} = 0$ for all $j$, which implies that the transformed state is completely filled, $\ket{0'} = \prod_j c_j^\dagger \ket{0}$.
In the latter equation, we understand some fixed ordering of fermion fields, e.g.,
\begin{equation}
	\ket{0'} \equiv \PH \ket{0} \equiv c_1^\dagger c_2^\dagger \dots c_{\Norb}^\dagger \ket{0} ~,
\end{equation}
where $\Norb$ is the total number of orbitals.

The key result of this section, which we will show momentarily, is
\begin{align}
	\fbox{$ \PH^2 = (-1)^{\Norb (\Norb - 1)/2} $} ~.
	\label{eq:PHsquare}
\end{align}


As a warm-up example, consider a simple case of one electron with two orbitals.
Call the two orbitals $c_1$, $c_2$, and label the states $\ket{1} \equiv c_1^\dagger \ket{0}$ and $\ket{2} \equiv c_2^\dagger \ket{0}$.
Now, under particle-hole we have 
\begin{align*}
	\ket{1} = c_1^\dagger \ket{0}  &\quad\Longrightarrow\quad  \PH\ket{1} = c_1 (c_1^\dagger c_2^\dagger \ket{0})  =  {+\ket{2}} ~,\\
	\ket{2} = c_2^\dagger \ket{0}  &\quad\Longrightarrow\quad  \PH\ket{2} = c_2 (c_1^\dagger c_2^\dagger \ket{0})  =  {-\ket{1}} ~.
\end{align*}
Thus $\PH^2 = -1$ for $\Norb = 2$.
Since $\PH$ is an anti-unitary operation, there is no way to gauge out the minus sign, and this implies a Kramers degeneracy of the two states. 

This can be readily extended to an arbitrary number of orbitals.
Consider a general basis state that can be represented as:
\begin{equation}
	\ket{\Psi} = \prod_{j=1}^{\Norb} \big( c_j^\dagger \big)^{n_j} \ket{0} ~,
\end{equation}
where $n_j \in \{0,1\}$.
Acting twice with particle-hole yields $\PH^2 |\Psi\rangle = \eta |\Psi\rangle$ for some number $\eta$.
We first argue that $\eta$ is independent of the number of creation operators $N_e = \sum_j n_j$.
Since $\PH^2 c_j^\dag \PH^{-2} = c_j^\dag$, one can commute the $\PH^2$ operator across all the creation operators:
\begin{align}
	\PH^2 \ket{\Psi} = \PH^2 \prod_{j=1}^{\Norb} \big( c_j^\dagger \big)^{n_j} \ket{0}
		= \prod_{j=1}^{\Norb} \big( c_j^\dagger \big)^{n_j} \PH^2 \ket{0}
		= \ket{\Psi} \times \braket{0 | \PH^2 | 0} ~.
\end{align}
Therefore $\ket{0}$ picks up the same sign $\eta$ under $\PH^2$ as that for every state with $\Norb$ orbitals.
To compute $\eta$, see that $\eta\ket{0} = \PH^2\ket{0} = c_1 c_2 \cdots c_{\Norb} c_1^\dagger c_2^\dagger \cdots c_{\Norb}^\dagger \ket{0}$.
Exchanging the raising/lowering operators until $c_j$ is next to $c_j^\dag$ takes $j-1$ swap operations, 
	for a total of $\sum_{j=1}^{\Norb} (j-1) = \Norb (\Norb - 1)/2$ negative signs.
Therefore, we have shown that $\eta = (-1)^{\Norb (\Norb - 1)/2}$ as advertised in Eq.~\eqref{eq:PHsquare}.
Again, we emphasize that the sign of $\PH^2$ is determined by $\Norb$ and not by the number of electrons.
The sign is reversed each time we add two orbitals (i.e., two flux quanta).

Particle-hole takes a state at filling $\nu$, where $0 \leq \nu \leq 1$, and transforms it into a state at filling $1 - \nu$.
A state can only be PH-symmetric if it is at half filling ($\nu = 1/2$); thus $\Norb = 2 N_e$ where $N_e$ is the total number of electrons.
In such case the sign depends only on whether $N_e$ is even or odd: $(-1)^{2N_e (2N_e-1)/2} = (-1)^{N_e}$.
For particle-hole-symmetric states, we claim that states with an odd number of electrons must be doubly degenerate, which can then be interpreted as arising from a Kramers-like degeneracy of composite fermions.
This may appear strange since time-reversal symmetry is broken by the magnetic field, and we always consider just one spin component of the electrons.
However, one can define $\TR' \equiv \PH$ as a new kind of time-reversal, which acts like regular time-reversal symmetry on the composite fermions leading to the Kramers degeneracy.

Given Eq.~\eqref{eq:PHsquare}, one may ponder on the cases when $\Norb$ is odd.
If we naively start with two orbitals with $\PH^2 = -1$ and take the square root, we may be mislead into thinking that $\PH^2 = \pm i$ for a single orbital.
Of course, this is false: Eq.~\eqref{eq:PHsquare} gives $\PH^2 = 1$ in this case, which we can also verify directly.  
Indeed,
$\PH \ket{0} = e^{i\alpha} \ket{1} = e^{i\alpha} c^\dag \ket{0}$ 
implies that
$\PH^2 \ket{0} = \PH (e^{i\alpha} c^\dag \ket{0}) = e^{-i\alpha} c (\PH \ket{0}) = \ket{0}$
and thus $\PH^2 = 1$ on a single orbital.
The resolution to this seeming paradox is that when acting on a single orbital $\PH$ is a fermionic operator---changing the fermion parity number---and thus does not have a trivial product structure when combining orbitals.
Labeling the two orbitals $A$, $B$, and defining operator $\PH \equiv \PH_A \PH_B$, we obtain
\begin{align}
	\PH^2 = (\PH_A \PH_B) (\PH_A \PH_B)
	= -\PH_A \underbrace{\PH_A \PH_B}_{\text{anticommute}} \PH_B
	= -\PH_A^2 \PH_B^2 = -1 ~,
\end{align}
where we used anticommutation of $\PH_A$ and $\PH_B$.
We recover the result Eq.~\eqref{eq:PHsquare} for $\Norb = 2$.

\roger{Maybe we should mention $\mathrm{U(1)} \times \mathbb{Z}_2^C$ instead?}
\lesik{I am not familiar with all details of the 1d classification.  From Max's comment it seems that the PH in LLL is actually different from 1d SPT, so maybe the question which one to use is moot?  His point is that PH is nonlocal and unlike any local symmetry, although I did not get why one needs to invoke translation symmetry - SPTs just need the on-site symmetry and end-states are stable to small disorder.  Leave it for Roger to judge.}

In fact, one can view $\PH$ on a single orbital as fractionalization (projective representation) of the complex-conjugation operator $\mathcal{K}$.
While $\mathcal{K}$ is a bosonic operator squaring to 1, the same does not hold true in general for $\PH$.
For $\Norb = 1$ (or 3), $\PH$ is a fermionic operator.
For $\Norb = 2$, $\PH^2 = -1$ and the system has a Kramers degeneracy.
Only for $\Norb$ multiple of 4 the $\PH$ operator behaves like complex-conjugation.
Equivalently, we can only find eigenstates (with definite fermion parity) of the $\PH$ operator when $\Norb \in 4\mathbb{Z}$.
When $\Norb \notin 4\mathbb{Z}$ and the Hamiltonian respects particle-hole symmetry, there will always be a ground state degeneracy.
(Note that at odd $\Norb$ all states break PH symmetry, so there is trivial two-fold degeneracy if the chemical potential is chosen such that the Hamiltonian is PH-symmetric.) 

The situation parallels the ``BDI-wire'' studied by Fidkowski and Kitaev,\cite{FidkowskiKitaev:10} where they considered a fermionic system which possesses a time-reversal symmetry squaring to $+1$.
There is a $\mathbb{Z}_8$ classification, i.e., phases described by an integer $n$ defined modulo 8.
When $n \neq 8$, the time-reversal operator also fractionalizes at each end of the 1D system.\cite{TurnerPollmannBerg:11, FidkowskiKitaev:11}
The case with $\PH$ operator acting on a single orbital is akin to the Fidkowski-Kitaev $n = 2$ case.
The $\PH$ operator acting on two orbitals corresponds to the $n = 4$ case.
When acting on four orbitals, the $\PH$ operator squares to $+1$, which is analogous to the $n = 8 \equiv 0 \pmod 8$ case in the BDI-wire.
In the latter case, the wire is in a trivial topological phase and has no ground state degeneracy.

\maxm{The weird $\PH^2$ is not directly related to non-locality. For instance, for spin Hamiltonians with $S  = 1/2$ we have $T^2 = -1$ on each spin, which is also a projective representation of $T$. Nevertheless, the symmetry action is local. Moreover, one can always group spin $1/2$'s into local pairs so that $T$ acts trivially on the pair. Putting the pairs into singlets gives a trivial product state. It is only in the presence of translational symmetry that the projective nature of representation on each site plays a role (leading to various versions of Lieb-Shultz-Mattis-Oshikawa-Hastings-Zaletel). However, the fact that no PH symmetric trivial state exists in the LLL does not rely on translational symmetry (one can break translations, as long as one preserves PH symmetry). In this way, PH symmetry in the LLL is analogous to symmetries on the surface of 3D topological phases, where again translational symmetry does not play a role (i.e. we are talking about ``strong" topological invariants).}

\section{Quantum Hall on a cylinder}
\label{app:cylinder}

To make the manuscript more self-contained, here we describe our setup for the DMRG studies of quantum Hall problems on the infinite cylinder.
We also describe symmetries in this setup, which are useful, e.g., in the discussion of putative effective field theories for the CFL at $\nu = 1/2$.
The electron kinetic energy Hamiltonian in the Landau gauge $(A_x, A_y) = (0, Bx)$ is
\begin{equation}
H_{\rm el.kin.} = -\frac{\hbar^2}{2m} \left[\nabla_x^2 + \left(\nabla_y - i\frac{x}{\ell_B^2} \right)^2 \right] ~,
\end{equation}
where $\ell_B \equiv \sqrt{\hbar/(e B)}$ is the magnetic length.
Orbitals in the lowest Landau level in this gauge are
\begin{equation}
\varphi_j(\br) = \frac{e^{i \frac{2\pi}{L_y} j y}}{\sqrt{L_y}} ~
\frac{e^{-\frac{(x - X_j)^2}{2 \ell_B^2}}}{\pi^{1/4} \ell_B^{1/2}} ~,
\label{phi_cyl}
\end{equation}
where $\br \equiv (x, y)$.  
The orbitals are labeled by an integer $j$; such an orbital is a plane wave in the $y$-direction with wavevector $k_y = (2\pi/L_y) j$ and is localized in the $x$-direction around position $X_j \equiv \ell_B^2 (2\pi/L_y) j$.
We can naturally view these orbitals as forming a 1D chain labeled by $j \in \mathbb{Z}$.
Going to the second-quantized language, we expand the continuum electron annihilation operator $c(\br)$ in terms of the annihilation operators $c_j$ for these orbitals, $c(\br) = \sum_j \varphi_j(\br) c_j$, cf.~Eq.~\eqref{eq:c_continuum}.

On the cylinder, the above electron kinetic energy has the following symmetries, which we readily translate to transformations of the $c_j$ fields.
$H_{\rm el.kin.}$ is invariant under translation by an arbitrary amount $\Delta y$ in the $y$-direction, which becomes multiplication by a $j$-dependent phase factor for the $c_j$ fermions:
\begin{equation}
T_y[\Delta y]: c_j \to e^{i \frac{2\pi}{L_y} j \Delta y} c_j ~.
\end{equation}
A generator for this symmetry is 
\begin{equation}
K_y \equiv \frac{2\pi}{L_y} \sum_j j n_j ~,
\label{Kychain}
\end{equation}
which is proportional to the ``center of mass'' position of the system of electrons on the 1D chain of orbitals.

$H_{\rm el.kin.}$ is also invariant under translation by a discrete amount $\Delta x = \ell_B^2 (2\pi/L_y)$ in the $x$-direction, accompanied by a gauge transformation that respects the periodic boundary conditions in the $y$-direction.
This discrete step is precisely such as to accommodate one flux quantum in the swept area of $L_y \Delta x$, and is also the spacing between centers of neighboring orbitals described above.
Naturally, this symmetry becomes translation of the 1D chain of orbitals,
\begin{equation}
T_x[\Delta x = \ell_B^2 (2\pi/L_y)]: c_j \to c_{j+1} ~.
\label{Txchain}
\end{equation}

The physics in an infinite 2D system is invariant under spatial rotations.
However, when restricted to the cylinder, we only have $180$ degree rotation symmetry left, which is the same as spatial inversion $\br \to -\br$.
In the above orbital language, this becomes
\begin{eqnarray}
I: c_j \to c_{-j} ~.
\end{eqnarray}
In some situations it is convenient to consider inversion about a midpoint between two neighboring orbitals, which can be also viewed as a combination of the above inversion and translation by one orbital,
\begin{eqnarray}
I': c_j \to c_{-j + 1} ~.
\label{Iprimechain}
\end{eqnarray}

Finally, $H_{\rm el.kin.}$ is invariant under mirror reflection in the $x$-axis, $(x, y) \to (x, -y)$, combined with time-reversal $\TR$ for spinless electrons (i.e., complex conjugation).
We denote this symmetry as $M_x \TR$, which becomes in the orbital language
\begin{equation}
M_x \TR: c_j \to c_j ~,~ \quad i \to -i ~.
\label{MxTchain}
\end{equation}
Thus, this symmetry is simply complex conjugation in the 1D chain, which ensures that the electron Hamiltonian including all interactions is real-valued in this basis.
DMRG can then use real numbers only, which simplifies numerical calculations and automatically maintains this symmetry.

We assume that all electron interactions also respect the above symmetries.  
In the standard procedure, we project the Coulomb interaction into the lowest Landau level and obtain
\begin{equation}
	H_{\rm el.int.} = \sum_j \sum_{n \geq 0, m > 0}
	\left[W_{mn} c_j^\dagger c_{j+n} c_{j+m+n} c_{j+m+2n}^\dagger + \Hc \right] ~,
	\label{eq:Helint}
\end{equation}
with calculable $W_{mn}$.
The $T_x$ and $T_y$ translation symmetries are manifestly present as chain translation symmetry and center-of-mass conservation respectively; the inversion $I$ is also already imposed in the above form, while the anti-unitary mirror $M_x \TR$ requires $W_{mn} \in \mathbb{R}$.
These details of the setup are all implemented in the DMRG but are not important for the rest of this Appendix.
The electronic Hamiltonian becomes more complicated when we allow tunneling into a second quantum Hall well or allow Landau level mixing, see Refs.~\onlinecite{ZaletelMixing} for details, but the symmetry analysis remains.

Let us now consider particle-hole transformation in the lowest Landau level.
It is expressed in terms of orbitals in Eq.~\eqref{eq:PHorb} and is a symmetry of $H_{\rm el.int.}$ at $\nu = 1/2$.
In fact, it is easy to see that any four-electron term $w_{jklm} c_j^\dagger c_k^\dagger c_l c_m + \Hc$, with distinct $j, k, l, m$, is invariant under PH.
Note that the interaction amplitudes $w_{jklm}$ can be arbitrary complex numbers and do not need to respect any of the spatial symmetries discussed earlier, as long as $j, k, l, m$ are distinct.  
If some indices coincide, e.g., $j = l$, a PH-symmetric form would have $c_j^\dagger c_j$ replaced with $c_j^\dagger c_j - \frac{1}{2}$, which can be interpreted as requiring specific relations between interactions and (possibly site-dependent) chemical potentials.
On the other hand, requiring $N_e = \Norb/2$ in a translationally invariant system automatically satisfies this.
Note also that any six-fermion interaction will violate PH symmetry, so generically it is not a symmetry of a truly microscopic electronic Hamiltonian that includes physical effects such as tunneling to another layer or Landau level mixing.

\section{Numerical methods}
\label{app:numerical}

\subsection{Infinite DMRG}
\label{app:DMRG}
Our implementation of infinite quantum Hall DMRG is described in Ref.~\onlinecite{ZaletelMixing}.
Here we provide some additional details specific to simulating the $\nu=\frac12$ state via DMRG.

DMRG is a 1D method.
Using the cylinder basis described in App.~\ref{app:cylinder}, we map the problem to a 1D fermion chain with a  basis labeled by the orbital occupations $n_j$.
Each DMRG run starts with some orbital product state (e.g., $\ket{\cdots 1010 \cdots}$) and then optimizes the state to reduce the variational energy.
As discussed above, the anti-unitary mirror symmetry $M_x \TR$ is maintained automatically by working with real-valued arithmetics.
Furthermore, the DMRG explicitly preserves $K_y$, so the initial state fixes a sector with definite $K_y$.
When started in one sector, the DMRG then resides solely in that sector.
We need to check multiple sectors to determine which one contains the ground state.
In the present study, we use a four-site unit cell when we set up MPS for the infinite-DMRG, and we can readily check that ``root'' configurations $1010$, $0101$, $1100$, and $0110$ generate four different such sectors.
They also exhaust all different sectors accessible with such a unit cell at half-filling, since $0011$ can be connected by center-of-mass-preserving terms to $1100$ (e.g., on a torus with $\Norb$ multiple of four), while $1001$ can be connected to $0110$.
Note also that the $1010$ and $0101$ sectors are related by $T_x$, and so are the $1100$ and $0110$ sectors, which gives exact two-fold degeneracy of eigenenergies.
Such a degeneracy in the half-filled Landau level is a consequence of the commutation relation $T_x T_y = -T_y T_x$ at $\nu = 1/2$. 
To be precise, the last equation is for an $L_x \times L_y$ torus, where we consider translations $T_x \equiv T_x[\ell_B^2 (2\pi/L_y)]$, which is the same as in Eq.~\eqref{Txchain}, and $T_y \equiv T_y[\ell_B^2 (2\pi/L_x)]$.  Using the generator $K_y$, Eq.~\eqref{Kychain}, the latter can be written as $T_y = \exp[i (2\pi/\Norb) \sum_j j n_j]$, where $\Norb = L_x L_y/(2\pi \ell_B^2)$ is the total number of orbitals, which is the same as the number of sites in the 1D chain representation.
In our numerical calculations, we then restrict to studying just two sectors, $0101$ and $0110$.

Note also that the $\PH$ transformation Eq.~\eqref{eq:PHorb} and $T_y$ anticommute (assuming even $\Norb$).
Hence, generically the $\PH$ will connect sectors with different $K_y$; however, there are exceptions where the $\PH$ acts within a sector.
For our four-site unit cell, the $\PH$ connects the $1010$ and $0101$ sectors.
On the other hand, the $\PH$ maps $0110$ to $1001$ which belongs to the same $K_y$ sector, so it acts within this sector; the PH is now a discrete symmetry for the Hamiltonian restricted to this sector.

The DMRG can spontaneously break the PH symmetry due to quantitative errors induced in the results by the finite bond dimension.
As bond dimension is increased, the energy approaches the infinite-bond-dimension limit from above as a power law in the bond dimension.\cite{ZaletelMixing}
The DMRG is trying to find the lowest energy state of the system, and it can choose between a ``cat state'' which is a superposition of the HLR and \aHLR{} states, or a symmetry-broken state which is only HLR (or \aHLR).
The exact energy splitting would be zero or exponentially small in the cylinder length.
However, for a finite bond dimension, the cat state must use half of the bond dimension to describe the HLR and half to describe the \aHLR.
Therefore each piece has its bond dimension effectively halved, which introduces an algebraic increase in its energy. 
This energy increase makes the cat state energetically less favorable than a pure state.
So if our DMRG-optimized state at finite bond dimension does not break the PH symmetry, we conclude that the true state in the infinite-length quasi-1D system does not break this symmetry.
Note that this argument is actually valid for any discrete symmetry breaking in quasi-1D.

Once we have the optimized DMRG state (matrix product state) at a given bond dimension $\chi$, we can calculate the correlation length $\xi$ in this state from the subdominant eigenvalue of the transfer matrix.
In a gapped system, $\xi$ converges to the true correlation length in the ground state.
In a critical system, it diverges with increasing $\chi$, while at a fixed $\chi$ it can be viewed as an effective cutoff length set by that bond dimension.
This is the length scale plotted in Fig.~\ref{fig:central_charge} vs entanglement entropy between left and right parts of the system, which is readily measured in the same MPS state.

We also study an overlap of the optimized DMRG state with its particle-hole conjugate.  
Using similar transfer matrix technique, we can naturally calculate an effective overlap per one orbital, $1 - \epsilon$, used in the main text.

\subsection{Correlation functions}
\label{subapp:corr}


Importantly for this work, we also measure density-density correlation function.
We first specify our correlation function in general.
We consider a fixed number of electrons $N_e$ moving in continuum in a region of volume ${\rm Vol} \equiv L_x L_y$ and define
\begin{equation}
D(\br - \br') \equiv \braket{\norder{ \delta\rho(\br) \delta\rho(\br') }} =  \braket{\norder{ \rho(\br) \rho(\br') }} - \bar{\rho}^2 ~.
\end{equation}
Here $\delta\rho(\br) = \rho(\br) - \bar{\rho}$ is the deviation of the density from its average value $\bar{\rho} \equiv N_e/{\rm Vol}$, and we are also assuming translational invariance.
The corresponding structure factor is
\begin{align}
	D(\bq) \equiv \int\! d\br\, D(\br - \br') e^{-i\bq \cdot (\br - \br')} 
	= \braket{\norder{\delta\rho_\bq \delta\rho_{-\bq}}}
        = \braket{\delta\rho_\bq \delta\rho_{-\bq}} - \bar{\rho} ~,
	\label{eq:Dq}
\end{align}
where $\delta\rho_\bq \equiv \int\!d\br\, \delta\rho(\br) e^{-i\bq \cdot \br}/\sqrt{\rm Vol}$.
The last expression follows from simple manipulations (e.g., in the first-quantized language), and one can use it to argue that thus defined $D(\bq)$ is a continuous function with limiting behaviors $D(\bq \to \infty) = 0$ and $D(\bq \to 0) = -\bar{\rho}$.
On the other hand, for numerical evaluations it is easier to use the original expression, which gives
\begin{equation}
	D(\bq) = \braket{\norder{ \rho_\bq \rho_{-\bq} }} - \bar{\rho}^2 \, {\rm Vol} \, \delta_{\bq, \mathbf{0}} ~,
\end{equation}
where $\rho_\bq \equiv \int\!d\br\, \rho(\br) e^{-i\bq \cdot \br}/\sqrt{\rm Vol}$.
(Note in this rather unusual convention, $\braket{\rho_{\textbf{0}}} = \bar\rho \sqrt{\rm Vol}$.)

We now specialize to the quantum Hall problem on the cylinder.
From the outset, we will assume that $L_x$ is very large and treat the corresponding wavevectors $q_x$ as continuous, while $L_y$ is finite and the wavevectors $q_y$ are discrete.
Using the basis of orbitals in Eq.~\eqref{phi_cyl}, the density operator becomes
\begin{align}\begin{split}
\rho(\br) = c^\dagger(\br) c(\br) 
	&= \sum_{j, j'} \frac{e^{i \frac{2\pi}{L_y} (j'-j) y}}{L_y} ~
	\frac{e^{-\frac{(x - X_j)^2 + (x - X_{j'})^2}{2 \ell_B^2}}}{\pi^{1/2} \ell_B}
	c_j^\dagger c_{j'} \\
 	&= \sum_{j, m} \frac{e^{i \frac{2\pi}{L_y} m y}}{L_y}
	\frac{e^{-\frac{(x - X_{j+m/2})^2}{\ell_B^2}} e^{-\frac{\ell_B^2}{4}(\frac{2\pi}{L_y} m)^2}}{\pi^{1/2} \ell_B}
        c_j^\dagger c_{j+m} 
	 ~.
\end{split}\end{align}
In the last line, $X_{j+m/2} \equiv \ell_B^2 (2\pi/L_y) (j+m/2)$ is a position half-way between orbitals $j$ and $j' = j+m$.
From the above expression, we immediately see that operator $O_m(j) \equiv c_j^\dagger c_{j+m}$ carries transverse momentum $q_y = (2\pi/L_y) m$.
We can now calculate the Fourier transform of the density operator,
\begin{equation}
\rho\left(q_x, q_y = \frac{2\pi}{L_y}m \right) 
= \frac{e^{-\frac{\ell_B^2 (q_x^2 + q_y^2)}{4}}}{\sqrt{\rm Vol}} \sum_j O_m(j) e^{-i q_x X_{j+m/2}} ~.
\end{equation}
Since $\rho_{-\bq} = \rho_\bq^\dagger$, we have
\begin{align}\begin{split}
\rho_\bq \rho_{-\bq}
	&= \frac{e^{-\frac{\ell_B^2 \bq^2}{2}}}{\rm Vol} \sum_{j, j'} O_m(j) O_m^\dagger(j') e^{-i q_x (X_j - X_{j'})}
\\	&= \frac{e^{-\frac{\ell_B^2 \bq^2}{2}}}{\rm Vol} \sum_{j, j'} c_j^\dagger c_{j+m} \left(c_{j'}^\dagger c_{j'+m} \right)^\dagger
	e^{-i q_x \ell_B^2 (2\pi/L_y)(j - j')}
\\	&= \frac{e^{-\frac{\ell_B^2 \bq^2}{2}}}{\rm Vol} \, \Norb \, \sum_\Delta c_0^\dagger c_m \left(c_\Delta^\dagger c_{\Delta + m} \right)^\dagger
	e^{i q_x \ell_B^2 (2\pi/L_y) \Delta} ~.
\end{split}\end{align}
In the last equation, we assumed that the state is invariant under translations of the chain, and $\Norb$ is the total number of sites in the chain, which is the same as the number of fluxes through the $L_x \times L_y$ region.

Now our correlation function becomes
\begin{equation}
D(\bq) = D\left(q_x, q_y = \frac{2\pi}{L_y}m \right) 
= \frac{\Norb}{\rm Vol} \left[ 
	e^{-\frac{\ell_B^2 \bq^2}{2}} \Braket{\sum_\Delta \norder{c_0^\dagger c_m \left(c_\Delta^\dagger c_{\Delta + m} \right)^\dagger} \, e^{i q_x \ell_B^2 (2\pi/L_y) \Delta} } 
        - \delta_{\bq, \mathbf{0}} \frac{N_e^2}{\Norb} \right] ~.
	\label{Dqfinal}
\end{equation}
We have $\Norb/{\rm Vol} = 1/(2\pi \ell_B^2)$.
Also, $N_e^2/\Norb = N_e \nu$, and we can verify that at $(q_x, q_y) = 0$ the term in the square brackets is equal to $-\nu$ and is actually independent of the number of sites in the chain (i.e., cylinder length).
While not obvious from the last expression, our earlier discussion after Eq.~\eqref{eq:Dq} implies that thus defined structure factor is a continuous size-independent function of $q_x$ near $q_x = 0$ at $q_y = 0$.
We have $D(q_x \to 0, q_y = 0) = -\nu/(2\pi\ell_B^2) = -1/(4\pi)$ in units used in the main text, and the DMRG measurements in Fig.~\ref{fig:cone} indeed give this value.

The above expression can then be used in the infinite cylinder DMRG setup.
The sum on $\Delta$ in principle runs from $-\infty$ to $\infty$, but in practice summing over a few hundred orbitals provides sufficient momentum resolution.
When plotting the results, we often omit the Gaussian factor, i.e., we show $\bar{D}(\bq) \equiv e^{\ell_B^2 \bq^2/2} D(\bq)$.

Finally, we note that when we work with a given unit cell and fixed root configuration, the state is guaranteed to be translationally invariant only under translations by unit cells.
The sites inside the unit cell need not be equivalent.  
In the limit of very large $L_y$ these sites are spaced by a very small amount $\ell_B^2 (2\pi/L_y)$, and such microscopic variation is irrelevant in the 2D limit.
In our finite $L_y$ cylinders, we simply average the position of the ``$0$'' site in Eq.~\eqref{Dqfinal} over sites in the unit cell.

\section{Additional numerical data}
\label{app:more_data}

Here we present a number of additional numerical results which support the conclusions drawn in the main text.

\begin{figure}
	\includegraphics[width=70mm]{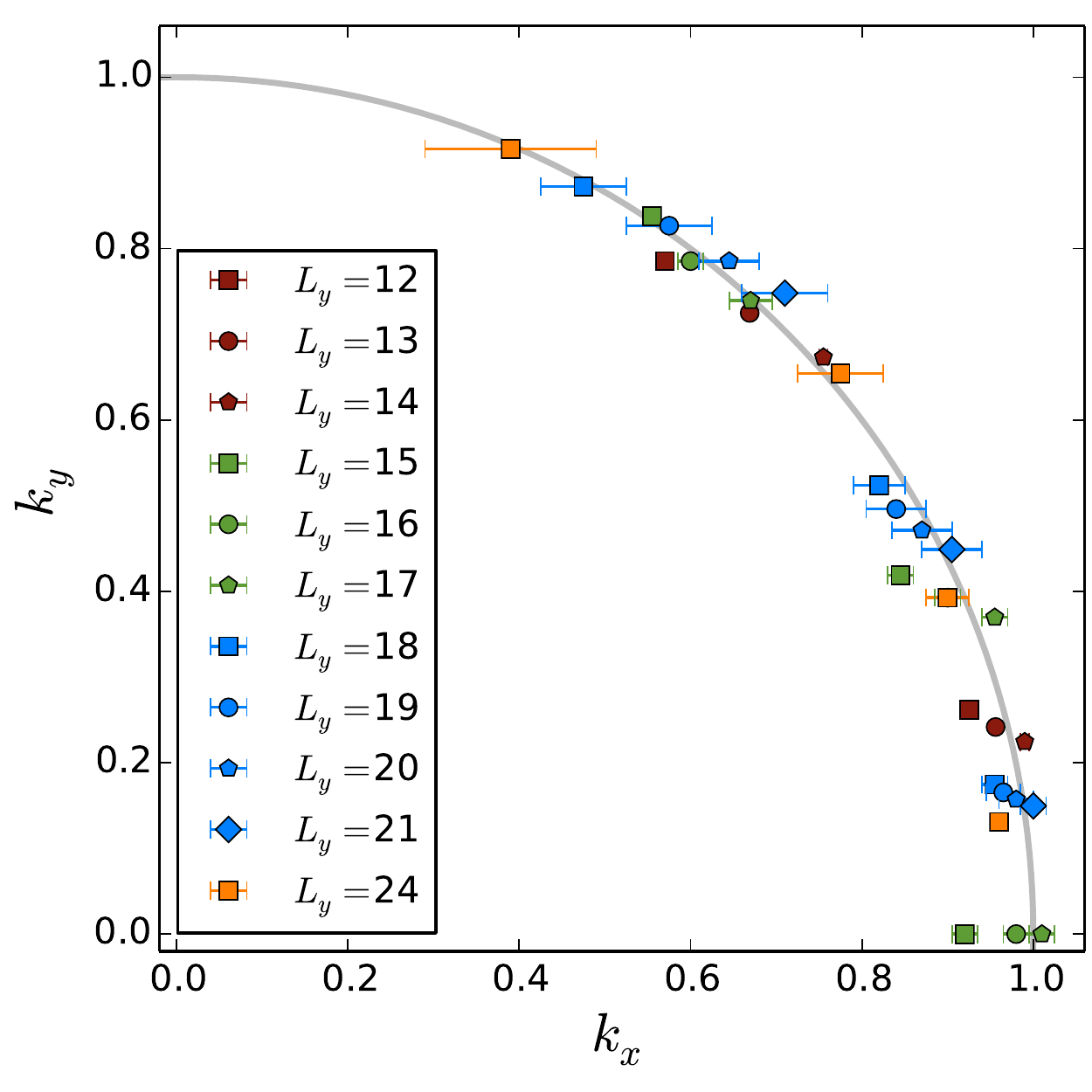}
	\caption{
		Map of the Fermi surface by collating data from various circumferences [same systems and color scheme as in~Figs.~\ref{fig:luttinger}~and~\ref{fig:central_charge}].
		Note that these points deviate slightly from the 2D circle, in order to satisfy Luttinger's theorem for the quasi-1D system with discrete wires, cf.~Fig.~\ref{fig:luttinger}.
	}
	\label{fig:fsurface}
\end{figure}
In Fig.~\ref{fig:fsurface} we map out the Fermi surface using the lengths $Q_m$ of the various ``wires.''
The $k_y$ coordinate of each point is given by $2\pi m/L_y$, with $m$ integer or half-integer depending on boundary conditions, and the $k_x$ coordinate is determined from the singular wavevectors $q_x = Q_m$ in the $q_y = 0$ structure factors (see Fig.~\ref{fig:wires}), namely $k_x = Q_m/2$.
We see that the resulting points lie very close to a Fermi surface with $k_F = 1$, with slight deviations which are needed to satisfy Luttinger's theorem in the quasi-1D system.

\begin{figure}
	\includegraphics[width=80mm]{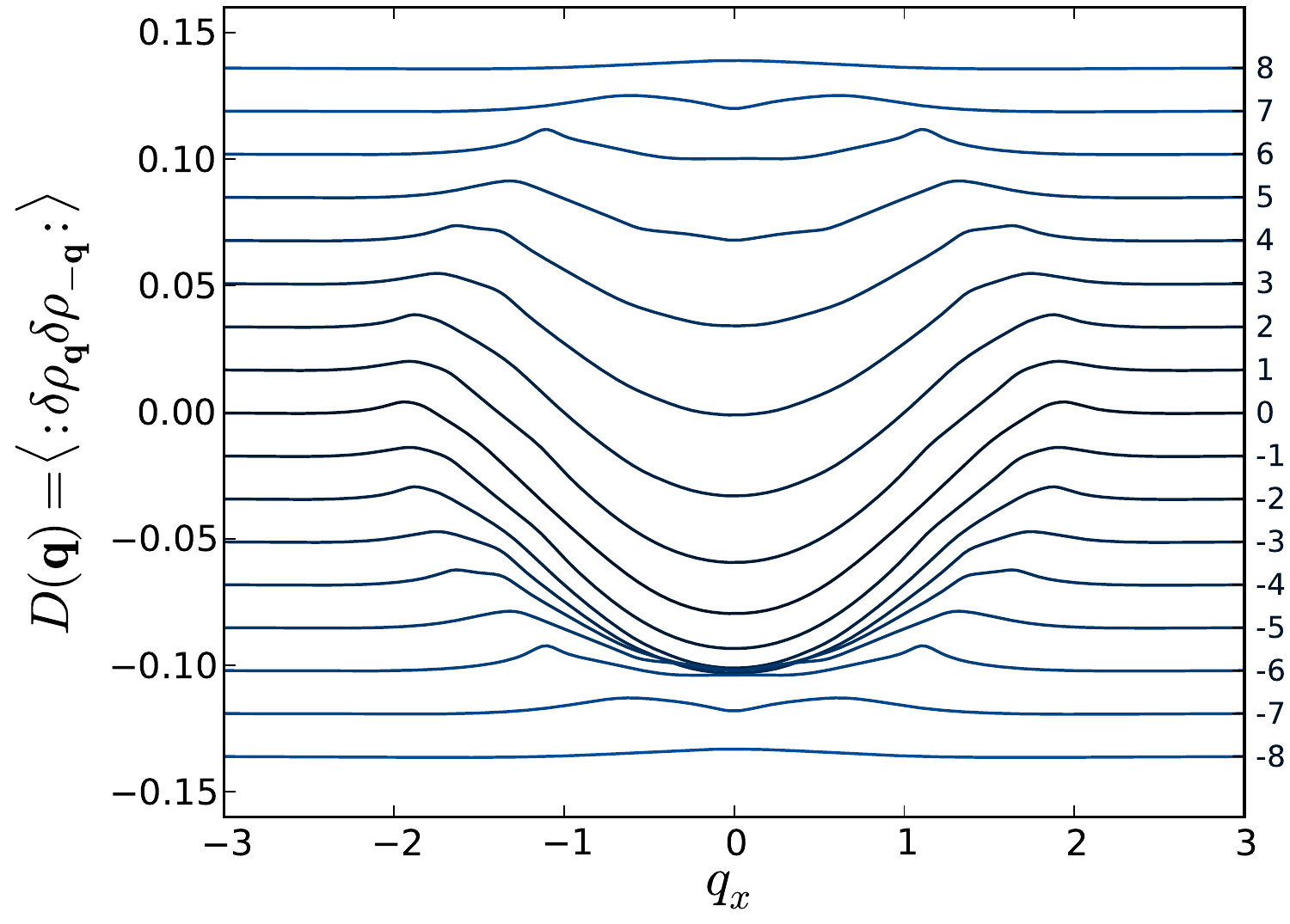}
	\caption{
		Alternative 3D view of the data in Fig.~\ref{fig:cone} for the $L_y = 24$ cylinder with eight-wire CFL.
		We show density structure factor measured at fixed $q_y$ ranging from $-8 (2\pi/L_y)$ to $8 (2\pi/L_y)$ from bottom to top (labeled on the right axis in units of $2\pi/L_y$).
		Each curve has a vertical offset proportional to $q_y$.
		While we see the appearance of the two-dimensional $2k_F$ accumulation circle from the low-energy composite fermion excitations, we also see finer features corresponding to singularities in the quasi-1D system with finite $L_y$.
	}
	\label{fig:boulder}
\end{figure}
In Fig.~\ref{fig:boulder} we show another view of the $L_y = 24$ density structure factor $D(\bq)$ shown in Fig.~\ref{fig:cone}. 
Here each line has a different $q_y$, with a vertical offset proportional to $q_y$. 
We can see both the accumulation circle at $2k_F$, and also other finer features representing other singularities present in the quasi-1D system.
Note that for this wide cylinder, our largest bond dimension used still effectively cuts off the correlations at a length scale of order $10$, see Fig.~\ref{fig:central_charge}, which explains some rounding of the features compared to cylinders with smaller $L_y$.
Note also that in this plot, as well as for all other sizes [cf.~Figs.~\ref{fig:two_body} and~\ref{fig:two_body16}], we do not see any features at $q_y = 0$ and small $q_x$.
This is consistent with a much weaker singularity $\sim |q_x|^3$ expected in this case (see App.~\ref{subapp:bosonizedQ1D}), which has continuous first and second derivatives.

\begin{figure*}
	\includegraphics[width=120mm]{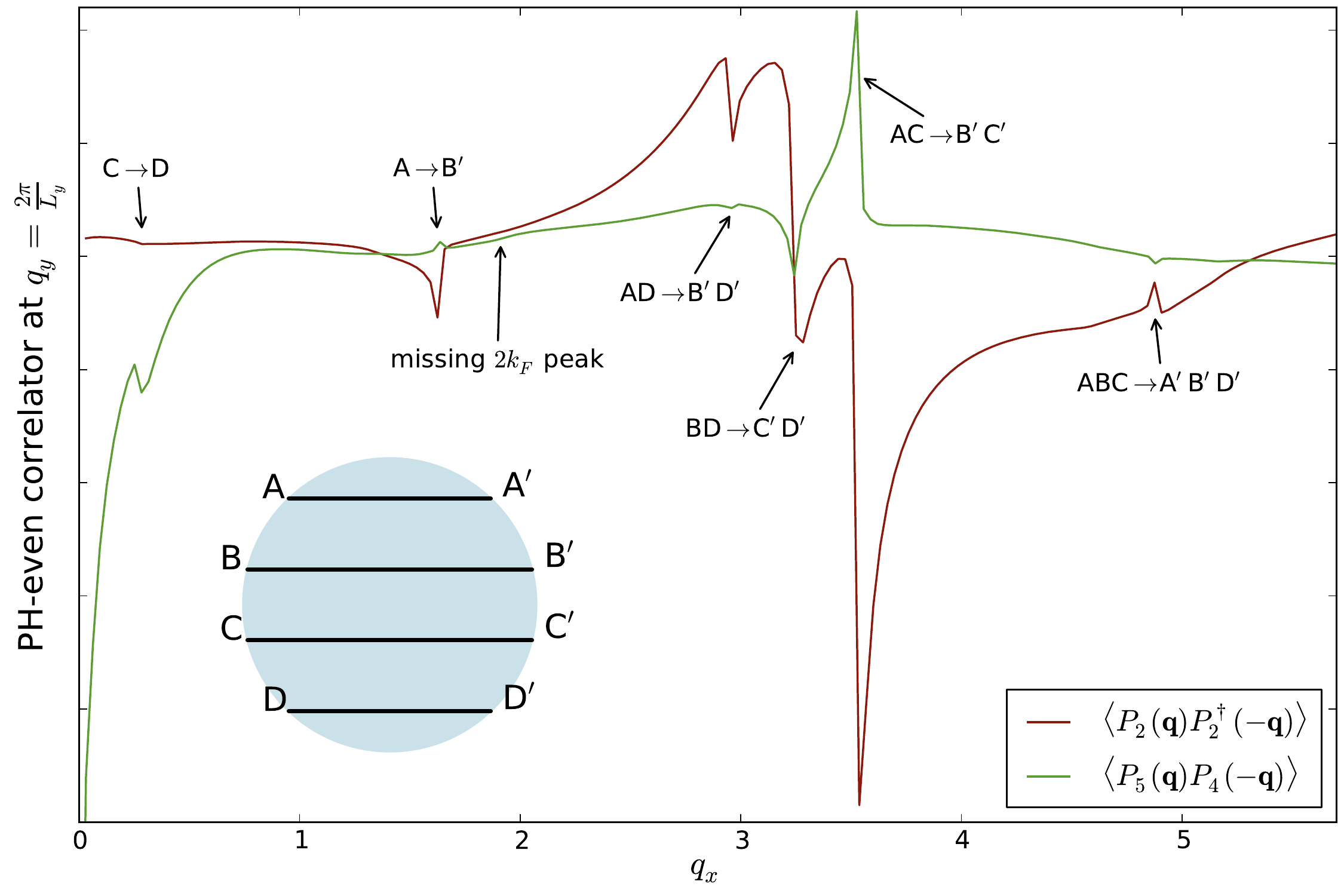}
	\caption{
		Derivative of correlation functions for PH-even operators as in Fig.~\ref{fig:bilayer} but for a single layer.
		The cylinder has $L_y = 13$ and realizes four-wire CFL (APBC; 0110 root configuration).
		The two curves correspond to different correlators used, $\braket{ P_5(\bq) P_4(-\bq) }$ (same as in Fig.~\ref{fig:bilayer}) and $\braket{ P_2(\bq) P_2(-\bq) }$, see text for details.
		We measure at $q_y = 2\pi/L_y$ since it can probe exact composite fermion backscattering with such boundary conditions.
		We can label all features by transfer processes contributing to such an operator $P(\br)$.
		We see one-fermion transfer $A \to B'$, while the exact backscattering $B \to C'$ is clearly missing, as expected for the Dirac CFL.
		Amazingly, we also see higher-order processes corresponding to transferring two left-moving composite fermions to the right-moving points, which we labeled $AD \to B'D'$, $BD \to C'D'$, $AC \to B'C'$ (in fact, these exhaust all distinct wavevectors for such four-fermion terms with $q_y=2\pi/L_y$).
		We even see a contribution from a six-fermion term transferring $ABC \to A'B'D'$;
			the only other six-fermion term that transfers three left movers to three right movers would be $ABD \to A'C'D'$,
			which however is a combination of three exact backscatterings and hence is odd under PH and cannot contribute to PH-even observables.
	}\label{fig:four_body13}
\end{figure*}

\begin{figure*}
	\includegraphics[width=120mm]{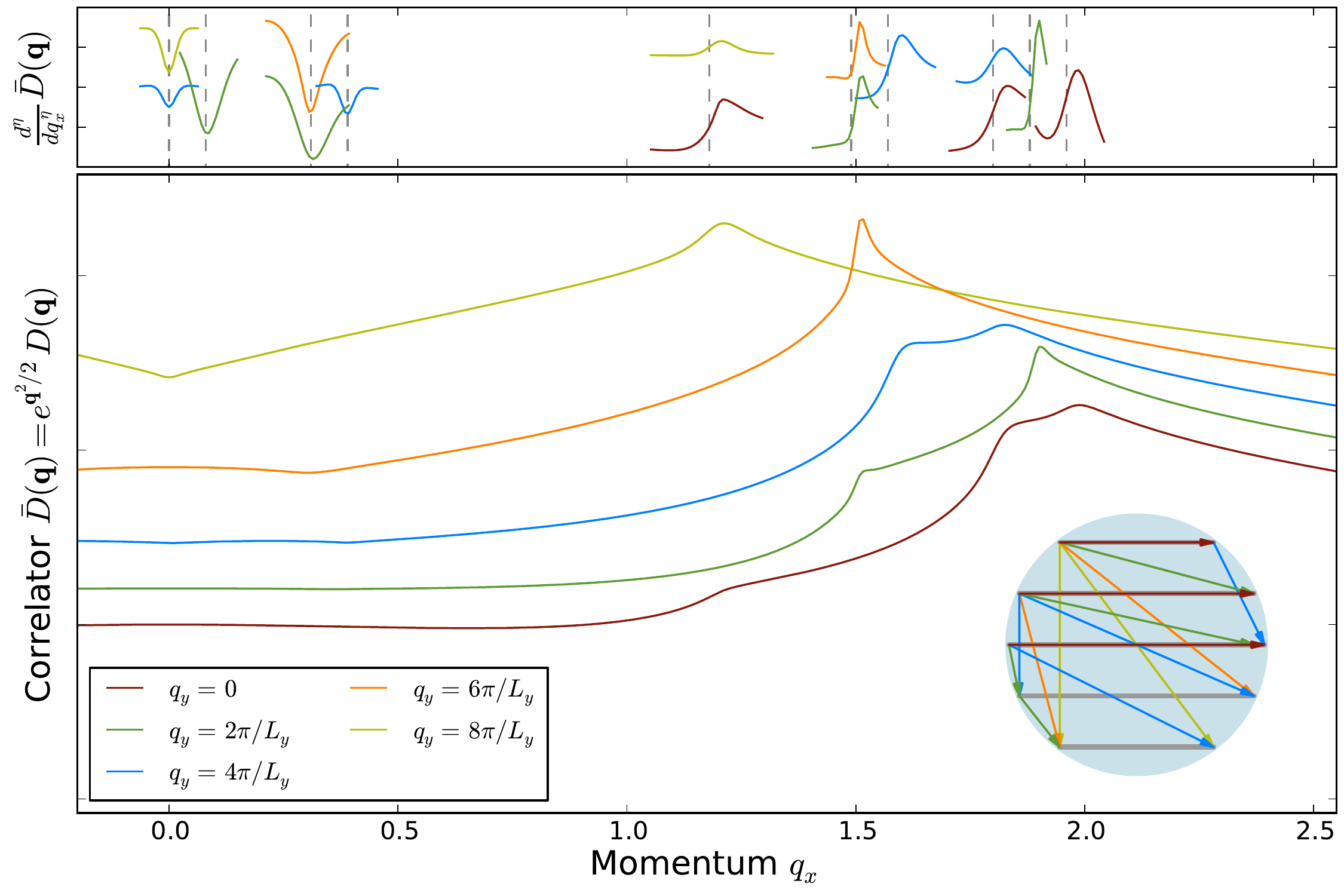}
	\caption{%
		Same as Fig.~\ref{fig:two_body} but for $L_y = 16$, which realizes five-wire CFL as shown (PBC for the composite fermions; root configuration 1010).
		Note that the singularities are more rounded here since the largest cutoff length $\xi$ accessible with our finite MPS bond dimensions is about four times smaller than in the $L_y = 13$ case, cf.~Fig~\ref{fig:central_charge}.
		The top panel shows fractional derivatives whose powers are chosen individually for each singularity, roughly to turn it into a step singularity. 
		We estimate the lengths of the wires from the $q_y=0$ data, and the dashed lines are determined from those estimates.
	}
	\label{fig:two_body16}
\end{figure*}

\begin{figure*}
	\includegraphics[width=120mm]{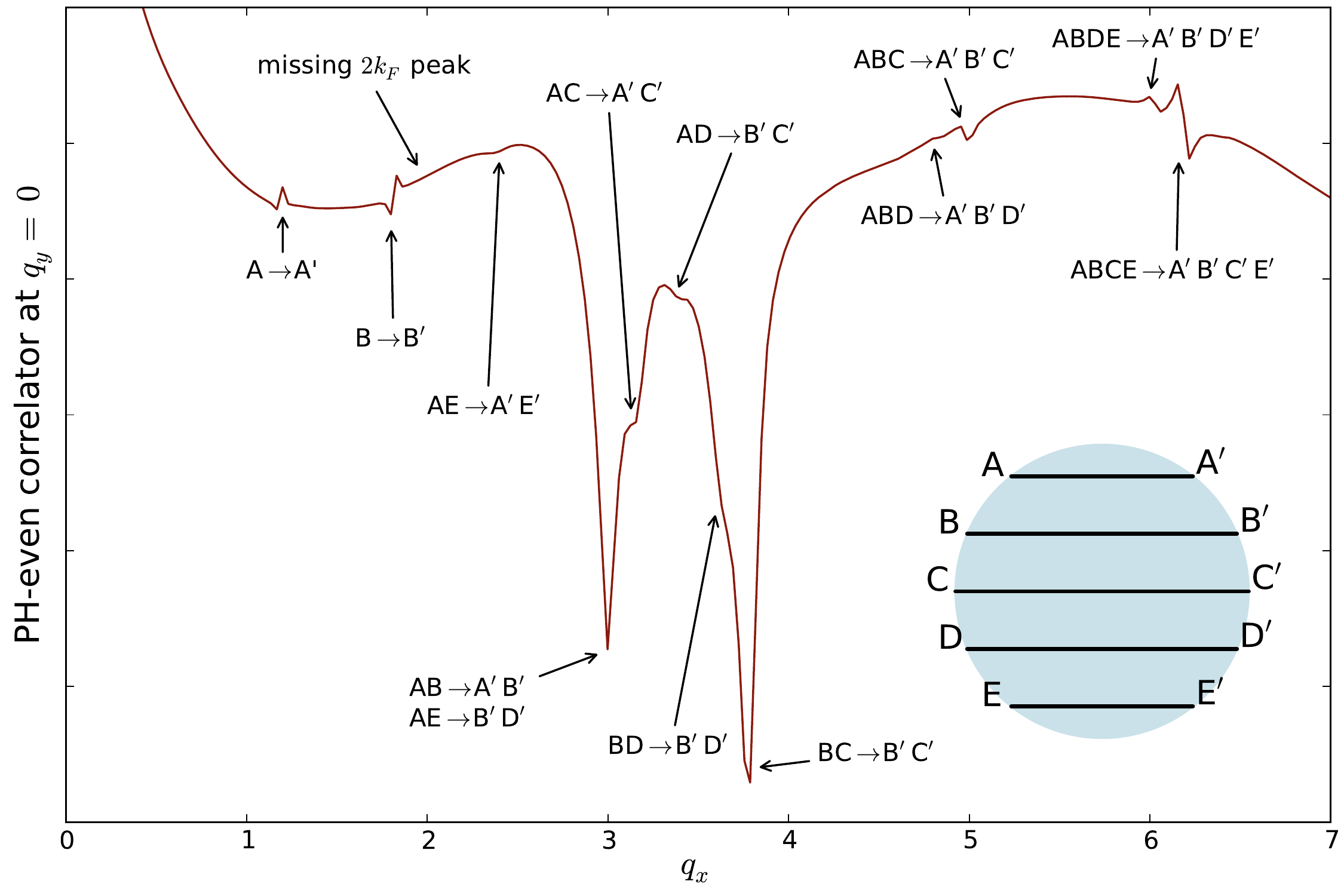}
	\caption{%
		Same as Fig.~\ref{fig:four_body13} but for $L_y = 16$ and transverse wavevector $q_y = 0$, plotting derivative of the correlation function of operator $P_1 \equiv \hat{n}_j \hat{n}_{j+1}$.
                This cylinder is the same as in Fig.~\ref{fig:two_body16} and realizes five wires as shown.
                PH-even observables cannot obtain contributions from precise $2k_F$ backscattering, which here corresponds to $C \to C'$ transfer of one composite fermion.
                On the other hand, $A \to A'$ and $B \to B'$ are not exact backscatterings in the 2D sense, so these processes do not have definite PH quantum number, and we do see features corresponding to these transfers.
                We see many other features, e.g., from four-fermion terms involving transferring two left movers to two right movers, and even from six and eight-fermion terms.
                In fact, we see all possible distinct wavevectors from such four-fermion processes and mark them on the plot, and we also mark the most visible higher-order processes. 
        }
    \label{fig:four_body16}
\end{figure*}

In the main text we measured correlators of the PH-even observable $P(\br)$. 
A natural observable to use for $P(\br)$ would be $\delta\rho(\br) \nabla^2 \rho(\br)$.
However, in the orbital basis each $\rho$ contains two summations over many sites on the chain, so the full correlator contains thousands of terms. 
\roger{We can just write the expression out....}
\lesik{just words are ok at this stage. we should at some point calculate this in DMRG honestly, as it is something good to know in the 2D limit. just need finite patience scaling :)}
Instead of calculating all of these, we calculated only a few dozen such terms, reasoning that all terms should contain the same singularities. 
None of the terms we calculated shows a singularity at the exact backscattering $2k_F$.
In Fig.~\ref{fig:bilayer} we plot $\langle P_5(\bq) P_4(-\bq)\rangle$, where $P_5 \equiv c_j^\dagger c_{j+1} n_{j+5}$, $P_4 \equiv n_j c_{j+3} c_{j+4}^\dagger$.
Note that $P_5$ carries $q_y = 2\pi/L_y$ and $P_4$ carries the opposite $q_y$, and the exact backscattering probed here is between the right-mover composite fermion at $k_y = \pi/L_y$ and left-mover at $k_y = -\pi/L_y$, see bottom picture in Fig.~\ref{fig:bc}.
We chose this correlator because it clearly shows the reemergence of the $2k_F$ exact backscattering peak when the particle-hole symmetry is broken by hand, but other correlators show this behavior as well. 
In Fig.~\ref{fig:four_body13} we take a detailed look at this correlator in the system with PH symmetry and identify all singularities with various processes, using information about the lengths of the wires determined from Fig.~\ref{fig:two_body} in the main text.
We have labeled many singularities, including those for all allowed two- and four-body processes transferring composite fermions from the left to right side of the Fermi surface (with total $q_y$ transfer of $2\pi/L_y$), but we do not see the exact backscattering $2k_F$ singularity, as expected in the presence of the PH symmetry. 
To show that our results do not depend on the particular PH-even observables presented, we also show the correlator $\langle P_2(\bq) P^\dagger_2(-\bq) \rangle$, where $P_2 \equiv c_j^\dagger c_{j+1} n_{j+2}$.

Most of the Fermi surface evidence in the main text was shown for $L_y = 13$ realizing four-wire Dirac-CFL, since here the finite-entanglement errors are the smallest.
However, we have similar evidence for several other sizes.
In Figs.~\ref{fig:two_body16} and \ref{fig:four_body16} we show the same data as in Figs.~\ref{fig:two_body} and \ref{fig:four_body13}, but for $L_y = 16$ realizing five-wire Dirac-CFL.
The singularities are more rounded at $L_y = 16$ compared to $L_y = 13$ due to more significant finite entanglement effects, i.e.~the effective cutoff length $\xi$ imposed by our finite MPS bond dimension is smaller at these sizes, see Fig.~\ref{fig:central_charge}.

In Fig.~\ref{fig:four_body16}, the PH-even operator whose correlator we measure is simply $P_1 \equiv n_j n_{j+1}$, and because of the PBC for the composite fermions, we probe the exact backscattering by measuring at $q_y = 0$.  
We clearly see singularities at wavevectors corresponding to processes transferring one composite fermion from $A$ to $A'$ and from $B$ to $B'$, which do not have definite PH quantum number and can hence contribute to both PH-even and PH-odd observables.
On the other hand, there is no singularity corresponding to transfer from $C$ to $C'$, in agreement with the absence of exact backscattering.
This figure also shows that our measurements can pick up signatures coming from contributions from higher-order processes.  
 In fact, we can identify all six distinct $q_x$ wavevectors corresponding to processes transferring two left-moving composite fermions to the right, with the total $q_y = 0$.
Each such distinct wavevector is labeled by one process with such $q_x$, e.g., $AB \to A'B'$, while there can be multiple processes with the same $(q_x, q_y=0)$.
We even see wavevectors corresponding to processes transferring three and four left-movers to the right, but we have not tried to identify all of them systematically.  [Note that six-fermion terms $ACE \to A'C'E'$ and $BCD \to B'C'D'$ are again combinations of three exact backscatterings and are not allowed in the PH-even operators.]
It is amazing how much information the DMRG gives us, and all is consistent with the proposed quasi-1D descendants of the Dirac CFL!

\begin{figure*}
	\includegraphics[width=72mm]{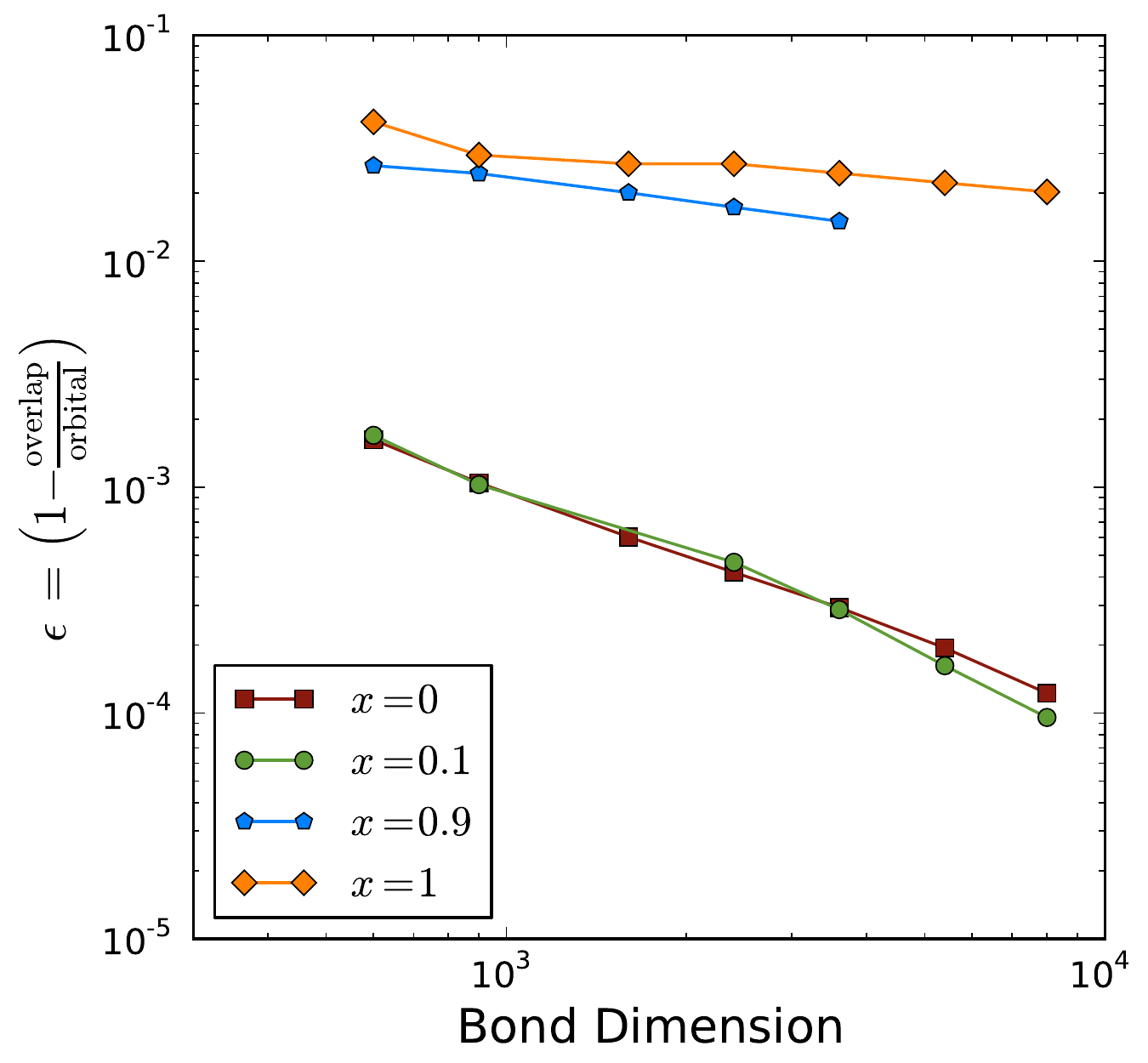}
	\caption{%
		A plot of $\epsilon$ vs.\ bond dimension, where the overlap (per orbital) between our wavefunction and its particle-hole conjugate is $1 - \epsilon$.
		Data was taken at $L_y=16$ using a modified Coulomb interaction $(1-x) V_0 + x V_1$, where $V_0$ is the Coulomb interaction in the lowest Landau level and $V_1$ is the interaction in the 1\textsuperscript{st} Landau level.
		When $x$ is small the system is in a CFL phase with particle-hole symmetry, while when $x\sim 1$ the system is in the Moore-Read phase and spontaneously breaks this symmetry.
		We see that that at the largest bond dimension used, when $x$ is small $\epsilon$ is two orders of magnitude lower than when the symmetry is broken, and it decays as a power law as the bond dimension is increased.
		The values quoted in the main text are obtained by fitting these curves to a power law.
\lesik{Should it be something like $\epsilon_\chi = \epsilon_\infty + C/\chi^p$, i.e., power law with an offset?  We quote $\epsilon_\infty = 0.022$ for $x=1$ and $\epsilon_\infty < 6 \times 10^{-5}$ for $x=0$?
Actually, it seems that the value quoted in the main text for x=1 was taken from the largest bond dimension $\chi = 8000$ and not from fits?  If that is the case, maybe do not mension any fitting and just quote values at the largest bond dimension, $< 2e-4$ for x=0 and 0.022 for x=1?
Refer to this figure from the main text, smth like (see App. for details).} 
}
	\label{fig:overlap}
\end{figure*}
In Fig.~\ref{fig:overlap} we show how the overlap behaves as a function of bond dimension, using a modified Coulomb interaction $(1-x) V_0 + x V_1$.
Here $V_0$ and $V_1$ are the (Gaussian-cutoff) Coulomb interaction projected into the lowest and 1\textsuperscript{st} Landau level, respectively.
At $x = 0$ the system is in the CFL phase, while at $x = 1$ the system is in the $\nu = 5/2$ state with spontaneous symmetry breaking.\cite{Eisenstein87, Morf1998, RezayiHaldane:PH:00, Papic2012}
We see that in the CFL phase the overlap is over one hundred times smaller than in the symmetry broken phase, and that it decays to zero as a power law.
\lesik{Should we explain in one or two sentence why $\epsilon$ is non-zero for finite $\chi$ even when we expect PH symmetry in the infinite cylinder, unlike finite-size ED with exact PH?
Here $L_y = 16$ is 5-wire PBC, so the root config is 0101 and breaks PH for any finite bond dimension, but I guess the overlap is with the translated state, so could ``restore'' the breaking in some sense?
What about $L_y$'s that have APBC, naively it seems that the PH could be preserved strictly also for finite $\chi$?  But could it be that Max's ``non-locality of the PH'' prevents this - after all, finite-$\chi$ MPS would be an example of ``local symmetry singlet''.  I am not sure about this, it seems likely you guys have much simpler explanations...  How would such plot look for $L_y = 13$ (4-wire APBC) or 19 (6-wire APBC)?
}
\roger{Scott, Lesik is asking too many questions!  I think he's on to us...}

\section{The effect of Landau level mixing on \twokF backscattering}
\label{app:LLmixing}
Several experimental perturbations break PH symmetry, including  disorder, incomplete polarization of the electron spin, and Landau level mixing.
Here we comment on the last two effects.
\begin{figure}
	\includegraphics[width=100mm]{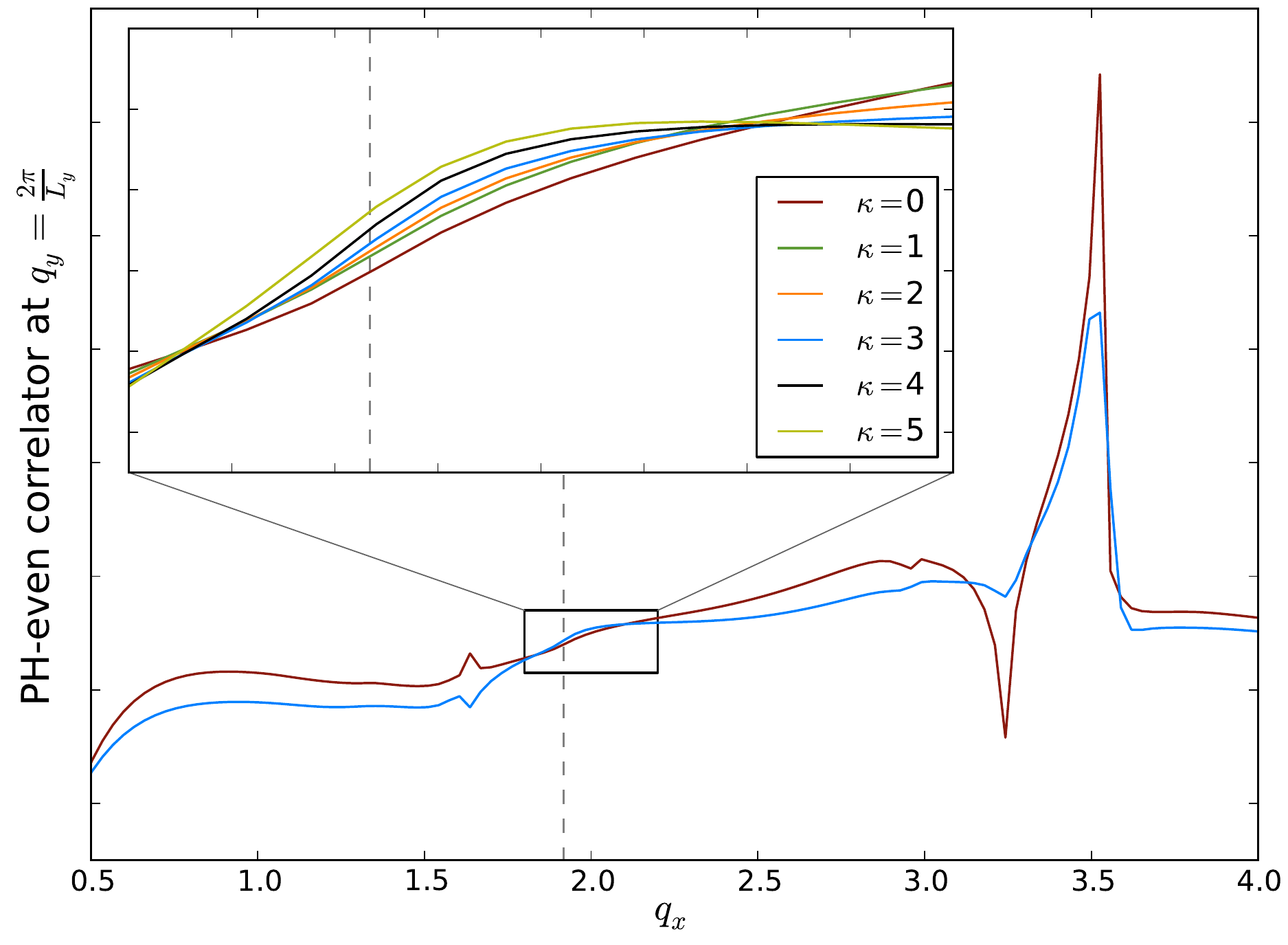}
	\caption{%
		Effect of Landau level mixing on the \twokF backscattering.
		Shown here is the derivative of the correlator $\braket{P_5(\bq)P_4(-\bq)}$ at $L_y = 13$.
		We see that including an additional Landau level at mixing parameter $\kappa$ does not lead to a reappearance of the \twokF singularity even for unrealistically large $\kappa$ (see text for details).
		The main plot shows data at $\kappa = 0$ and $\kappa = 3$, and we see that the two plots have the same features.
		The inset focuses near the backscattering wavevector and shows data up to the value $\kappa = 5$, above which there is a transition to a gapped phase.
	}
	\label{fig:LLmixing}
\end{figure}

At many of the gapped plateaus, a combination of a Zeeman energy and interaction-driven ferromagnetism fully polarize the electrons.
However, at $\nu = 1/2$, the polarization in fact depends sensitively on the electron density and the strength of the $B$-field used to achieve half-filling.
For spin-unpolarized electrons, LLL particle-hole transformation exchanges $\nu \leftrightarrow 2 - \nu$, as it must act on both spin species, so the particle-hole symmetry discussed in the main text is broken.
In the CF picture, the CFs are spinful, and will polarize depending on the ratio of the CF Fermi energy to Zeeman energy, $E_F / E_Z$. 
Note that $E_Z \propto B$, while $E_F \sim e^2/ (4 \pi \epsilon \ell_B) \propto \sqrt{B}$, since the CF Fermi surface is due entirely to Coulomb repulsion.
Thus at high fields the Zeeman energy dominates and the CFs fully polarize.
This transition is accessible experimentally;\cite{Kukushkin99} thus in order to preserve PH symmetry, we assume the high-field side.

Particle-hole symmetry is exact only when the Hamiltonian is projected into the LLL, which is justified when the cyclotron gap $E_\omega = \hbar \omega_c \propto B$ is large.
However, the Coulomb scale $E_C \equiv e^2 /(4 \pi \epsilon \ell_B)$  is not negligible, so the LLL projection is violated by terms of order $\kappa =  E_C / E_\omega \propto B^{-1/2}$, with $\kappa \sim 0.4$ in typical LLL experiments.

We can ask whether the $2 k_F$ backscattering returns for experimentally relevant $\kappa$, as it did in the bilayer setup in the main text.
To investigate, we project the Coulomb Hamiltonian into the two lowest Landau levels, $n = 0$ and $1$, and tune $\kappa$ by changing the cyclotron energy.
The result is shown in Fig.~\ref{fig:LLmixing}.
At $\kappa = 0$, the measurement is identical to the $\mu = \infty$ curve of the main text, Fig.~\ref{fig:bilayer}.

The singularities \emph{away} from $2 k_F$ change quantitatively with $\kappa$, but the $2 k_F$  backscattering does not reappear, in contrast to the bilayer numerical experiment in the main text.
Presumably there is some $2 k_F$ singularity with an amplitude too small for us to resolve.
Thus, according to this operational definition, PH-breaking is ``weak.'' 
Furthermore, numerically we find that at $\kappa = 1$ the occupation per flux in the $n = 1$ LL is $\nu_1 \approx 0.003$.
We observe such ``weakness'' of the effective PH-breaking over the entire range of the gapless CFL phase in this numerical experiment, including unrealistically larger $\kappa$, while for even larger $\kappa \geq 5$ our model goes into a gapped phase (likely the Moore-Read phase).


\section{The Dirac CFL}
\label{app:reviewDCFL}
Here we briefly review the proposed Dirac-CFL theory in Ref.~\onlinecite{DSon:CFL2015}.
Like in the TI, the composite fermion $\psi_{\textrm{CF}; s}$ is a two-component fermion ($s = 1, 2$) on which PH acts as $\PH \psi_{\textrm{CF}} = i \sigma^2 \psi_{\textrm{CF}}$.
\roger{The `CF' subscript is cumbersome and unnecessary.}
The component ``$s$'' is \emph{not} the electron spin, which is polarized, but is rather an emergent degree of freedom which can be pictured as the dipole moment of the CF.
The new ingredient in the Dirac-CFL, compared to the surface of a TI, is the emergent gauge field $a_\mu$.
The $\nu = 1/2$ phase is then proposed to be equivalent (dual) to (2+1)D quantum electrodynamics:
\begin{align}
	\mathcal{L}_{\textrm{QED}_3} &= \bar{\psi}_{\textrm{CF}} \gamma^\mu (i \partial_\mu + a_\mu) \psi_{\textrm{CF}} - \frac{1}{4 e^2} f_{\mu \nu} f^{\mu \nu} -  \frac{B}{4 \pi} a_0 + \dots
	\label{eq:QED3}
\end{align}
Here $\bar{\psi} = \psi^{\dagger} \gamma^0$, with $\gamma^0 = \sigma^3, \gamma^0 \gamma^1 = \sigma^1, \gamma^0\gamma^2 = \sigma^2$; $f_{\mu\nu} = \partial_\mu a_\nu - \partial_\nu a_\mu$; we have also set velocity of the Dirac fermions to 1 and used common writing of the coupling constant in the Maxwell term $f^2$.
In addition to the Dirac and Maxwell Lagrangians, the last term ``dopes'' the CFs away from the Dirac point to finite density, as it is equivalent to a static background charge $B / 4 \pi$.
Thus, like in HLR, the CFs have a Fermi surface of radius $k_F = \ell_B^{-1}$ and are also coupled to a dynamical gauge field.

To probe the physical electron, we introduce the physical vector potential $A$.
It appears in Eq.~\eqref{eq:QED3} via the replacement $\frac{B}{4\pi} a_0 \Rightarrow \frac{1}{4 \pi} \epsilon^{\mu \nu \lambda} (a_{\mu} + \frac{A_{\mu}}{2}) \partial_{\nu} A_{\lambda}$.
The following relation for the electron density follows:
$n_{\textrm{e}} = \frac{1}{4\pi} \nabla \times \mathbf{a} + \frac{1}{2} \frac{1}{2 \pi \ell_B^2}$.

The Dirac CFL theory maintains the particle-hole symmetry and thus differs from the original HLR construction.
However, some crude properties are quite similar in the two theories.
For example, both theories predict that in the case of Coulomb interactions, the leading non-analyticity in the long-wavelength density structure factor is $|\bq|^3 \log(1/|\bq|)$.
While this is a very weak singularity and difficult to see in numerical simulations, it would be interesting to attempt different scaling analyses of the DMRG data to study such non-Fermi-liquid aspects of the CFL in 2D.

\section{Locking of momentum with CF boundary conditions}
\label{app:mom_bc}
We observe that the BC `$\Phi_{\rm int}$' of the CF  is locked to the momentum $K_y = \frac{2 \pi}{L_y} \sum_j j n_j$.
While the absolute value of $K_y$ is difficult to define on an infinite cylinder, we can safely consider the momentum per electron \emph{relative} to (say) the orbital configuration $\cdots 0101 \cdots$.
Our observation of PBC for $0101$ and APBC for $0110$ is consistent with the relation
\begin{align}
\Delta K_y = N_e  \frac{2 \pi}{L_y}\frac{\Phi_{\rm int}}{2 \pi},
\label{eq:bc_P}
\end{align}
since the difference between the $\Phi_{\rm int}  = 0$ or $\pi$ orbital configurations is such that every \emph{second} electron is bumped by one orbital, with $\Delta K_y = \frac{2 \pi }{L_y}$.
We now derive this relation within the Dirac-CFL theory.

Note that  any discussion of momentum at finite field is subtle as translations obey a magnetic algebra:
\begin{align}
T_x( x) T_y( y)  T_x(- x)  T_y(- y)  = e^{i  \frac{x y}{\ell_B^2} \hat{N} }.
\end{align}
On a cylinder of circumference $L_y$, translations along the length of the cylinder are  broken down to a discrete subgroup $T_x(x = \frac{2 \pi}{L_y} \ell_B^2 )$.
On a torus, translation is also broken along $y$, and we have
\begin{align}
T_x \left( \frac{2 \pi}{L_y} \ell_B^2 \right) T_y \left( \frac{2 \pi}{L_x} \ell_B^2 \right) =  T_y \left( \frac{2 \pi}{L_x} \ell_B^2 \right) T_x\left( \frac{2 \pi}{L_y} \ell_B^2 \right) e^{i  2 \pi  \nu} ~.
\label{eq:mag_alg}
\end{align}
Thus when the dual theory, finite density QED$_3$, is placed on a cylinder / torus, we expect that translation symmetry must be broken in some subtle fashion. 
We will see this is indeed the case.

Let us begin with the dual theory on a torus in the presence of a uniform background magnetic field $B$. 
At the end of the calculation we can take a limit to recover the infinite cylinder result.
\maxm{Maybe drop the rest of this paragraph, (\ref{eq:L}) and text below it, and simply reference Eq.~(\ref{eq:QED3}).}
We also take the temporal direction to be compact.
The Lagrangian is
\begin{align}
	L = \bar{\psi} i \gamma^{\mu} (\partial_{\mu} - i a_{\mu}) \psi - \frac{1}{4 \pi} a_{0} B  - \frac{1}{4e^2} f_{\mu \nu} f^{\mu \nu}. \label{eq:L}
\end{align}
Note that there are several ways to rewrite the CS term (related by integration by parts) and various boundary conditions one may place on $a_{\mu}$ and $\psi$.
Let us stick to the form of the Lagrangian in Eq.~\eqref{eq:L}.

We discuss the physics in a Hamiltonian formalism.
We consider the situation when the physical Landau level is half-filled, so there is no net flux of $(\partial_x a_y - \partial_y a_x)$ through the spatial torus.
We take all fields ($a_{\mu}$, $\psi$) to be periodic along $x$ and $y$ directions.
We have large gauge transformations,
\begin{align}\begin{split}
	U_x a_x U^{\dagger}_x &= a_x + \frac{2 \pi}{L_x},\quad U_x a_y U^{\dagger}_x = a_y, \quad U_x \psi(x,y) U^{\dagger}_x = e^{i 2 \pi x/L_x} \psi(x,y) ~,
\\	U_y a_x U^{\dagger}_y &= a_x,\quad U_y a_y U^{\dagger}_y = a_y  + \frac{2 \pi}{L_y}, \quad U_y \psi(x,y) U^{\dagger}_y = e^{i 2 \pi y/L_y} \psi(x,y) ~.
	\label{eq:Large}
\end{split}\end{align}
As discussed in Ref.~\onlinecite{MetlitskiVishwanath2015}, for a single Dirac cone regularized in a time-reversal invariant manner, 
\begin{align}
	U_x U_y = - U_y U_x ~. \label{eq:anti}
\end{align}
This is the parity anomaly.
We require the ground state wavefunction to carry $U_x = 1$ and $U^2_y = 1$; equivalent results are produced if we exchange the role of $x, y$.

Now, let us discuss the translational symmetry.
The key issue has been previously identified as a feature of QED$_2$ (the Schwinger model) at finite density: translation symmetry is anomalous, and the $\theta$-vacua are permuted by translations.\cite{FKS79, Metlitski07}
When QED$_3$ is placed on a cylinder, an analogous effect occurs.
When $B = 0$, we have the  conserved gauge-invariant energy momentum tensor,
\begin{align}
	T^{\mu \nu} = \bar{\psi} i \gamma^{\mu} D^{\nu} \psi - \frac{1}{4 e^2} F^{\mu \lambda}F^{\nu}_{\, \lambda} - g^{\mu \nu} \left(\bar{\psi} i \gamma^{\lambda} (\partial_{\lambda} - i a_{\lambda}) \psi - \frac{1}{4e^2} f_{\lambda \sigma} f^{\lambda \sigma} \right) .
\end{align}
\roger{The equations below are missing some punctuations.}
We have not symmetrized the energy momentum tensor as symmetrization will not be important for our purposes.
When $B  =0$, we have $\partial_{\mu} T^{\mu \nu} = 0$.
If one couples the theory to an external current, $j^{\mu}_{ext}$,
\begin{align}
	L = \bar{\psi} i \gamma^{\mu} (\partial_{\mu} - i a_{\mu}) \psi - \frac{1}{4e^2} f_{\mu \nu} f^{\mu \nu} + a_{\mu} j^{\mu}_{ext}
\end{align}
then the energy momentum tensor is no longer conserved:
\begin{align}
	\partial_{\mu} T^{\mu \nu} = f^{\nu \lambda} j_{\lambda, ext}
\end{align}
This result follows from classical equations of motion.
It does not rely on the Dirac dispersion of the fermion (would be still true for a non-relativistic fermion, provided we use an appropriately modified $T^{\mu \nu}$).
Now, let's consider a uniform background charge density $j_{ext} = (-\rho, 0)$.
In our case, $\rho = \frac{B}{4 \pi}$. Then, 
\begin{align}\begin{split}
	\partial_{\mu} T^{\mu 0} &= 0 ,
\\	\partial_{\mu} T^{\mu i} &= \rho f_{i 0} .
\end{split}\end{align}
Thus, energy is conserved, but momentum is no longer locally conserved.
We may try to fix the problem by redefining,
\begin{align}
	\tilde{T}^{0 i} = T^{0 i} + \rho a_i, \quad \tilde{T}^{ij} = T^{i j} - \rho \delta^{i j} a_0
\end{align}
We now have $\partial_{\mu} \tilde{T}^{\mu i} = 0$.
However, $\tilde{T}^{\mu i}$ is now no longer gauge-invariant.
The situation appears to be a bit better for the global momentum built out of $\tilde{T}^{0i}$,
\begin{align}
	\tilde{P}^i = \int d^2x \tilde{T}^{0i} = \int d^2 x \left(T^{0i} + \rho a_i\right)
\end{align}
The momentum $\tilde{P}^i$ is conserved, moreover, it is invariant under small gauge-transformations ($e^{i \alpha(x)}$ with $\alpha(x)$ periodic in space).
However, it is not invariant under large gauge transformations \eqref{eq:Large},
\begin{align}\begin{split}
	U_x \tilde{P}^x U^{\dagger}_x = \tilde{P}^x + 2 \pi \rho L_y = \tilde{P}^x + \frac{\pi \Norb}{L_x}, \quad \quad U_x \tilde{P}^y U^{\dagger}_x = \tilde{P}^y ,
\\
	U_y \tilde{P}^x U^{\dagger}_y = \tilde{P}^x , \quad \quad U_y \tilde{P}^y U^{\dagger}_y = \tilde{P}^y  + 2 \pi \rho L_x = \tilde{P}^y + \frac{\pi \Norb}{L_y} ,
\end{split}\end{align}
where $\Norb = B L_x L_y/2\pi$ is the number of flux-quanta.
Thus, there are no infinitesimal translations which commute with the Hamiltonian and with the large-gauge transformations $U_x$ and $U^2_y$.
However, there are discrete translations which do.
Define
\begin{align}
	T_x(b_x) = \exp(i \tilde{P_x} b_x), \quad T_y(b_y) = \exp(i \tilde{P}_y b_y)
\end{align}
$T_y(b_y)$ commutes with $U_x$ for any $b_y$.
It commutes with $U^2_y$ provided that 
\begin{align}
	b_y = \frac{L_y}{\Norb} m_y, \quad m_y \in \mathbb{Z}
\end{align}
This is the correct quantization of translations in the Landau level, Eq.~\eqref{eq:mag_alg}.
As for $T_x(b_x)$ it commutes with $U^2_y$ for any $b_x$, and with $U_x$ when 
\begin{align}
	b_x = 2 \frac{L_x}{\Norb} m_x, \quad m_x \in \mathbb{Z}
\end{align}
We know that in the Landau level, smaller translations by $b_x = \frac{L_x}{\Norb}$ should also be a symmetry.
However, we have,
\begin{align}
	T_x\left(\frac{L_x}{\Norb}\right) U_x = -U_x T_x\left(\frac{L_x}{\Norb}\right)
\end{align}
To get rid of the minus sign, consider a modified translation operator,
\begin{align}
	\hat{T}_x\left(\frac{L_x}{\Norb}\right) = U_y T_x\left(\frac{L_x}{\Norb}\right) .
\end{align}
We know that $U_y$ is a symmetry of the Hamiltonian, so $\hat{T}_x$ is also a symmetry.
Moreover, due to Eq.~\eqref{eq:anti}, $\hat{T}_x$ now commutes with $U_x$ (and with $U^2_y$).
Note that on the physical Hilbert space $\hat{T}^2_x\left(\frac{L_x}{\Norb}\right) = T_x\left(2\frac{L_x}{\Norb}\right)$ since $U^2_y = 1$ on the physical Hilbert space.

Finally, let us compute the commutator of $\hat{T}_x\left(\frac{L_x}{\Norb}\right)$ and $T_y\left(\frac{L_y}{\Norb}\right)$.
Defining the bare, unconserved momenta as
\begin{align}
	P^i = \int d^2 x \, T^{0i}
\end{align}
for any local operator $O(x)$ we have,
\begin{align}
	\left[ P^i, O(x)\right] = \partial_i O(x) .
\end{align}
Moreover, since the action contains no Chern-Simons term for $a_{\mu}$, $a_x$ and $a_y$ commute. Therefore,
\begin{align}
	\left[ \tilde{P}^x, \tilde{P}^y \right] = -i \rho \int d^2 x (\partial_x a_y - \partial_y a_x) = -i B (N_e - \Norb/2) \label{eq:Pcomm}
\end{align}
where $N_e$ is the number of physical electrons in the Landau level.
Strictly speaking we are considering the case where there is no background magnetic flux of $a$ ($N_e = \Norb/2$), so the commutator is $0$.
(However, further consideration indicates that Eq.~\eqref{eq:Pcomm} continues to hold even away from half-filling).
Then,
\begin{align}
	\hat{T}_x\left(\frac{L_x}{\Norb}\right) \hat{T}_y\left(\frac{L_y}{\Norb}\right) = e^{2 \pi i \nu} \hat{T}_y\left(\frac{L_y}{\Norb}\right) \hat{T}_x\left(\frac{L_x}{\Norb}\right)
\end{align}
as required by the magnetic algebra, Eq.~\eqref{eq:mag_alg}.
In particular, at $\nu = 1/2$, $\hat{T}_x$ and $\hat{T}_y$ anti-commute.

Now we are ready to understand the change in momentum when we have an ``even" vs ``odd" number of CF modes on the cylinder, i.e., when $a_y = 0$ vs $a_y = \frac{\pi}{L_y}$.
In this case, the expectation value of the un-corrected momentum is $\langle P^y \rangle = 0$, since the Fermi sea fills in a rotationally symmetric fashion.
Thus the corrected momentum is
\begin{align}
	\braket{ \tilde{P}^y } = \rho \int d^2 x  \, a_y
\end{align}
Hence, the difference in $\tilde{P}^y$ between $a_y = 0$ and $a_y = \frac{\pi}{L_y}$ is,
\begin{align}
	\Delta \langle \tilde{P}^y \rangle = \frac{2 \pi}{L_y} \frac{N_e}{2}
\end{align}
in agreement with the hypothesis of Eq.~\eqref{eq:bc_P}.

One may ask if this difference is meaningful on a torus (rather than a cylinder).
Consider the operator $T_y(2 L_y/\Norb)$.
This is the double of the minimal allowed translation operator on the torus.
This operator acting on our two states gives results differing by a $-$ sign.
So, indeed, the difference in momenta is meaningful on a torus.
Further note that since we are using the double of the minimal translation to distinguish between the two states here, the subtlety involved in the definition of $\hat{T}_x$ does not enter here (i.e., there is no difference between the two directions).

\section{Half-filled Landau level and topological superconductor in class AIII}
\label{app:TSC}
In this appendix, we elaborate on the connection between the half-filled Landau level and the surface state of a 3D topological superconductor (TSc) in class AIII.\cite{DSon:CFL2015, MetlitskiVishwanath2015}
TSc's in class AIII in 3D are close cousins of the familiar 3D topological insulators.
Like the TIs, they are phases protected by a combination of $U(1)$ symmetry and time-reversal $\TR$.
However, unlike in the TI, here the (anti-unitary) time-reversal symmetry acts on the charged fermion $\psi$ in a particle-hole manner, inverting the $U(1)$ charge of $\psi$. 
One may physically think of the $U(1)$ symmetry as conservation of the $z$-component of spin in a superconductor, thus the nomenclature.
Unlike non-interacting 3D TIs, which have a ${\mathbb Z}_2$ classification, non-interacting 3D TSc's in class AIII have an integer classification. 
The surface of a phase in class $\nu \in {\mathbb Z}$ supports $|\nu|$ Dirac cones.
We focus on the $\nu = 1$ phase with a single surface Dirac cone,
\begin{align}
H = v \psi^{\dagger} \left[-i (\partial_x - i e A_x) \sigma^1 - i(\partial_y - i e A_y) \sigma^2 \right] \psi
\end{align}
Here, $\psi$ is a 2-component fermion, which transforms as 
$\TR: \psi \to \sigma^3 \psi^{\dagger}$, 
$U(1): \psi \to e^{i \alpha} \psi$. 
For future convenience, we have included the coupling to a $U(1)$ gauge field $\vec{A}$. 

Let us place the surface in a magnetic field $B = \partial_x A_y - \partial_y A_x$. 
Unlike in a TI, here the magnetic field does not violate the time-reversal symmetry, due to the particle-hole nature of the latter. 
The surface spectrum breaks up into Landau levels with energies, 
$E_n = {\rm sgn}(n) v \sqrt{2 e |B| |n|}$, $n = 0, \pm 1, \pm 2, \ldots$.
$\TR$-symmetry forces the $n = 0$ LL to be half-filled. 
If one takes the limit where the Landau levels with $|n| \ge 1$ decouple, then the surface theory becomes identical to the 2DEG in the LLL level. 
In particular, the $\TR$ symmetry of the surface maps to the PH symmetry of the LLL. 
However, unlike in a 2DEG, here the entire spectrum is $\TR$-symmetric, i.e., there is no need to project into the $n = 0$ LL for the PH symmetry to emerge. 
Notably, the time-reversal symmetry acts on the full bulk Hilbert space of the 3D TSc in a local manner, unlike the non-local action of the PH symmetry in the lowest Landau level of a 2DEG.
Conventional wisdom states that surfaces of 3D topological phases cannot be imitated in a purely 2D setting.
However, this wisdom assumes that the 2D implementation has a local symmetry action, while the PH symmetry of the 2DEG in the LLL is non-local, as explained in the main text and in App.~\ref{app:PH}, which allows it to mimic the $\TR$-symmetric surface of a class AIII TSc.

\section{Field theory for the quasi-1D descendant of Son's Dirac CFL}
\label{app:DCFLquasi1D}

\subsection{Fields near the Fermi surface for Dirac composite fermions at finite chemical potential}

In this Appendix, we use Son's Dirac CFL theory\cite{DSon:CFL2015} (reviewed also in App.~\ref{app:reviewDCFL}) to propose an effective field theory for the quasi-1D descendant states on the infinite cylinder.
In the same spirit as in Son's paper, we postulate that low-energy degrees of freedom are some fermions (loosely referred to as ``composite fermions'') coupled to a dynamical gauge field.
Here we do not derive these fermion fields microscopically (but see ideas in Appendices~\ref{app:TSC}~and~\ref{app:2wire}).
Instead, we follow Son and postulate transformation properties of the fields under all microscopic symmetries.
This constitutes in principle a complete specification of the theory and allows in particular identification of low-energy field contributions to any observable.

Son's theory is formulated as a Dirac theory at a finite chemical potential.
Here we will focus on fields residing near the Fermi surface where the Fermi level cuts through the Dirac cone.
We encode the Dirac cone nature in the symmetry transformation properties of the fermions near the Fermi surface.
We do this rather than keeping the two-component Dirac field, since this approach translates immediately to the quasi-1D setting.

The Dirac Hamiltonian $v(\sigma^1 k_x + \sigma^2 k_y) - \mu$ has eigenergies $\pm v|\bk| - \mu$.
Assuming $\mu > 0$, the low-energy excitations come from the ``+''~branch and reside near the Fermi surface at $|\bk| = k_F = \mu/v$.
We take the corresponding eigenvector as
$\frac{1}{\sqrt{2}} \begin{pmatrix} 1 \\ e^{i\alpha_\bk} \end{pmatrix}$,
where $\alpha_\bk$ denotes an angle formed by the vector $\bk$ with the $k_x$-axis.
This fixes our choice of the phases of the wavefunctions for different $\bk$, which in turn fixes the transformation properties of the low-energy fields residing near the Fermi surface (note in particular discussion of the electronic particle-hole symmetry below).
Specifically, we can expand Son's two-component Dirac fermion field $\Psi_{\rm CF}(\br)$ in terms of the low-energy fields residing near the Fermi surface,
\begin{equation}
\Psi_{\rm CF}(\br) \sim \sum_{\bk~\textrm{near the Fermi surface}}
\begin{pmatrix} 1 \\ e^{i\alpha_\bk} \end{pmatrix}
e^{i \bk \cdot \br} f_\bk(\br) + \dots ~,
\end{equation}
where $f_\bk(\br)$ is a slowly varying field describing fermions near a Fermi surface patch at $\bk$, and we have also omitted contributions from high-energy fields.
The electronic quantum Hall problem has spatial symmetries and the anti-unitary particle-hole symmetry (in the lowest Landau level) described in App.~\ref{app:cylinder}.
We have in mind the same cylinder geometry in the Landau gauge and focus only on symmetries that are present in this geometry.
For all symmetries, we use Son's postulated transformations of the Dirac field $\Psi_{\rm CF}(\br)$ to obtain transformation properties of the fields near the Fermi surface $f_\bk(\br)$ defined above.

In the quasi-1D setting where the cylinder is infinite in the $x$-direction but finite in the $y$-direction, the low-energy fermions reside near Fermi points where the ``wires'' at discrete $k_y$ cut through the 2D Fermi surface, cf.~Fig.~\ref{fig:bc}.
Let $j$ label these Fermi points; $\bk_j$ is the wavevector at $j$; and $v_j \equiv v_x(\bk_j)$ is the group velocity along the cylinder and can be positive or negative corresponding to right- or left-moving fermions.
The kinetic energy of the fermions is
\begin{equation}
H_{f, {\rm kin.}} = \sum_j \int dx f_j^\dagger(x) (-i v_j \partial_x) f_j(x) ~.
\label{Hkin}
\end{equation}

As described in App.~\ref{app:cylinder}, the electronic model on the cylinder in the chosen gauge is invariant under arbitrary translations $\Delta y$ in the $y$-direction.
It is also invariant under a discrete translation $\Delta x = \ell_B^2 (2\pi/L_y)$ in the $x$-direction supplemented by a gauge transformation.
When we say that we work with a long-wavelength field carrying momentum $\bk$, we implicitly understand the following transformation properties under the above symmetries:
\begin{eqnarray}
&& T_y[\Delta y]: f_\bk \to e^{i k_y \Delta y} f_\bk ~, \\
&& T_x[\Delta x]: f_\bk \to e^{i k_x \Delta x} f_\bk ~.
\end{eqnarray}

The microscopic model has the anti-unitary mirror symmetry combining mirror $(x, y) \to (x, -y)$ and time reversal, which we denoted $M_x \TR$ in App.~\ref{app:cylinder}.
Son postulates its action on the two-component Dirac field as $M_x \TR: \Psi_{\rm CF}(x, y) \to \sigma^3 \Psi_{\rm CF}(x, -y)$ (note that he denotes this symmetry as ``$\mathcal{P}\TR$,'' while we preserve notation from App.~\ref{app:cylinder}).
Translated to our long-wavelength fields near the 2D Fermi surface, it reads 
\begin{equation}
M_x \TR: f_{(k_x, k_y)} \to f_{(-k_x, k_y)} ~, 
\quad i \to -i ~.
\end{equation}
In particular, this symmetry implies that the Fermi surface is invariant under reflections in the $k_y$-axis and $v_x(-k_x, k_y) = -v_x(k_x, k_y)$.

The physics in the 2D system is invariant under spatial rotations.
However, as in App.~\ref{app:cylinder}, on the cylinder we only have $180$ degree rotation symmetry left, which is the same as spatial inversion $\br \to -\br$.
It transforms the Dirac fermions as $I: \Psi_{\rm CF}(\br) \to \sigma^3 \Psi_{\rm CF}(-\br)$, hence for the long-wavelength fields near the Fermi surface
\begin{equation}
I: f_\bk \to f_{-\bk} ~.
\end{equation}
This symmetry implies that the Fermi surface is invariant under inversions in $k$-space and $v_x(-\bk) = -v_x(\bk)$.  
Combined with the $M_x \TR$ symmetry, we see that four Fermi points $(\pm k_x, \pm k_y)$ are symmetry-related, with group velocities satisfying $v_x(k_x, k_y) = v_x(k_x, -k_y) = -v_x(-k_x, k_y) = -v_x(-k_x, -k_y)$.
This justifies our implicit assumptions about the composite fermion Fermi surface in the main text, cf.~Fig.~\ref{fig:bc}.

\subsection{Particle-hole symmetry and absence of \twokF backscattering}

We now consider the anti-unitary particle-hole symmetry present in the lowest Landau level at $\nu = 1/2$.
Son postulates its action on the Dirac composite fermions as $\PH: \Psi_{\rm CF}(\br) \to -i\sigma^2 \Psi_{\rm CF}(\br)$, which translates for the long-wavelength fields near the Fermi surface as
\begin{equation}
\PH: f_\bk \to e^{i\alpha_\bk} f_{-\bk} ~, 
\quad i \to -i ~.
\end{equation}
(Note that Son denotes this symmetry as ``$\CC\TR$'', while we preserve label ``$\PH$'' used everywhere else in our paper.)
Note in particular that since $\alpha_{-\bk} = \alpha_\bk + \pi$, applying this symmetry two times takes any $f_\bk$ to $-f_\bk$, i.e., $\PH$ squares to $-1$ when acting on an odd number of fermions.  
This connects with our discussion of formal aspects of particle-hole symmetry in the main text and App.~\ref{app:PH}.

Furthermore, we observe that
\begin{equation}
\PH: A f_\bk^\dagger f_{-\bk} \to -(A f_\bk^\dagger f_{-\bk})^\dagger ~
\end{equation}
for any complex coefficient $A$. 
This has an immediate consequence that particle-hole-even observables cannot have contributions from such ``$2k_F$'' fermion bilinears.
In the context of the surface of the 3D TI, similar considerations of a Dirac cone protected by time reversal symmetry imply absence of exact back-scattering by time-reversal-invariant perturbations.
While the DMRG setup allows us to measure only properties of the ground state, we can nevertheless detect such physics by studying correlation functions of particle-hole-even observables and noting the absence of the corresponding $2k_F$ features, which we loosely refer to as the absence of the exact back-scattering.
As we will see shortly, such $2k_F$ bilinears are enhanced and are prominent operators in the CFL theory, so their absence in the particle-hole-even observables is a dramatic feature.
(Note that we do not claim complete absence of any singularity at $2k_F$. One can construct particle-hole-even contributions carrying such momenta, but only using more fermion fields, e.g., by multiplying $f_\bk^\dagger f_{-\bk}$ by a particle-hole-odd operator carrying zero momentum such as $f_{\bk'}^\dagger f_{\bk'} - f_{-\bk'}^\dagger f_{-\bk'}$.
Such higher-order contributions are of course less important than bilinears would be.)

\subsection{Bosonized quasi-1D theory, gauge fluctuations, Amperean-enhanced \twokF bilinears, and density structure factor at long wavelengths}
\label{subapp:bosonizedQ1D}

The postulated Fermi points and the above symmetries fix the quadratic part of the Hamiltonian to the form in Eq.~\eqref{Hkin} with the symmetries of the Fermi surface as described above.
Anticipating eventual multi-mode Luttinger liquid description in the quasi-1D limit, we want to use hydrodynamic bosonization approach to represent the fermionic system, and this will also allow us to readily treat gauge fluctuations.
For the discussion of the particle-hole symmetry, it is convenient to group Fermi points at $\bk$ and $-\bk$ into right- and left-moving fields.
Let $m$ label such a pair of fields, which form together one gapless mode; henceforth $m$ labels such ``modes'' before including effects of gauge field fluctuations in the CFL theory.
We denote the right-mover of this pair as $j = Rm$ and left-mover as $Lm$, and bosonize
\begin{equation}
	f_{Pm} = \eta_m e^{i (\phi_m + P \theta_m)} ~,
	\text{~for~} P = R/L = {+/-} ~,
\end{equation}
with canonically conjugate boson fields $[\phi_m(x), \theta_{m'}(x')] = i\pi \delta_{m, m'} \Theta(x-x')$, where $\Theta(x)$ is the Heaviside step function.
Here $\eta_m$ are Klein factors taken to be Majorana fermions, $\{\eta_m, \eta_{m'} \} = 2\delta_{m, m'}$, which ensure that all fermion fields anticommute.
The slowly varying fermionic densities are simply
\begin{equation}
\rho_{f, Pm} \equiv f_{Pm}^\dagger f_{Pm} = \partial_x (P \phi_m + \theta_m)/(2\pi) ~.
\end{equation}
The kinetic energy in Eq.~\eqref{Hkin} now has sum over cumulative index $j = Pm$, with $v_{Rm} = -v_{Lm} \equiv v_m$ (and we assume $v_m > 0$).
[Note that, alternatively, we could have grouped points $(k_x, k_y)$ and $(-k_x, k_y)$ into a right- and left-moving pair, since these also have exactly opposite group velocities along the cylinder.
The physics does not depend on such choices, but the discussion of the particle-hole symmetry is slightly simpler with the made choice grouping $\bk$ and $-\bk$ used from here on.]

Working in the imaginary time path integral, a bosonized Lagrangian corresponding to such $H_{f, {\rm kin}}$ in Eq.~\eqref{Hkin} is
\begin{equation}
\mathcal{L}_{f, {\rm kin}} = \sum_m \frac{v_m}{2\pi} \left[(\partial_x \theta_m)^2 + (\partial_x \phi_m)^2 \right] + \frac{i}{\pi} \partial_x \theta_m \partial_\tau \phi_m ~.
\end{equation}
In the CFL theory (either HLR or Dirac-CFL), the fermions are coupled to a dynamical gauge field, whose net effect in the quasi-1D system is to pin the overall ``gauge charge'' mode, thus reducing the total number of gapless modes by one.\cite{KimLee99, Sheng2008_2legDBL, Sheng2009_zigzagSBM}
Indeed, working in the (1+1)D space-time continuum theory in the gauge with only temporal component $a_\tau$ of the gauge field, the Lagrangian for the gauge field has an ``energy'' term proportional to $(\partial_x a_\tau)^2$ and is coupled to the fermions via $i a_\tau \rho_{f, {\rm tot}}$, where $\rho_{f, {\rm tot}} \equiv \sum_{Pm} \rho_{f, Pm} = \sum_m \partial_x \theta_m/\pi$.
Integrating out the field $a_\tau$ produces a mass term for the overall gauge charge mode $\theta_{f, {\rm tot}} \equiv \frac{1}{\sqrt{N_w}} \sum_m \theta_m$, where $N_w$ is the number of $R$-$L$ pairs (same as the number of wires).
The remaining gapless modes can be obtained as linear combinations of $\theta_m$ that are orthogonal to $\theta_{f, {\rm tot}}$, but we will not need details of these for the general discussion below.

One consequence of the above physics is that fermion bilinears that involve transfer from right-movers to left-movers are enhanced compared to the ``free-fermion'' theory.
Indeed, consider such a bilinear
\begin{equation}
f_{Rm}^\dagger f_{Lm'} \sim \eta_m \eta_{m'}
e^{-i [\phi_m - \phi_{m'} + \theta_m + \theta_{m'}]} ~.
\end{equation}
The combination of the $\theta$ fields in the exponent has a component onto $\theta_{f, {\rm tot}}$, and since the latter is pinned, the fluctuating content in the exponent is reduced.
This is quasi-1D manifestation of so-called Amperean enhancement of the $2k_F$ bilinears in the 2D gauge theory: CF particle and hole from the opposite sides of the Fermi surface have parallel gauge currents which experience Amperean attraction.
It explains our DMRG observations in the main text and App.~\ref{app:more_data} where bilinears (and also quartic terms) involving such transfers from right-movers to left-movers are prominent in the structure factors.


Finally, we consider the behavior of the density structure factor at $q_y = 0$ and small $q_x$.
Consider an operator in the long-wavelength theory
$\partial_x [\sum_j k_{j, y} f_j^\dagger(x) f_j(x)]$,
where $k_{j, y}$ is the $y$-component of the Fermi momentum $\bk_j$ at Fermi point $j$.
This operator can be thought of as a derivative of a local momentum in the $y$-direction (i.e., local piece of the conserved $K_y$).
Since according to Eq.~\eqref{Kychain}, $K_y$ also gives the $x$-component of the total electron polarization operator on the cylinder, we see that the above operator gives effectively the electron density at $q_y = 0$ (i.e., averaged over the circumference of the cylinder).
[We remark that while it is easy to verify that the proposed operator indeed has correct transformation properties under $I$, $\PH$, and $M_x {\cal T}$, it is important to have the above connection to a microscopically conserved current expressed in the long-wavelength theory.
Note also that unlike the 2D, there is no gapless gauge field left in the quasi-1D system, and the above contribution from the CF fields appears to be the dominant one.] 
We can now calculate contribution to the density structure factor and obtain $D(q_x, q_y = 0) \sim |q_x|^3$ at small $q_x$.
This is a very weak singularity, where only the third derivative becomes discontinuous.
The density structure factors measured in the DMRG do not show any features at $q_y = 0$ and small $q_x$, cf.~Figs.~\ref{fig:cone},~\ref{fig:two_body},~\ref{fig:boulder}, and~\ref{fig:two_body16}, which is consistent with such weakness of the singularity.

\subsection{Stability of the Dirac CFL phase and irrelevance of particle-hole-breaking perturbations}
\label{subapp:DCFLstability}

Besides the fermion kinetic energy, the full theory also includes quartic fermion interactions (terms with more fermion fields or with derivatives are expected to be less important).
The momentum-conserving quartic terms can be divided into two groups.

The first group is forward scattering interactions
\begin{equation}
F_{\bk, \bk'} f_\bk^\dagger f_\bk f_{\bk'}^\dagger f_{\bk'} \sim F_{Pm, P'm'} \rho_{f, Pm} \rho_{f, P'm'} ~,
\end{equation}
with real-valued $F_{Pm, P'm'}$ as required by the Hermiticity.
Under the anti-unitary mirror symmetry $\rho_{f, Pm} \to \rho_{f, -P, -m}$ (using convention where Fermi point $-P, -m$ has the same $k_y$ as $Pm$ but opposite $k_x$), while under the spatial inversion $\rho_{f, Pm} \to \rho_{f, -P, m}$.
We can easily write down corresponding conditions on $F_{Pm, P'm'}$.
The particle-hole acts on $\rho_{f, Pm}$ identically to the inversion, and since $F_{Pm, P'm'}$ are real-valued, it does not introduce new conditions.
In any case, the forward scattering interactions are strictly marginal and only modify Luttinger parameters but cannot gap out the modes.

The second group is Cooper (or backscattering) chanel interactions,
\begin{equation}
V_{\bk, \bk'} f_\bk^\dagger f_{-\bk}^\dagger f_{-\bk'} f_{\bk'} \sim V_{m, m'} f_{Rm}^\dagger f_{Lm}^\dagger f_{Lm'} f_{Rm'} ~,
\end{equation}
with in general complex-valued $V_{m, m'}$.
The Hermiticity imposes $V_{m', m} = V_{m, m'}^*$.
The anti-unitary mirror symmetry imposes $V_{-m, -m'} = V_{m, m'}^*$, while the inversion symmetry is automatically satisfied for any $V_{m, m'}$.  
On the other hand, the anti-unitary particle-hole symmetry imposes additional conditions 
$V_{m, m'} = V_{m, m'}^* e^{-i 2(\alpha_\bk - \alpha_{\bk'})}$, where $\bk = \bk_{Rm}$ is the wavevector of the right-mover in our $m$-th mode, and similarly for $\bk'$.
Thus, the particle-hole symmetry essentially fixes the complex phase of $V_{m, m'}$ to be $e^{-i (\alpha_\bk - \alpha_{\bk'})}$.

Focusing on the CFL phase, we can make several interesting predictions.  
First, we note that the Cooper channel interactions can be all irrelevant.
For example, in the theory with only the bare $\mathcal{L}_{f, {\rm kin}}$ supplemented by an infinitely strong gapping out condition on the overall gauge charge mode, i.e., $\theta_{f, {\rm tot}} = {\rm const}$, and no forward scattering interactions, the Cooper channel interactions are always irrelevant.  
We can see this by noting that such interactions bosonize to
\begin{equation}
V_{m, m'} f_{Rm}^\dagger f_{Lm}^\dagger f_{Lm'} f_{Rm'} 
\sim V_{m, m'} e^{-i 2(\phi_m - \phi_{m'})} ~.
\end{equation}
When $\theta_{f, {\rm tot}}$ is pinned, the fluctuations of the $\phi$ fields are increased compared to the free fermion case and the scaling dimensions of the $V_{m,m'}$ terms are increased.
This is a quasi-1D manifestation of the suppression of the Cooper channel by the Amperean interaction effects in the gauge theory: oppositely oriented currents repel, hence two oppositely moving fermions in a Cooper pair repel and do not like to form such a pair.
This stability extends also over a finite range of forward scattering interactions, as long as all $V_{m, m'}$ terms remain irrelevant.

The second observation is that if we start with a particle-hole-symmetric and stable fixed point, it means that the scaling dimensions of operators $e^{-i 2(\phi_m - \phi_{m'})}$ are all greater than 2.
The fact that in the particle-hole-symmetric case $V_{m, m'}$ have definite phases does not affect such stability considerations.
Hence we predict that introducing explicit particle-hole symmetry breaking will only add irrelevant perturbations that do not destabilize the phase.
This is consistent with our numerical DMRG observations on the cylinders where we broke the particle-hole symmetry by allowing tunneling to another layer and observed that the central charge of the gapless phase remained unchanged.
We also expect such stability to hold in the 2D Dirac CFL as well.
The absence of the particle-hole symmetry microscopically does manifest itself in observables even though the system flows to a fixed point with emergent particle-hole symmetry, since there are no microscopic restrictions for observables to pick up components of either formally particle-hole-even or odd combinations in the fixed point theory.
Nevertheless, the emergence of an effective particle-hole symmetry even when one is not present microscopically makes our numerical studies of strictly particle-hole-symmetric models only more important, as a more clear way to access the fixed point theory of the general CFL phase.

\section{Exactly solvable model corresponding to two-wire Dirac CFL}
\label{app:2wire}

It is instructive to consider a descendant phase with only two wires cutting through the Fermi sea, which can happen for small $L_y$.
We do not find this phase in the model with the screened Coulomb interactions used in the main text; instead, the system appears to develop a charge density wave for $L_y \lesssim 8$.
The two-wire CFL phase does occur for different more short-range interactions and in fact was discovered by Bergholtz and Karlhede in Ref.~\onlinecite{Bergholtz2005}.
However, their interpretation does not connect with the CFL picture, and we also point out the particle-hole symmetry aspect.

Bergholtz and Karlhede studied the following electronic model, formulated in the same orbital basis as in Appendix~\ref{app:cylinder}:
\begin{eqnarray}
H = \sum_j [V_{10} n_j n_{j+1} + V_{20} n_j n_{j+2}
- V_{21}(c_j^\dagger c_{j+1} c_{j+2} c_{j+3}^\dagger + \Hc) ] ~,
\label{HBK}
\end{eqnarray}
with real and positive $V_{10}$, $V_{20}$, and $V_{21}$.
The above form taken from Ref.~\onlinecite{Bergholtz2005} uses different conventions from our Eq.~\eqref{eq:Helint}, but this is not important.
Of all the electron interaction terms, this model keeps only nearest- and next-nearest-neighbor repulsions plus the simplest four-fermion term that does not reduce to density-density interaction, and is a natural model near so-called thin torus limit.
Focusing on the half-filled case, Bergholtz and Karlhede observed that if one considers a subspace spanned by all configurations where each pair of sites $(2J, 2J + 1)$ has precisely one electron, the above Hamiltonian acts within this subspace.
They also showed that, in the regime of interest here, the global ground state resides in this subspace.
(More precisely, there are two exactly degenerate ground states, and the second one is obtained from the first one by a translation by one lattice spacing, see also discussion in Apps.~\ref{app:cylinder}-\ref{app:numerical}.)
Working in this subspace, we can associate a spin-1/2 degree of freedom with each pair of sites $(2J, 2J + 1)$, where $10$ and $01$ configurations correspond to spin up and down respectively.  
We can then identify spin operators as $S_J^z \equiv n_{2J} - 1/2 = 1/2 - n_{2J + 1}$, $S_J^+ = c_{2J}^\dagger c_{2J + 1}$, and the Hamiltonian acting in this subspace is simply an XXZ chain,
\begin{equation}
H = \sum_J [(2 V_{20} - V_{10}) S_J^z S_{J + 1}^z + V_{21} (S_J^+ S_{J + 1}^- + \Hc)] ~.
\end{equation}
Note that $J$ here refers to the pair $(2J, 2J + 1)$ of the original electronic orbitals, so the spacing between the ``sites'' in the spin chain is twice that in the electronic chain.
For $2 V_{21} > |2 V_{20} - V_{10}|$, the spin chain is in a critical phase with one gapless mode.
The dominant power law correlations are staggered along the spin chain,
$\langle S_J^+ S_{J'}^- \rangle \sim (-1)^{J-J'}/|J - J'|^{1/(2g)}$ and 
$\langle S_J^z S_{J'}^z \rangle \sim (-1)^{J-J'}/|J - J'|^{2g}$,
where $g$ is Luttinger parameter varying between $1$ for the XX chain and $1/2$ for the XXX chain.
There is also a non-oscillating $\sim 1/|J-J'|^2$ contribution to the $\langle S_J^z S_{J'}^z \rangle$ correlation, whose power is fixed by the conservation of total $S^z$.

We now translate these results to the electrons.  
First, we point out that the $S^z$ correlations correspond to the electron density correlations at $q_y = 0$, while the $S^+$ correlations correspond to the density correlations at $q_y = 2\pi/L_y$, cf.~Apps.~\ref{app:cylinder}-\ref{app:numerical}.
Furthermore, since each spin site corresponds to two electronic sites, the staggered correlations in the spin chain correspond to correlations at wavevector $\pi/2$ in the electronic chain, in units where the spacing between orbitals in Eq.~\eqref{HBK} is $1$.
In units used in the main text, this wavevector corresponds to $Q = L_y/4$.
This immediately allows us to make interpretation in terms of a two-wire CFL with wires at $k_y = \pi/L_y$ and $-\pi/L_y$.
Indeed, by the Luttinger theorem discussed in the main text, the length of each wire must be $Q_{1/2} = L_y/4$, which is precisely the above $Q$.
Furthermore, our analysis in the main text predicts singularities in the density structure factor at $(q_x, q_y) = (Q_{1/2}, 0)$ (right to left transfer within a wire) and $(Q_{1/2}, 2\pi/L_y)$ (right to left transfer between the wires), in agreement with the spin chain solution.
Also, from the microscopic expression (in chain units) $\sum_j (n_j - 1/2) e^{-i q_x j} = (1 - e^{-iq_x}) \sum_J S_J^z e^{-i q_x 2J}$, we can readily deduce that the electron structure factor at $q_y = 0$ and small $q_x$ behaves as $|q_x|^3$, in agreement with the prediction at the end of App.~\ref{subapp:bosonizedQ1D} for the general quasi-1D case.
Finally, the exact ground state resides in the sector with the $0110$ root configuration,\cite{Bergholtz2005} which is what we found for all cases with even number of wires (APBC).
Thus, we can match properties of the exact solution with those expected from the two-wire Dirac-CFL.

It is instructive to identify the microscopic symmetries of the electronic problem discussed in App.~\ref{app:cylinder} with symmetries of the above spin chain.
This will also lead us to a more microscopic identification of the composite fermions.
The anti-unitary mirror symmetry Eq.~\eqref{MxTchain} translates to
\begin{equation}
M_x \TR: S_J^z \to S_J^z ~,~ \quad S_J^+ \to S_J^+ ~,~ \quad S_J^- \to S_J^- ~,~ 
\quad i \to -i ~,
\end{equation}
and hence acts like a ``boson time reversal'' when the spin model is interpreted as a hard-core boson model.

For the spatial inversion, we consider inversion in the bond center (midpoint between two neighboring orbitals), Eq.~\eqref{Iprimechain}.
This preserves the above restricted Hilbert space and becomes in the spin model
\begin{equation}
I': S_J^z \to -S_{-J}^z ~,~ \quad S_J^+ \to S_{-J}^- ~,~ \quad S_J^- \to S_{-J}^+ ~.
\end{equation}

Most interestingly, the anti-unitary particle-hole symmetry Eq.~\eqref{eq:PHorb} becomes 
\begin{equation}
\PH: S_J^z \to -S_J^z ~,~ \quad S_J^+ \to -S_J^- ~,~ \quad S_J^- \to -S_J^+ ~,
\end{equation}
i.e., $\PH: \vec{S}_J \to -\vec{S}_J$, which acts like familiar spin time reversal.

We can now guess what the composite fermions are in the quasi-1D descendant Dirac-CFL theory in this case.
They can be viewed as arising from a slave particle description of the spin chain, where we write $S_J^+ = f_{J\up}^\dagger f_{J\dn}$, with a constraint $f_{J\up}^\dagger f_{J\up} + f_{J\dn}^\dagger f_{J\dn} = 1$ on each site, and postulate a mean field where the spinons $f_\up$ and $f_\dn$ hop independently.
Since $S_J^+$ carries transverse momentum $2\pi/L_y$, we can interpret $f_\up$ as carrying momentum $k_y = \pi/L_y$ and $f_\dn$ as carrying $-\pi/L_y$, similar to the two-wire picture with APBC in the transverse direction.
Furthermore, we can implement all of the above symmetries in terms of the spinons and match these with the proposed transformation properties in the quasi-1D Dirac-CFL in App.~\ref{app:DCFLquasi1D}.
In particular, we can implement the anti-unitary particle-hole symmetry in the familiar way as time reversal symmetry on the spinons, $f_{J\up} \to f_{J\dn}, f_{J\dn} \to -f_{J\up}$, which immediately gives us the Kramers doublet physics.
[More precisely, one needs to modify the spinon right- and left-moving fields with appropriate phase factors to match with the transformation properties of the Dirac-CFL fields.
When matching, one needs to remember that the inversion symmetry $I'$ in this Appendix is relative to a mid-point between two orbitals, while in App.~\ref{app:DCFLquasi1D} the inversion $I$ is relative to the origin which coincides with one orbital; $I$ and $I'$ are related by translation by one orbital.]

Interestingly, since microscopically $S_J^+ = c_{2J}^\dagger c_{2J + 1}$ and we are considering subspace with $c_{2J}^\dagger c_{2J} + c_{2J+1}^\dagger c_{2J+1} = 1$, we can interpret the spinons as the original fermions in the LLL orbitals on the cylinder that develop spontaneous coherence along the chain independently on the even and odd sublattices, writing schematically $f_{J\up} \sim c_{2J}$, $f_{J\dn} \sim c_{2J + 1}$.  
Note that while microscopically the $\PH$ maps $c_{2J} \to c_{2J}^\dagger$ and $c_{2J + 1} \to c_{2J + 1}^\dagger$, in the considered subspace this action coincides with the above familiar time reversal transformation of the spinons.
While the above is special to the two-wire case, it would be interesting to find a microscopic derivation of the quasi-1D Dirac-CFL fields also for cases with more wires.

We can also examine what happens when we break the particle-hole symmetry and the interplay with the other symmetries.  
For example, we can consider a perturbation
\begin{equation}
\delta H = \sum_j v_6 (c_j^\dagger c_{j+1} c_{j+2} c_{j+3}^\dagger + \Hc)(1 - n_{j-1} - n_{j+4}) ~,
\end{equation}
which preserves the anti-unitary mirror and the inversion symmetries but breaks the particle-hole symmetry.  
This perturbation also acts in the above restricted Hilbert space and becomes in the spin model
\begin{equation}
\delta H = \sum_j v_6 (S_J^+ S_{J + 1}^- + \Hc)(S_{J+2}^z - S_{J-1}^z) ~.
\end{equation}
We can understand its effects using bosonization.  
Let us use a hydrodynamic description of the XXZ chain where the $S^z$ spin component is represented as $S_J^z = \partial_x \theta/\pi + A (-1)^J \sin(2\theta)$ and the bond energy $[J, J+1]$ is represented as $B (-1)^J \cos(2\theta)$.
In these variables, additional $S^z S^z$ interactions in the XXZ chain that can gap it out (a.k.a.\ ``umklapp'' terms in half-filled boson or fermion chains) contribute $\lambda \cos(4\theta)$;
this term of course respects all the symmetries of the XXZ chain, but in the critical phase it must be irrelevant.
Now we see that when we break the particle-hole symmetry by the above $\delta H$, it contributes a term $\lambda' \sin(4\theta)$, which must also be irrelevant as long as the unperturbed critical phase is stable.
This observation connects with our discussion of particle-hole symmetry conditions on for four-fermion terms in the general Dirac-CFL theory in App.~\ref{subapp:DCFLstability} (with different bosonization conventions between the two sections, since in App.~\ref{subapp:DCFLstability} we effectively grouped right spin-up with left spin-down fields).

As another example, we can consider a perturbation 
\begin{eqnarray}
\delta H' = \sum_j v_6' (c_j^\dagger c_{j+1} c_{j+2} c_{j+3}^\dagger + \Hc)(1/2 - n_{j-1}) = -\sum_j v_6' (S_J^+ S_{J+1}^- + \Hc) S_{J-1}^z ~.
\end{eqnarray}
This perturbation preserves the anti-unitary mirror symmetry but breaks both the inversion and particle-hole symmetry.  
In the spin model, we see that the effect of this perturbation is roughly similar to adding a field in the $S^z$ direction on the XXZ spin chain.
The effect on the correlations is that the $S^+$ correlations remain staggered, while the $S^z$ correlations are shifted to an incommensurate wavevector.
This is consistent with developing an imbalance between the spin-up and spin-down spinons---correspondingly, between the $k_y = \pi/L_y$ and $-\pi/L_y$ wires, which is expected when both the inversion and particle-hole symmetries are broken.
We thus see that the Bergholtz and Karlhede model realizing the simplest quasi-1D descendant of the CFL provides useful playground for examining the interplay of symmetries in the half-filled Landau level problem.

\end{widetext}
\end{document}